%% file: main.tex
\documentclass[12pt,a4paper]{article}
\usepackage[a4paper,top=3cm,bottom=3cm,left=2cm,right=3cm,bindingoffset=5mm]{geometry}
\usepackage[bottom]{footmisc}

\usepackage{jheppub}

\usepackage{xcolor}

\usepackage[utf8]{inputenc}

\usepackage{amsmath}
\usepackage{amsthm}
\usepackage{amssymb}
\usepackage{physics}
\usepackage{comment}
\usepackage{graphicx}
\usepackage{float}
\usepackage{hyperref}
\usepackage{amsfonts}
\usepackage{mathrsfs}
\usepackage{xcolor}

\usepackage{simpler-wick}
\usepackage{mathtools}
\usepackage{makecell}
\usepackage{tabu}

\hypersetup{
colorlinks=true,         
linkcolor=blue,          
citecolor=red,        
urlcolor=blue            
}

\newcommand{\bz}{{\overline{z}}}
\newcommand{\bw}{{\overline{w}}}
\newcommand{\bn}{{\overline{n}}}
\newcommand{\bu}{{\overline{u}}}

\newcommand{\bbz}{{\bar{z}}}
\newcommand{\bbw}{{\bar{w}}}
\newcommand{\bbn}{{\bar{n}}}
\newcommand{\bbu}{{\bar{u}}}

\newcommand{\bell}{\overline{\ell}}
\newcommand{\bL}{\overline{\mathsf{L}}}
\newcommand{\mL}{\mathcal{L}}
\newcommand{\bmL}{\overline{\mathcal{L}}}
\newcommand{\bZ}{{\bar{Z}}}
\newcommand{\sL}{\mathsf{L}}

\newcommand{\bPhi}{\bar{\Phi}}
\newcommand{\bphi}{\bar{\phi}}
\newcommand{\mT}{\mathcal{T}}
\newcommand{\bt}{\mathbf{t}}
\newcommand{\bT}{\mathbf{T}}
\newcommand{\mR}{\mathcal{R}}
\newcommand{\mX}{\mathcal{X}}
\newcommand{\sW}{W}
\newcommand{\sF}{F}
\newcommand{\ww}{\text{w}}
\newcommand{\Bell}{\bar{\ell}}

\DeclareFontFamily{U}{jkpmia}{}
\DeclareFontShape{U}{jkpmia}{m}{it}{<->s*jkpmia}{}
\DeclareFontShape{U}{jkpmia}{bx}{it}{<->s*jkpbmia}{}
\DeclareMathAlphabet{\mathfrakalt}{U}{jkpmia}{m}{it}
\SetMathAlphabet{\mathfrakalt}{bold}{U}{jkpmia}{bx}{it}

\newcommand{\myw}{\mathfrakalt{w}}

\newcommand{\Lw}{\text{L}\myw_{1+\infty}}
\newcommand{\Ls}{\text{L}\mathfrak{s}}

\newcommand{\bigPhi}{\raisebox{-0.1\baselineskip}{\Large\ensuremath{\Phi}}}
\newcommand{\bigphi}{\raisebox{-0\baselineskip}{\Large\ensuremath{\phi}}}

\title{Spacetime $\Lw$ Symmetry and Self-Dual Gravity in Plebański Gauge}

\author{Noah Miller}

\affiliation{School of Natural Sciences, Institute for Advanced Study,
1 Einstein Drive, Princeton, NJ 08540 USA}
\affiliation{Princeton Gravity Initiative, Princeton University,
Jadwin Hall, Washington Road, Princeton NJ 08544, USA}

\emailAdd{noahmiller@ias.edu}

\preprint{}

\abstract{
The space of self-dual Einstein spacetimes in 4 dimensions is acted on by an infinite dimensional Lie algebra called the $\Lw$ algebra. In this work we explain how one can ``build up'' self-dual metrics by acting on the flat metric with an arbitrary number of infinitesimal $\Lw$ transformations, using a convenient choice of gauge called Plebański gauge. We accomplish this through the use of something called a ``perturbiner expansion,'' which will perturbatively generate for us a self-dual metric starting from an initial set of quasinormal modes called integer modes. Each integer mode corresponds to a particular $\Lw$ transformation, and this perturbiner expansion of integer modes will be written as a sum over ``marked tree graphs,'' instead of momentum space Feynman diagrams.

We find that a subset of the $\Lw$ transformations act as spacetime diffeomorphisms, and the algebra of these diffeomorphisms is $\myw_{\infty} \ltimes \mathfrak{f}$. We also show all analogous results hold for the $\Ls$ algebra in self-dual Yang Mills.
}

\begin{document}

\maketitle

\section{Introduction and main idea}

In the last few years, there has been a great deal of interest in self-dual gravity and its relation to the so-called $\Lw$ symmetry algebra. The purpose of this paper is to explain how the $\Lw$ algebra acts on 4-dimensional self-dual spacetime metrics, everywhere in the bulk, in a particular gauge we'll call ``Plebański gauge.'' In order to explain what we mean by all of this, let us first back up and explain what self-dual gravity is, what the $\Lw$ algebra is, what Plebański gauge is, and why people have become interested in this subject as of late.

``Self-dual gravity'' (SDG) refers to a subsector of Einstein gravity in which the Riemann curvature tensor is required to be self-dual when regarded as a 2-form:
\begin{equation}\label{eq1}
    R_{\mu \nu \rho \sigma} = \frac{i}{2} \epsilon_{\mu \nu \alpha \beta} R^{\alpha \beta}_{\;\;\;\; \rho \sigma}.
\end{equation}
Any metric which satisfies \eqref{eq1} automatically obeys the vacuum Einstein equation $R_{\mu \nu} = 0$. In $(1,3)$ signature, \eqref{eq1} can only hold for a non-flat metric if the metric is complex, while in $(2,2)$ or $(0,4)$ signature the metric can be real. Because the Hodge star sends $2$-forms to $(d-2)$-forms, SDG can only be defined in $d = 4$ spacetime dimensions. SDG can intuitively be understood as a theory of gravity which only contains positive helicity gravitons, with no negative helicity gravitons.

SDG is well known to be an integrable system \cite{mason1990h, mason1989connection, mason1996integrability, dunajski2009solitons, hitchin2013integrable, ward1990twistor, dunajski19982d, dunajski1998nonlinear}, is the main object of study of twistor theory \cite{penrose1976nonlinear, penrose1976nonlinear2}, and also arises in the low energy effective field theory of the N = 2 string \cite{Ooguri:1991fp, Ooguri:1990ww, Ooguri:1991ie}. The existence of this integrable theory underlying 4-dimensional Einstein gravity implies the existence of many miraculous formulae within Einstein gravity itself, such as the simplicity of the all-multiplicity graviton MHV amplitude \cite{Mason:2009afn, Adamo:2021bej, mypaper}.

Twistor theorists long ago observed that self-dual spacetimes could be acted on, in a natural way, by an infinite dimensional Lie algebra called the $\Lw$ algebra \cite{penrose1976nonlinear2,boyer1985infinite,Popov:1996uu}. The $\Lw$ algebra deforms the complex structure of the twistor space associated to spacetime which ultimately has the effect of slightly perturbing the spacetime metric $g_{\mu \nu} \mapsto  g_{\mu \nu} + \delta g_{\mu \nu}$ in some way. Roughly speaking, any self-dual spacetime can be produced by acting with a finite $\Lw$ transformation on Minkowski space (modulo issues of topology change, boundary conditions at infinity, or convergence).

Recently, the $\Lw$ algebra has become a source of renewed interest due to its appearence in the field of celestial holography \cite{Ball:2021tmb,Mason:2022hly,Himwich:2021dau,Bittleston:2022jeq,Mago:2021wje,Bu:2024cql,Bhardwaj:2024wld,Melton:2024jyq,Himwich:2023njb,Taylor:2023ajd,Guevara:2023tnf,Banerjee:2023jne,Ren:2023trv,Bu:2022iak,Bittleston:2023bzp,Bittleston:2024rqe,Costello:2022wso,Costello:2022upu,Lipstein:2023pih,Pano:2023slc,Crawley:2023brz,Crawley:2021auj,Banerjee:2023zip,Hu:2022txx,Monteiro:2022xwq,Melton:2022fsf,Monteiro:2022nqt,Mol:2024etg,Mol:2024qct,Guevara:2022qnm,Schwarz:2022dqf,Monteiro:2022lwm,Ahn:2022oor,Bu:2021avc,Adamo:2021zpw,Jiang:2021csc,Banerjee:2021dlm,Bogna:2024gnt,Guevara:2024vlc,Adamo:2025mqp,Bu:2024wnf}. In this context, the algebra was initially encountered unexpectedly in the context of non-self-dual tree-level scattering amplitudes in Einstein gravity, emerging from the holomorphic OPEs of positive helicity gravitons \cite{Guevara:2021abz,Strominger:2021mtt}. When two positive helicity gravitons in a scattering amplitude are brought close together (in a holomorphic way), the collinear splitting function reproduces a Kac-Moody-current-like OPE formula where the color group of the Kac-Moody symmetry is the $\myw_{1+\infty}$ algebra.

The appearance of the $\Lw$ algebra in \cite{Guevara:2021abz,Strominger:2021mtt} seemed to be entirely unrelated to its original twistorial incarnation, but a later work by Adamo, Mason, and Sharma \cite{Adamo:2021lrv} bridged the gap, explaining the connection between the two. This appearance of the $\myw_{1+\infty}$ algebra in the splitting function was also observed earlier in the self-dual sector within the context of the double copy by Monteiro and O’Connell in \cite{Monteiro:2011pc}. See also \cite{Doran:2023cmj,Monteiro:2022nqt,Monteiro:2022lwm,Chacon:2020fmr,Armstrong-Williams:2022apo,Cheung:2022mix,Monteiro:2013rya}.

The reason that this discovery prompted a flurry of research into the $\Lw$ algebra is as follows. A main goal of the celestial holography program has been to try to reconstruct gravitational scattering amplitudes on a 4d flat background using some kind of 2d boundary theory, where asymptotic symmetries and soft theorems must play a fundamental role \cite{Kapec:2016jld,Donnay:2018neh,Strominger:2017zoo}. However, soft theorems alone (or even sub-leading soft theorems, or sub-sub-leading soft theorems) cannot on their own constrain scattering amplitudes strongly enough to reconstruct the whole amplitude. Because the $\Lw$ algebra turns out to be much larger than the analogous algebras constructed out of only the leading soft theorems, and there is a general sense that this symmetry algebra should be powerful enough to perhaps constrain a self-dual theory uniquely in some way. This can more or less be understood as the overarching goal of the subject currently, and any deeper understanding of the $\Lw$ algebra potentially gets us closer to that goal. In fact, celestial duals utilizing self-duality have already been constructed in different modified contexts, such as in \cite{Costello:2022wso,Bittleston:2024efo,Costello:2023vyy,Costello:2022jpg,Adamo:2021bej,Melton:2024akx,Costello:2023hmi,Bu:2024wnf,Bu:2023vjt,Bu:2022dis}, so the idea seems to be a fruitful one.

While the action of this algebra is well understood at the level of twistor space and at the level of scattering amplitudes, it is more difficult to understand how this algebra acts on the spacetime metric itself. In principle one can simply act with the algebra on twistor space and then project that action down onto spacetime, although this is a somewhat painful procedure. Nevertheless, this was accomplished on linearized fields at null infinity $\mathcal{I}^\pm$ in a work by Donnay, Freidel, and Herfray \cite{Donnay:2024qwq}. The expression they get for the action on the metric is non-local, in the sense that it involves inverse derivatives along $\mathcal{I}^\pm$. A different approach for writing the action on linearized fields at $\mathcal{I}^\pm$ was also undertaken by Freidel, Pranzetti, and Raclariu in a series of works \cite{Freidel:2021dfs,Freidel:2021ytz,Freidel:2023gue} where a set of canonical charges charges for the symmetry algebra was proposed. It was later understood that these charges are valid only in the self-dual sector \cite{Geiller:2024bgf,Kmec:2024nmu}, and their relation to twistor theory was clarified in \cite{Kmec:2024nmu}. See also \cite{Mason:2023mti,Cresto:2024fhd,Cresto:2024mne,Compere:2022zdz} for other $\Lw$ spacetime investigations.

The goal of this present paper is to answer the basic question of how the $\Lw$ algebra acts on the spacetime metric everywhere in the bulk, i.e. beyond linearized fields at null infinity, and without using twistor theory at an intermediate step.

\begin{figure}
    \centering
    \input{figures/pic_that_goes_hard.tex}
    \caption{Here we depict the space of all self-dual spacetimes, with the special point $g_{\mu \nu} = \eta_{\mu \nu}$, corresponding to $\phi = 0$, depicted. One can then repeatedly linearly perturb the metric by $\Lw$ transformations and, in principle, get to any other SD metric through this procedure.}
    \label{fig:mainpic}
\end{figure}
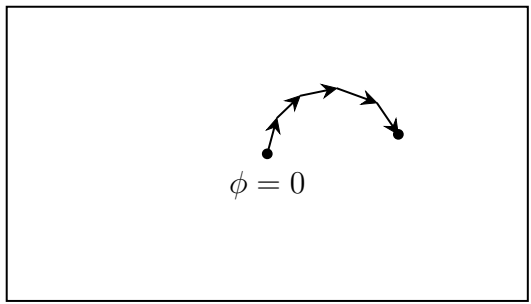

Here is what we accomplish, more precisely. Imagine one starts with the Minkowski metric $g_{\mu \nu} = \eta_{\mu \nu}$ and then performs one particular infinitesimal $\Lw$ transformation to it, linearly perturbing the metric. One can then take this metric and linearly perturb it again by a second infinitesimal $\Lw$ transformation. One can then take \textit{this} metric and linearly perturb again, etc., as depicted in figure \ref{fig:mainpic}. We will give an exact formula for the metric at the $N^{\rm th}$ $\Lw$ perturbation of the Minkowski metric, for any $N$, explaining how the metric can be varied as many times as desired. Note however that we really are working with the full non-linear theory of SDG in this set-up, as any metric can be built out of an arbitrary number of linear perturbations in the sense of a path-ordered exponential. That is, the Lie algebra action should in principle be able to be exponentiated to construct finite transformations of the group for which $\Lw$ is the Lie algebra.

Note, then, that this work is answering a somewhat different question than previous works. In particular, we are not providing a formula for how one can act a $\Lw$ transformation on pre-existing metric data, but are instead giving an iterative procedure for how one can act $\Lw$ repeatedly on an initially-flat metric, outputting an arbitrary SD metric at the end. One benefit we achieve by reframing the task in this way is that we never need to perform a large-$r$ asymptotic expansion of the metric, and our final expression for the metric holds everywhere in the bulk.

In order to tackle this subject, we will use a formulation of SDG which is much easier to work with than \eqref{eq1}. In 1975, Jerzy Plebański showed that for any SD metric, there always exists a coordinate system $(u, \bbu, w, \bbw)$ in which the metric can be written as \cite{Plebanski1975some} 
\begin{equation}\label{eq2}
    ds^2 = 4 \big( d u \, d \bbu - d w \, d \bbw + (\partial_{w}^2 \phi ) \, d \bbu^2 + (2 \partial_u \partial_w \phi) \, d \bbu d \bbw + (\partial_u^2 \phi) \, d \bbw^2  \big)
\end{equation}
where $\phi$ is a scalar field that satisfies an equation of motion known as ``Plebański's second heavenly equation''
\begin{equation}\label{pleb_1}
    \Box \phi - \{ \partial_u \phi, \partial_w \phi \} = 0.
\end{equation}
Here, $\Box$ is the flat space wave operator and $\{ \cdot, \cdot \}$ is a spacetime Poisson bracket which acts on the two spacetime coordinates $u, w$, as
\begin{equation}\label{pois_def_1}
    \Box \equiv \partial_u \partial_\bu - \partial_w \partial_\bw, \hspace{1 cm} \{ f, g \} \equiv \pdv{f}{u} \, \pdv{g}{w} - \pdv{f}{w} \, \pdv{g}{u}.
\end{equation}
When $\phi = 0$ the metric reduces to the vacuum state, i.e. flat Minkowski space. An important non-flat exact solution to \eqref{pleb_1} is given by $\phi = C e^{i p \cdot X}$ for $p^2 = 0$ and any $C$. This solution should be thought of as a finite-amplitude positive-helicity gravitational wave of momentum $p^\mu$. Of course, as \eqref{pleb_1} is non-linear, the sum of two such plane waves would not also solve the equations of motion.

The equation of motion \eqref{pleb_1} explicitly has a linear ``free'' part and a non-linear ``interacting'' part. The free part of the equation is just the scalar wave equation, implying that linear perturbations away from the vacuum can be characterized by a basis of solutions to the wave equation. Plane waves provide one such basis, but there are others as well. In particular, in a recent paper \cite{Cotler:2023qwh} we found a particular basis of solutions to the wave equation we called the ``integer mode basis,'' which is particularly well suited for studying $\Lw$ symmetry in SDG. Elements of the integer mode basis, denoted as $\phi^\Delta_{\bn, n}$, transform under discrete representations of the Lorentz algebra $\mathfrak{sl}(2,\mathbb{C})$ and satisfy $\Box \, \phi^\Delta_{\bn, n} = 0$. $\Delta$, $\bbn$, and $n$ are independent integer labels. (We will give the exact expression for $\phi^\Delta_{\bn, n}$ later.) Therefore, a particular linear perturbation of the vacuum will be of the form
\begin{equation*}
    \phi = 0 \hspace{0.5 cm} \longmapsto \hspace{0.5 cm} \phi = \epsilon_1 \phi^{\Delta_1}_{\bn_1, n_1}
\end{equation*}
where $\epsilon_1$ is an infinitesimal number. Because $(\epsilon_1)^2 = 0$, the above expression for $\phi$ solves \eqref{pleb_1}.

\begin{figure}
    \centering
    \input{figures/pic_that_goes_hard2}
    \caption{The commutator of our action on the space of self-dual metrics exactly reproduces the $\Lw$ commutation relation.}
    \label{fig:mainpic2}
\end{figure}

What if we now want to add a second linear perturbation? This can be accomplished by adding another linear solution to the wave equation, as well as a second term required to solve the equation of motion. Schematically,
\begin{equation*}
    \phi = \epsilon_1 \phi^{\Delta_1}_{\bn_1, n_1} \hspace{0.5 cm} \longmapsto \hspace{0.5 cm} \phi = \epsilon_1 \phi^{\Delta_1}_{\bn_1, n_1} + \epsilon_2 \phi^{\Delta_2}_{\bn_2, n_2} + \epsilon_1 \epsilon_2 (\ldots)
\end{equation*}
where the exact form of $(\ldots)$ will be given later. We could also add a third perturbation which would roughly look like
\begin{equation*}
    \longmapsto \hspace{0.4 cm} \phi = \epsilon_1 \phi^{\Delta_1}_{\bn_1, n_1} + \epsilon_2 \phi^{\Delta_2}_{\bn_2, n_2} + \epsilon_3 \phi^{\Delta_3}_{\bn_3, n_3} + \epsilon_1 \epsilon_2 (\ldots)  + \epsilon_1 \epsilon_3 (\ldots) + \epsilon_2 \epsilon_3 (\ldots) + \epsilon_1 \epsilon_2 \epsilon_3 (\ldots).
\end{equation*}
This sort of expansion is known in the literature as a ``perturbiner expansion'' \cite{Monteiro:2011pc,Mizera:2018jbh,Boulware:1968zz,Kim:2023qbl}.

Now, crucially, the perturbiner expansion in integer modes we described above depends in a particular way on the \textit{order} in which the modes are added. In other words, if we have some SD metric, add $\phi^{\Delta_1}_{\bn_1, n_1}$ to it, and then add $\phi^{\Delta_2}_{\bn_2, n_2}$ to it, we will end up with a different final metric than if we added them in the opposite order, as shown in figure \ref{fig:mainpic2}. This implies that this algebra has a non-zero commutator on the space of self-dual spacetimes, and this turns out to be nothing more than the $\Lw$ commutation relation!

In this way, we arrive at the following interpretation for the $\Lw$ algebra: it is the algebra of all possible self-dual metric perturbations, where these linear perturbations are expressed in a convenient basis of scalar integer modes.

This paper depends heavily on the results of another paper by the author \cite{mypaper}, which solved for the perturbiner expansion of Plebański's second heavenly equation using a set of combinatorial objects known as ``marked tree graphs,'' which are different from Feynman diagrams. In that paper, these tree graphs were used to calculate the all-multiplicity graviton MHV amplitude. This present paper, while not interested in MHV amplitudes, will nonetheless use many of the results of \cite{mypaper}. We provide a ``reader's manual'' for \cite{mypaper} in appendix \ref{app_manual} so that the reader can more easily extract the results and proofs of \cite{mypaper} which are relevant to this paper.

With an explicit formula for the action of the $\Lw$ algebra in hand, we then provide two applications of the formula. In the first application, we explicitly show what subset of the $\Lw$ transformations in Plebański gauge are pure diffeomorphisms in spacetime, and give the expressions for the corresponding vector fields. Perhaps surprisingly, we find that the algebra of diffeomorphisms is equivalent to a semidirect product algebra ${\myw}_{\infty} \ltimes \mathfrak{f}$, where $\mathfrak{f}$ is a certain abelian algebra we will define later. The ${\myw}_{\infty}$ part can be thought of as a generalization of anti-holomorphic superrotations, and the $\mathfrak{f}$ part can be thought of as a generalization of the anti-holormorphic supertranslations. These pure-diffeo transformations, in the perturbiner framework, can be understood as the scattering of one extra particle with a linearized diff wavefunction of the form $h_{\mu \nu} = \partial_{\mu} \xi_{\nu} + \partial_{\nu} \xi_{\mu}$. The full change in the perturbiner expansion from adding this particle, with all interaction terms included, ends up reducing to $\mathcal{L}_\xi g_{\mu \nu}$. (In addition to explaining which $\Lw$ transformations are pure diffeo, we also show that some of the $\Lw$ transformations do not act on the metric at all in Plebański gauge.) In our second application, we explain how the algebra interacts with a mathematical object associated with the integrability of SDG, called the ``recursion operator'' $\mR$. In particular, we give a spacetime account of how this recursion operator generates the ``$\text{L}$'' part of $\Lw$, as was originally studied in twistor space by Dunajski and Mason \cite{Dunajski:2000iq, dunajski1996heavenly}. See also \cite{Kmec:2024nmu}.

As a warm up, we begin the paper by studying self-dual Yang Mills theory (SDYM) instead of self-dual gravity, which is an integrable subsector of Yang Mills theory that possesses its own infinite dimensional symmetry algebra called the $\Ls$ algebra \cite{Popov:1998pc,Mansfield:2009ra,Adam:2008jx,Schiff:1992wg,Popov:1995qb,Grant:2008fr,la1992symmetries,Papachristou:1990fw,Papachristou:1991au,Papachristou:2008kg}. (Aside from being a warm-up for SDG, the study of SDYM is interesting in its own right \cite{atiyah1977instantons, atiyah1994construction}. It is in some sense the ``only'' classical integrable system as all other known classical integrable systems, including SDG, are dimensional reductions of it \cite{ward1985integrable,mason1990h, mason1989connection}.)

This paper is organized as follows, where analogous sections for SDYM and SDG are given in parallel.
\begin{itemize}
    \item In sections \ref{sec21} and \ref{sec31} we review some common formulations of SDYM and SDG.
    \item In sections \ref{sec22} and \ref{sec32} we review the definitions of the $\Ls$ and $\Lw$ algebras.
    \item In sections \ref{sec23} and \ref{sec33} we review the definitions of the spin-1 and spin-2 conformal primary modes, the half-descendant modes, and the full-descendant ``integer modes.'' 
    \item In sections \ref{sec24} and \ref{sec34} we review the perturbiner expansions of the Chalmers-Siegel scalar in SDYM and the perturbiner expansion of the Plebański scalar in SDG. See \eqref{psi_k} and \eqref{gravity_perturbiner}. The SDYM perturbiner is written as a sum over color ordering permutations, while the SDG perturbiner is written as a sum over marked tree graphs. 
    \item In sections \ref{sec25} and \ref{sec35} we use perturbiners to write the action of the $\Ls$ and $\Lw$ algebras on the Chalmers-Siegel and Plebański scalars, respectively, and confirm that these actions reproduce the correct commutation relations. See equations \eqref{proposed_action_1}, \eqref{Ls_alg_confirmation} and \eqref{lw_action_phi}, \eqref{check_lw_comm}. 
    \item In section \ref{sec26} we discuss the residual gauge symmetry of SDYM in Chalmers-Siegel gauge, and in section \ref{sec36} we discuss the residual diffeomorphism symmetry of SDG in Plebański gauge.
    \item In section \ref{sec27} and \ref{sec37} we explain which descendants of the conformal primary modes are pure gauge and pure diffeo in the spin-1 and spin-2 cases. We also discuss some descendants that vanish identically.
    \item In section \ref{sec28} and \ref{sec38} we explicitly show that some of our spacetime $\Ls$ and $\Lw$ actions exactly reduce to residual gauge transformations and diffeomorphisms of Chalmers-Siegel and Plebański gauge, respectively. These transformations arise when the integer mode seed functions are linearized pure gauge / pure diffeo. We also show that some of the $\Ls$ and $\Lw$ algebra elements do not act on the Chalmers-Siegel and Plebański scalars. Their corresponding seed functions vanish identically.
    \item In section \ref{sec29} and \ref{sec39} we discuss the action of the recursion operator $\mR$ in SDYM and SDG. $\mR$ generates the loop part of the loop algebras.
\end{itemize}

Finally, let us discuss the relation of this paper to another previous work. In \cite{Campiglia:2021srh}, Campiglia and Nagy studied a certain set of spacetime symmetries of SDG, also using Plebański's second heavenly equation, and explored these symmetries in the context of the double-copy with SDYM. In particular, they solved for the set of residual diffeomorphisms of Plebański gauge and abstractly discussed the action of the recursion operator $\mR$. These residual diffeomorphisms were also discussed in \cite{Dunajski:2000iq}. In a sense, our paper can be thought of as a follow-up to Campiglia and Nagy's paper where we use perturbiners to explicitly solve for the transformations they studied more abstractly. Also see \cite{Nagy:2022xxs}.

In this work we use $(1,3)$ signature with the $(+,-,-,-)$ convention. This means that, writing our metric as $g_{\mu \nu} = \eta_{\mu \nu} + h_{\mu \nu}$, the perturbation $h_{\mu \nu}$ will be complex.

\section{SDYM and the $\Ls$ algebra}

\subsection{Equations of motion for SDYM}\label{sec21}

For a finite dimensional Lie algebra $\mathfrak{g}$ with generators $\bT^{a} \in \mathfrak{g}$, consider a $\mathfrak{g}$-valued connection 1-form $A_\mu$ with curvature tensor
\begin{equation}
    F_{\mu \nu} = \partial_\mu A_\nu - \partial_\nu A_\mu + [A_\mu, A_\nu].
\end{equation}
The equation of SDYM is $\star F = i F$, which can also be expressed as\footnote{We define the anti-symmetric pseudo-tensor to be $\varepsilon_{\mu \nu \rho \sigma} \equiv \sqrt{-g} [\mu \nu \rho \sigma]$, $\varepsilon^{\mu \nu \rho \sigma} = \frac{-1}{\sqrt{-g}} [\mu \nu \rho \sigma]$, with the symbol $[\mu \nu \rho \sigma]$ being the totally antisymmetric object $[0 1 2 3] = 1$.}
\begin{equation}\label{self_duality_F}
    F_{\mu \nu} = \frac{i}{2} \varepsilon_{\mu \nu \rho \sigma} F^{\rho \sigma}.
\end{equation}
Instead of using the standard flat space Minkowski coordinates $(X^0, X^1, X^2, X^3)$, let us use a more convenient set of convenient coordinates $(u, \bbu, w, \bbw)$ called ``lightcone coordinates,''
\begin{equation}
\begin{aligned}
    u &= (X^0 - X^3)/2 \, , \\
    \bbu &= (X^0 + X^3)/2 \, , \\
    w &= (X^1 + i X^2)/2\, ,  \\
    \bbw &= (X^1 - i X^2)/2\, .
\end{aligned}
\end{equation}
In these coordinates the flat metric is
\begin{equation}
    ds^2 = 4 d u \, d \bbu - 4 d w \, d \bbw.
\end{equation}
Notice that $u$ and $\bbu$ are both real independent coordinates with $\bbu \neq (u)^*$. If we were to analytically continue to $(2,2)$ signature via $X^3 \to i X^3$, then $\bbu$ really would be the complex conjugate of $u$.

In lightcone coordinates, $\varepsilon_{u\bu w\bw} = 4i$ and \eqref{self_duality_F} reduces to
\begin{equation}\label{three_sdym_eq}
    F_{u w} = 0, \hspace{0.5 cm} F_{u \bu} = F_{w\bw}, \hspace{0.5 cm} F_{\bu \bw} = 0.
\end{equation}
If we use lightcone gauge $A_u = 0$, then $F_{uw} = 0$ implies $A_w = 0$. $F_{u \bu} = F_{w \bw}$ then implies $\partial_u A_\bu = \partial_w A_\bw$ which can be solved by
\begin{equation}
    A_\bu = \partial_{w} \Phi, \hspace{1 cm} A_{\bw} =  \partial_u \Phi,
\end{equation}
where $\Phi$ is a $\mathfrak{g}$-valued scalar called the Chalmers-Siegel scalar. We have therefore shown that a self-dual connection can be expressed in the form
\begin{equation}\label{A_Phi}
    A[\Phi]_\mu = \begin{pmatrix} A[\Phi]_u \\ A[\Phi]_\bu \\ A[\Phi]_w \\A[\Phi]_\bw \end{pmatrix} \equiv \begin{pmatrix}
        0 \\ \partial_w \Phi \\ 0 \\ \partial_u \Phi
    \end{pmatrix}
\end{equation}
which we'll refer to as ``Chalmers-Siegel gauge.'' Note that the harmonic gauge condition $\partial^\mu A[\Phi]_\mu = 0$ is also satisfied.

Note that if a function of $(\bbu,\bbw)$ alone is added to $\Phi$, the connection $A[\Phi]_\mu$ will not change, so we call such functions ``trivial.''

The only equation from \eqref{three_sdym_eq} we haven't yet used is
\begin{equation}\label{eom_siegel}
\begin{aligned}
    F_{\bu \bw} = \Box\, \Phi - [\partial_u \Phi, \partial_w \Phi ] = 0
\end{aligned}
\end{equation}
where $\Box$ was defined in \eqref{pois_def_1}. This is an equation of motion for $\Phi$, and can also be derived from the ``Chalmers-Siegel action'' \cite{Chalmers:1996rq,Parkes:1992rz}
\begin{equation}\label{Ssdym}
    S_{\rm SDYM}( \Phi, \bPhi ) = \int d^4 x \Tr( \bPhi \big( \Box \Phi - [\partial_u \Phi, \partial_w \Phi] \big) )
\end{equation}
where $\bPhi$ is a Lagrange multiplier that enforces the e.o.m.\! \eqref{eom_siegel}. The e.o.m.\! for $\bPhi$ is
\begin{equation}
\begin{aligned}
    \Box \bPhi - [\partial_u \bPhi, \partial_w \Phi] - [\partial_u \Phi, \partial_w \bPhi] = 0
\end{aligned}
\end{equation}
which one may notice is also the equation for a linearized perturbation of $\Phi$, meaning $\Phi + \epsilon \, \bPhi$ also solves \eqref{eom_siegel} to the first order in $\epsilon$. Furthermore, while $\bPhi$ has been introduced here as an unphysical Lagrange multiplier, it turns out that it should really be thought of as a degree of freedom corresponding to a linearized anti-self-dual (ASD) perturbation on the non-linear SD background. For example, if a single linearized ASD gluon is added to a SD background $\Phi$, then one can show that ASD part of the field strength tensor $F_{\mu \nu}^-$ is proportional to $\bPhi$. (See C.44 in \cite{mypaper}.)

There is also a second formulation of SDYM due to Yang \cite{Yang:1977zf}. Using the first and third equations of \eqref{three_sdym_eq}, we see that our curvature is flat in the $uw$ plane and in the $\bbu \bbw$ plane. This implies that there exist matrices $J$ and $\widetilde{J}$ such that
\begin{equation}\label{AJ}
\begin{aligned}
    A_u = \widetilde{J}^{-1} \partial_u \widetilde{J}, \hspace{0.5cm} A_\bu = J^{-1} \partial_\bu J, \hspace{0.5cm} A_w = \widetilde{J}^{-1} \partial_w \widetilde{J}, \hspace{0.5cm} A_\bu = J^{-1} \partial_\bu J.
\end{aligned}
\end{equation}
We are then free to choose a gauge where we set $\widetilde{J} = 0$, which again is lightcone gauge $A_u = A_w = 0$. The not-yet-used equation $F_{u \bu} = F_{w \bw}$ becomes the constraint
\begin{equation}\label{Jeom}
\begin{aligned}
    \partial_u ( J^{-1} \partial_\bu J ) - \partial_w (  J^{-1} \partial_\bw J) = 0.
\end{aligned}
\end{equation}
$J$ is called the ``Yang $J$-matrix'' and the above equation is its equation of motion.

\subsection{Abstract definitions of the $\mathfrak{s}$ and $\Ls$ algebras}\label{sec22}

Here we review the definition of the $\mathfrak{s}$-algebra, its wedge subalgebra $\mathfrak{s}_{\wedge} \subset \mathfrak{s}$, and the associated loop algebra $\Ls$.

If the generators $\bT^a \in \mathfrak{g}$ have the commutation relations
\begin{equation}
    [\bT^a, \bT^b] = f^{abc} \, \bT^c
\end{equation}
then we can abstractly define the $\mathfrak{s}$-algebra as the tensor product of $\mathfrak{g}$ with polynomials in two variables $(u,w)$, where the polynomials commute with the $\mathfrak{g}$-bracket. The generators of this algebra $\text{s}^{\Delta, a}_{\bn} \in \mathfrak{s}$ can be expressed as
\begin{equation}\label{s_monomial}
    \text{s}^{\Delta, a}_{\,\bn} \equiv \mathbf{T}^a u^{1 - \Delta - \bn} w^{\bn}
\end{equation}
and the commutation relations can straightforwardly be computed to be\footnote{If we make the substitutions $\Delta = - 2 q + 3$, $\bbn = -m -\frac{\Delta-1}{2}$, $n_{\rm here} = - n_{\rm there} - \frac{\Delta + 1}{2}$, we recover the conventions of \cite{Melton:2022fsf}.}
\begin{equation}\label{s_alg}
    [\text{s}^{\Delta_1, a}_{\,\bn_1}, \text{s}^{\Delta_2, b}_{\,\bn_2} ] = f^{a b c} \text{s}^{\Delta_1 + \Delta_2 - 1, c}_{\,\bn_1 + \bn_2}.
\end{equation}
Here $\Delta, \bbn \in \mathbb{Z}$. If $\Delta$ and $\bbn$ both range over all integers, these generators are contained within $\mathfrak{s}$.
\begin{equation}
    \text{s}^{\Delta, a}_{\, \bn} \in \mathfrak{s} \hspace{0.5 cm} \text{if} \hspace{0.25 cm} \Delta,  \bbn \in \mathbb{Z}.
\end{equation}

When $\Delta \leq 1$ and $1-\Delta \geq \bbn \geq 0$, the monomials \eqref{s_monomial} have non-negative degree and we say our generators are elements wedge subalgebra $\mathfrak{s}_{\wedge} \subset \mathfrak{s}$.
\begin{equation}
    \text{s}^{\Delta, a}_{\, \bn} \in \mathfrak{s}_{\wedge} \hspace{0.5 cm} \text{if} \hspace{0.25 cm} \Delta \leq 1 ,\;\;\; 0 \leq \bbn \leq 1-\Delta.
\end{equation}

Furthermore, it is possible to ``loopify'' any Lie algebra by simply appending an extra integer onto the generators and summing said integers together in the commutation relation (and adding +1 in the conventions of this paper). In particular, the loop algebra of $\mathfrak{s}$ is denoted $\Ls$, and its generators are denoted $\text{s}^{\Delta, a}_{\; \bn, n}$ where we have appended the integer index $n$ to $\text{s}^{\Delta, a}_{\; \bn}$.
\begin{alignat}{3}
    \text{s}^{\Delta, a}_{\; \bn, n} &\in \Ls & \hspace{1 cm} & \Delta, \bbn, n \in \mathbb{Z}
\end{alignat}
We define the commutation relations of $\Ls$ to be
\begin{align}\label{Lscomm}
    [\text{s}^{\Delta_1, a}_{\,\bn_1,n_1 }, \text{s}^{\Delta_2, b}_{\,\bn_2,n_2} ] &\equiv f^{a b c} \text{s}^{\Delta_1 + \Delta_2 - 1, c}_{\,\bn_1 + \bn_2, \; \; n_1 + n_2 + 1}.
\end{align}
Notice that we are using a non-standard convention for the loop algebra commutator, as $n_1 + n_2+1$ appears on the RHS of the above expression instead of the usual $n_1 + n_2$. We do this for later convenience, as with this choice acting with $\text{s}^{\Delta, a}_{\; \bn, n}$ on our classical solution will add in a gluon with wave function $\Phi^{\Delta, a}_{\; \bn, n}$.

\subsection{Spin-1 conformal primary modes, half-descendants,  and full-descendants of the Chalmers-Siegel scalar}\label{sec23}

Because SDYM contains only a single $\mathfrak{g}$-valued scalar degree of freedom $\Phi$, we are interested in bases of solutions to the free scalar wave equation. In this section we will define three related bases of solutions: the conformal primary modes, the half-descendant modes, and the full-decendant modes.

The first set, the ``conformal primary mode'' functions \cite{Pasterski:2017kqt,Pasterski:2016qvg}, are given by
\begin{equation}\label{conf_prim}
    \Phi^{\Delta, a}_{\bz, z}(X) \equiv \bT^a (q(\bbz, z) \cdot X)^{1 - \Delta}.
\end{equation}
These functions are parameterized by a single integer $\Delta \in \mathbb{Z}$, two independent complex numbers $\bbz, z \in \mathbb{C}$, and a color index $a$. $q(\bbz, z)$ is a null vector defined, in raised Minkowski coordinates, by
\begin{equation}
\begin{aligned}
    q^\mu(\bbz, z) &\equiv (q^0(\bbz, z), q^1(\bbz, z), q^2(\bbz, z), q^3(\bbz, z)) \\
    &= \frac{1}{2}(1 + z \bbz, z + \bbz, - i(z - \bbz), 1 - z \bbz).
\end{aligned}
\end{equation}
In lowered lightcone coordinates the components are
\begin{align}
    (q_u(\bbz, z), q_\bu(\bbz, z), q_w(\bbz, z), q_\bw(\bbz, z)) = (1, z \bbz, -\bbz , -z)
\end{align}
meaning that if we use the $X^\mu = (u, \bbu, w, \bbw)$ lightcone coordinates, then
\begin{align}
    q(\bbz, z) \cdot X = u + z \bbz \bbu - \bbz w - z \bbw.
\end{align}

The conformal primary modes transform in a special way under Lorentz transformations. The $\mathfrak{so}(1,3)$ Lorentz generators can be rewritten as $\mathfrak{sl}(2,\mathbb{C})$ generators as follows. If $j_i, k_i \in \mathfrak{so}(1,3)$, for $i=1,2,3$, are the usual $4 \times 4$ rotation and boost generators with commutation relations
\begin{equation}
    [j_i, j_j] = \epsilon_{ijk} j_k, \hspace{1 cm} [j_i, k_j] = \epsilon_{ijk} k_k, \hspace{1 cm} [k_i, k_j] = - \epsilon_{ijk} j_k ,
\end{equation}
then the combinations
\begin{equation}\label{sl2c_generators}
  \begin{split}
    \ell_0  &= \frac{1}{2} (-k_3 - i j_3)\,, \\
    \ell_{1} &= \frac{1}{2} (-k_1 + j_2 - i (k_2 + j_1) )\,,  \\
    \ell_{-1} &= \frac{1}{2} (k_1 + j_2 - i (k_2 - j_1) )\,,
  \end{split}
\quad \quad \quad
  \begin{split}
    \Bell_0  &= \frac{1}{2} (-k_3 + i j_3)\,,  \\
    \Bell_{1} &= \frac{1}{2} (-k_1 + j_2 + i (k_2 + j_1 ) )\,,  \\
    \Bell_{-1}  &= \frac{1}{2} (k_1 + j_2 + i (k_2 - j_1) )\,,
  \end{split}
\end{equation}
satisfy
\begin{equation}
    [\ell_m, \ell_n] = (m-n) \ell_{m+n}\,, \hspace{1 cm} [\Bell_m, \Bell_n] = (m-n) \Bell_{m+n}\,,  \hspace{1 cm} 
    [\ell_m, \Bell_n] = 0\,,
\end{equation}
for $m, n = -1, 0, 1$.

In $(u,\bbu,w,\bbw)$ coordinates, the matrices $\ell_n$ and $\Bell_n$ are
\begin{equation}\label{ell_matrices}
\begin{aligned}
    (\ell_0)_{\mu}^{\;\;\; \nu} = \frac{1}{2} \begin{pmatrix} 
    -1 & 0 & 0 & 0 \\ 
    0 & 1 & 0 & 0 \\ 
    0 & 0 & -1 & 0 \\
    0 & 0 & 0 & 1
    \end{pmatrix}, \hspace{0.25 cm} (\ell_1)_{\mu}^{\;\;\; \nu} =  \begin{pmatrix} 
    \,0\, & \,0\, & \,0\, & \,1\, \\ 
    \,0\, & \,0\, & \,0\, & \,0\, \\ 
    \,0\, & \,1\, & \,0\, & \,0\, \\
    \,0\, & \,0\, & \,0\, & \,0\,
    \end{pmatrix}, \hspace{0.25 cm}
    (\ell_{-1})_{\mu}^{\;\;\; \nu} =  \begin{pmatrix} 
    0 & 0 & 0 & 0 \\ 
    0 & 0 & -1 & 0 \\ 
    0 & 0 & 0 & 0 \\
    -1 & 0 & 0 & 0
    \end{pmatrix}, \\
    (\bell_0)_{\mu}^{\;\;\; \nu} = \frac{1}{2} \begin{pmatrix} 
    -1 & 0 & 0 & 0 \\ 
    0 & 1 & 0 & 0 \\ 
    0 & 0 & 1 & 0 \\
    0 & 0 & 0 & -1
    \end{pmatrix}, \hspace{0.25 cm} (\bell_1)_{\mu}^{\;\;\; \nu} = \begin{pmatrix} 
    \,0\, & \,0\, & \,1\, & \,0\, \\ 
    \,0\, & \,0\, & \,0\, & \,0\, \\ 
    \,0\, & \,0\, & \,0\, & \,0\, \\
    \,0\, & \,1\, & \,0\, & \,0\,
    \end{pmatrix}, \hspace{0.25 cm}
    (\bell_{-1})_{\mu}^{\;\;\; \nu} =  \begin{pmatrix} 
    0 & 0 & 0 & 0 \\ 
    0 & 0 & 0 & -1 \\ 
    -1 & 0 & 0 & 0 \\
    0 & 0 & 0 & 0
    \end{pmatrix}.
\end{aligned}
\end{equation}

The spacetime vector fields corresponding to these matrices are
\begin{equation}
    \sL_n \equiv (\ell_n)_\mu^{\;\; \nu} X^\mu \partial_\nu \, , \hspace{1 cm} \bL_n \equiv (\Bell_n)_\mu^{\;\; \nu} X^\mu \partial_\nu \,,
\end{equation}
with Lie brackets
\begin{equation}
    [\sL_m, \sL_n] = (m-n) \sL_{m+n}\,, \hspace{1 cm} [\bL_m, \bL_n] = (m-n) \bL_{m+n}\,,  \hspace{1 cm} 
    [\sL_m, \bL_n] = 0\,,
\end{equation}
and in coordinates read
\begin{equation}\label{list_lorentz_gen}
\begin{aligned}
    \sL_0 =  \frac{1}{2} ( - u \partial_u +  \bbu \partial_\bu - w \partial_w + \bbw \partial_{\bw}) \, , & & & & \sL_1 = w \partial_\bu + u \partial_{\bw} \, , & & & &
    \sL_{-1} = -\bbw \partial_u - \bbu \partial_w \, , \\
    \bL_0 = \frac{1}{2}(- u \partial_u +  \bbu \partial_\bu + w \partial_w- \bbw \partial_{\bw}) \, ,& & & &
    \bL_1 = \bbw \partial_\bu + u \partial_{w} \, , & & & &
    \bL_{-1} = -w \partial_u - \bbu \partial_{\bw}\, .
\end{aligned}
\end{equation}
We can also define the associated Lie derivatives $\mL_n$ and $\bmL_n$, which act on $A_\mu$ by
\begin{equation}\label{lie_der_A}
\begin{aligned}
    \mL_n A_\mu &= (\ell_n)_\mu^{\;\;\; \alpha} A_\alpha + \sL_n A_\mu \, ,\\
    \bmL_n A_\mu &= (\Bell_n)_\mu^{\;\;\; \alpha} A_\alpha + \bL_n A_\mu \, . \\
\end{aligned}
\end{equation}
These satisfy
\begin{equation}
    [\mL_m, \mL_n] = (m-n) \mL_{m+n}\,, \hspace{1 cm} [\bmL_m, \bmL_n] = (m-n) \bmL_{m+n}\,,  \hspace{1 cm} 
    [\mL_m, \bmL_n] = 0\,.
\end{equation}

The special property that the conformal primary modes \eqref{conf_prim} 
 satisfy in Chalmers-Siegel gauge is that they transform exactly like 2d conformal primaries with weights $h = \frac{\Delta+1}{2}$, $\bar{h} = \frac{\Delta-1}{2}$, up to a pure-gauge term for $\mL_1$:
\begin{align}
    \mL_n \, A[\Phi^{\Delta,a}_{\,\bz,z}]_\mu &= \left( \frac{\Delta +1}{2} (n+1) z^n + z^n \partial_z \right) A[\Phi^{\Delta,a}_{\,\bz,z}]_\mu  - \delta_{n,1}  \partial_\mu \Phi^{\Delta,a}_{\,\bz,z} \, , \\
    \bmL_n \, A[\Phi^{\Delta,a}_{\,\bz,z}]_\mu &= \left( \frac{\Delta -1}{2} (n+1) \bbz^n + \bbz^n \partial_\bz \right) A[\Phi^{\Delta,a}_{\,\bz,z}]_\mu \, .
\end{align}
Now, because Chalmers-Siegel gauge \eqref{A_Phi} breaks Lorentz symmetry, a priori we should not expect that acting with $\mL_n$ or $\bmL_n$ on $A[\Phi]_\mu$ should be equivalent to acting with $\sL_n$ and $\bL_n$ on $\Phi$. However, as we explain in appendix \ref{app_lorentz}, 4 out of 6 of the Lorentz generators are preserved in this gauge, namely $\bmL_{-1}$, $\bmL_0$, $\bmL_1$, and $\mL_{-1}$. In particular, both $\mL_{-1}$ and $\bmL_{-1}$ are preserved, meaning
\begin{equation}
\begin{aligned}
    \mL_{-1} A[\Phi]_\mu = A[ \sL_{-1}  \Phi ]_\mu, \hspace{1 cm} \bmL_{-1} A[\Phi]_\mu = A[  \bL_{-1} \Phi ]_\mu.
\end{aligned}
\end{equation}
This means $\sL_{-1}$ and $\bL_{-1}$ simply act by differentiation of $\partial_z$ and $\partial_\bz$ on $\Phi^{\Delta,a}_{\bz, z}$:
\begin{align}
    \sL_{-1} \Phi^{\Delta,a}_{\bz, z} = \partial_z  \Phi^{\Delta,a}_{\bz, z} \, & \hspace{1.5 cm} \bL_{-1} \Phi^{\Delta, a}_{\bz, z} = \partial_\bz \Phi^{\Delta,a}_{\bz, z}  \, .
\end{align}
Therefore, we can ``Taylor expand'' the conformal primary modes in powers of $\bbz$ and $z$ by repeatedly acting $\bL_{-1}$ and $\sL_{-1}$ on a mode with $\bbz = 0$ and $z = 0$. These are called ``descendant'' modes.

The ``half-descendant modes'' are formed by only performing an anti-holomorphic Taylor expansion, trading the continuous parameter $\bbz \in \mathbb{C}$ for the integer $\bbn \in \mathbb{Z}$:

\begin{equation}\label{half_desc_spin1_first}
    \Phi^{\Delta, a}_{\bn, z} \equiv \left(\frac{\Gamma(\Delta-1+\bbn)}{\Gamma(\Delta-1)} \right) \, (\bL_{-1})^{\bn} \Phi^{\Delta, a}_{\bz = 0, z}.
\end{equation}
The ratio of gamma functions above was chosen so that $\Phi^{\Delta, a}_{\bn, z}$ has the simple functional form
\begin{align}\label{half_desc_spin1}
    \Phi^{\Delta, a}_{\bn, z} &= \bT^a (u - \bbw z )^{1 - \Delta - \bn} (w - \bbu z)^\bn.
\end{align}
Note that the expression \eqref{half_desc_spin1} can be defined for any $\bbn$, including negative $\bbn$, whereas \eqref{half_desc_spin1_first} cannot be. We will therefore implicitly use the expression \eqref{half_desc_spin1} if we wish to take $\bn$ negative.

In any case, note that the half-descendant modes are monomials of $(u - \bbw z)$ and $(w - \bbu z)$. This implies the following fact, which we will need to use later. Plugging two gluon half-descendant modes \textit{with the same} $z$ into the matrix commutator, we get
\begin{equation}\label{sdym_same_z}
    [\Phi^{\Delta_1,a}_{\bn_1,z}, \Phi^{\Delta_2,b}_{\bn_2,z}] = f^{abc} \Phi^{\Delta_1 + \Delta_2 - 1, c}_{\,\bn_1 + \bn_2, z}.
\end{equation}
From \eqref{Lscomm}, we see this relation is that of the $\mathfrak{s}$ algebra! This observation will come in handy in section \ref{sec25}.

Finally, we can define the ``full-descendant modes'' by taking holomorphic descendants of the half-descendant modes. We could do this simply by acting with $\sL_{-1}$ or $\partial_z$ repeatedly on the half-descendants, but will use a different method which will be more useful later on. Noting that $(\partial_z)^n$ can be replaced with a contour integral around the origin using Cauchy's integral formula, we define the full-descendant modes by
\begin{equation}\label{full_descendant_contour_spin1}
\Phi^{\Delta,a}_{\bn, n} \equiv \oint \frac{dz}{2 \pi i} \frac{\Phi^{\Delta,a}_{\bn,z}}{z^{1 + n}} \, .
\end{equation}

In appendix \ref{app_full_descendant}, we calculate the functional form of the full-descendant modes and find them to be
\begin{equation}\label{phi_spin_1_full_first}
    \Phi^{\Delta,a}_{\bn, n} = \mathbf{T}^a \frac{1}{(1 - \Delta - n)!} ( - u  \partial_\bw - w  \partial_\bu)^{1 - \Delta - n} \, \bbu^\bn \, ( - \bbw)^{1 - \Delta - \bn}.
\end{equation}
This formula will become useful later on. It should be noted that in the case that $n > 1 - \Delta$, the full-descendant modes actually vanish:
\begin{equation}
    \Phi^{\Delta,a}_{\bn,n} = 0 \hspace{0.5 cm} \text{ if } \hspace{0.5 cm} n > 1-\Delta.
\end{equation}
This is simply because $\Phi^{\Delta,a}_{\bn,z}$ is a polynomial in $z$ with degree $1 -\Delta$ (see \eqref{half_desc_spin1}) and so if we act on it with $(\partial_z)^n$, for $n > 1 - \Delta$, the polynomial will be annihilated. 

\subsection{Review of perturbiner expansion in SDYM}\label{sec24}

The half-descendant modes $\Phi^\Delta_{\bn, z}$ solve the free e.o.m.,
\begin{equation}
    \Box \, \Phi^{\Delta,a}_{\bn, z} = 0,
\end{equation}
but not the full non-linear e.o.m.\! \eqref{eom_siegel}. However, we can use these modes to perturbatively generate solutions to the full e.o.m.\! using something called a ``perturbiner expansion.'' For brevity in the following discussion, we introduce the notation
\begin{equation}
    \Phi_i \equiv \epsilon_i \Phi^{\Delta_i,a_i}_{\bn_i, z_i}
\end{equation}
where $i$ is an arbitrary index and each $\epsilon_i$ is an independent infinitesimal parameter.

The idea of the perturbiner expansion is to start with a solution to the free e.o.m.\! written as a sum of non-interacting particle wavefunctions, say
\begin{equation*}
    \Phi = \sum_{i=1}^N \Phi_i \hspace{0.5 cm} \text{(solves free e.o.m.)}
\end{equation*}
and then recursively solve for the higher order terms in the parameters $\epsilon_i$. The functions $\Phi_i$ are referred to as ``seed functions'' in the expansion. Usually, the seed functions are taken to be plane waves $e^{i p \cdot X}$ and the perturbiner expansion can be solved using tree-level Feynman diagrams.

In the perturbiner expansion, one must consider the square of any particular infinitesimal parameter to be 0, but the product of distinct infinitesimal parameters to \textit{not} be zero:
\begin{equation}\label{epsilon_instruction}
    (\epsilon_i)^2 = 0, \hspace{1 cm} \epsilon_i \epsilon_j \neq 0 \hspace{0.2 cm} \text{ if } i \neq j.
\end{equation}

The exact expression for the perturbiner expansion in SDYM has long been known \cite{Bardeen:1995gk,Selivanov:1996gw,Rosly:1996vr,selivanov1997selfdual,Cangemi:1996pf,Cangemi:1996rx,Korepin:1996mm} due to its close relation to the Parke-Taylor formula for the gluon MHV amplitude \cite{Berends:1987me, Parke:1985ax, Parke:1986gb}.

Let us now review the perturbiner expansion for the Chalmers-Siegel scalar. We define
\begin{equation}\label{bigpsi_1}
\begin{aligned}
    \bigPhi ( \Phi_1, \ldots, \Phi_N )  \equiv \;\; \begin{matrix}\text{full perturbiner expansion of Chalmers-Siegel} \\ \text{scalar $\Phi$ with seed functions }\Phi_1, \ldots, \Phi_N \end{matrix}
\end{aligned}
\end{equation}
(note the large font) to be the full expression for the perturbiner expansion. If $\Phi$ equals the above function, it will solve the non-linear e.o.m.\! \eqref{eom_siegel} assuming each of the infinitesimal parameters squares to zero, as in \eqref{epsilon_instruction}.

The perturbiner expansion \eqref{bigpsi_1} will be equal to a sum of terms, and each term will only depend on a subset of the seed functions. For $\{i_1, \ldots, i_k\} \subset \{1, \ldots, N\}$, let us define the extra function
\begin{equation}
    \bigPhi^{(k)}( \Phi_{i_1}, \ldots, \Phi_{i_k}) \equiv \; \begin{matrix}\text{sum of terms in } \bigPhi(\Phi_1, \ldots, \Phi_N ) \\ \text{containing } \Phi_{i_1}, \ldots, \Phi_{i_k} \end{matrix}
\end{equation}
so that we can clearly notate the dependence of each collection of seed functions in the full perturbiner expansion as
\begin{equation}\label{perturbinerPhi}
    \bigPhi(\Phi_1, \ldots, \Phi_N) = \sum_{k = 1}^N \sum_{ \substack{\{i_1, \ldots, i_k\}   \subset \{1, \ldots, N\} }  } \bigPhi^{(k)}( \Phi_{i_1}, \ldots, \Phi_{i_k}).
\end{equation}
For a simple example, when $N = 2$, the perturbiner expansion turns out to be
\begin{equation}
\begin{aligned}
    \bigPhi( \Phi_1, \Phi_2) &= \bigPhi^{(1)} (\Phi_1) + \bigPhi^{(1)} (\Phi_2) + \bigPhi^{(2)} (\Phi_1, \Phi_2) \\
    &= \Phi_1 + \Phi_2 + \frac{1}{z_{12}} [\Phi_1, \Phi_2]
\end{aligned}
\end{equation}
where
\begin{equation}
    z_{ij} \equiv z_i - z_j.
\end{equation}

Having defined all the necessary components, we are now able to quote the expression for the perturbiner expansion. $\Phi^{(k)}$ for general $k$ can be written as a sum over color-orderings as
\begin{equation}\label{psi_k}
    \bigPhi^{(k)} ( \Phi_1, \ldots, \Phi_k ) = \sum_{\sigma \, \in \, \mathrm{Sym}(k)} \left( \prod_{i = 1}^{k-1} \frac{1}{z_{\sigma(i) \sigma(i+1) }} \right) \Phi_{\sigma(1)} \Phi_{\sigma(2)} \Phi_{\sigma(3)} \ldots \Phi_{\sigma(k)}
\end{equation}
where $\mathrm{Sym}(k)$ is the permutation group of $k$ elements and $\sigma$ is a permutation.

There is also a nice recursive way to compute $\Phi^{(k)}$ in terms of $\Phi^{(k-1)}$, as
\begin{equation}\label{sdym_recursive_1}
    \bigPhi^{(k)} ( \Phi_1, \ldots, \Phi_k ) = \sum_{i = 1}^{k-1} \bigPhi^{(k-1)}  ( \Phi_1, \ldots, \Phi_{k-1} ) \bigg\rvert_{\displaystyle{  \Phi_i \mapsto \frac{1}{z_{ik}} [\Phi_i, \Phi_k] }  }.
\end{equation}
A proof of the equations \eqref{psi_k} and \eqref{sdym_recursive_1} are provided in Appendix C.1 of \cite{mypaper}.

The above equation suggests that SDYM has the character of a 2d conformal field theory. This is because \eqref{sdym_recursive_1} manifestly looks like a Kac-Moody Ward identity, as one can see the OPE in the structure of the equation. This key observation has been used for instance in \cite{Costello:2022wso,Bittleston:2024efo,Costello:2023vyy,Costello:2022jpg}. In the next section we will see why this implies that there is an action of the $\Ls$ algebra on the Chalmers-Siegel scalar.

\subsection{$\Ls$ action on spacetime gauge field}\label{sec25}

Now we have defined all of the ingredients necessary to explain how to act the $\Ls$ algebra on the Chalmers-Siegel scalar $\Phi$. 

Consider the perturbiner expansion built from the seed functions $\Phi_1, \ldots, \Phi_{N}$. Each seed function $\Phi_i$ contains four parameters: $\Delta_i$, $a_i$, $\bbn_i$, and $z_i$. All of the $z_i$'s are points on the complex plane.

Now consider adding one more seed function to the pre-existing list of seed functions, called $\Phi_{I}$, which has an associated $z_{I}$.

As a useful piece of notation, let us notate the linear change in the perturbiner expansion due to the addition of a single extra seed function with a vertical line, by
\begin{align}
    \bigPhi( \Phi_1, \ldots, \Phi_N \rvert \Phi_{I} ) &\equiv \bigPhi(\Phi_1, \ldots, \Phi_N, \Phi_{I} ) - \bigPhi(\Phi_1, \ldots, \Phi_N) \, .
\end{align}

For this added seed function, we are going to \textit{integrate its position $z_I$ around in a contour} $C$ which surrounds all the pre-existing $z_i$'s, and include in the integrand a factor of $1/z_I^{\, n_I +1}$. This weighted contour integral has the effect of adding a full-descendant mode $\Phi^{\Delta_I}_{\bn_I, n_I}$ to the list of seed functions instead of a half-descendant mode, as per \eqref{full_descendant_contour_spin1}.

Let us denote $ \delta^{\Delta_{I}, a_I}_{\bn_I, n_I}$ to be the infinitesimal change in the perturbiner expansion which results from this procedure. We may then write
\begin{equation}\label{proposed_action_1}
    \delta^{\Delta_{I},a_I }_{\bn_I, n_I} \bigPhi( \Phi_1, \ldots, \Phi_{N} ) \equiv \frac{1}{2 \pi i} \oint_{C} \frac{d z_I}{z_I^{\, n_I+1}} \bigPhi\left(\Phi_1, \ldots, \Phi_{N} \rvert \Phi_I \right) ,
\end{equation}
or equivalently
\begin{equation}
    (1 + \delta^{\Delta_{I},a_I }_{\bn_I, n_I} ) \bigPhi( \Phi_1, \ldots, \Phi_{N} ) =  \bigPhi\left(\Phi_1, \ldots, \Phi_{N}, \frac{1}{2 \pi i} \oint_{C} \frac{d z_I}{z_I^{\, n_I+1}} \Phi_I \right).
\end{equation}

What we now want to check is that the above definition of $\delta^{\Delta,a}_{\bn, n}$ does indeed satisfy the $\Ls$ commutation relation \eqref{Lscomm}.

Let us begin the proof. To be explicit, we define
\begin{equation}
    I \equiv N+1, \hspace{1 cm} J \equiv N +2.
\end{equation}
We start with the elementary identity
\begin{equation}
    [ \delta^{\Delta_{I},a_I }_{\bn_I, n_I}, \delta^{\Delta_{J},a_J }_{\bn_{J}, n_{J} } ] = (1 + \delta^{\Delta_{I},a_I }_{\bn_I, n_I} )( 1 + \delta^{\Delta_{J},a_J }_{\bn_{J}, n_{J} } ) - ( 1 + \delta^{\Delta_{J},a_J }_{\bn_{J}, n_{J} } ) (1 + \delta^{\Delta_{I},a_I }_{\bn_I, n_I} )
\end{equation}
which implies
\begin{align}\label{comm_intermediate_11}
    [ \delta^{\Delta_{I},a_I }_{\bn_I, n_I}, \delta^{\Delta_{J},a_J }_{\bn_{J}, n_{J} } ] \bigPhi( \Phi_1, \ldots, \Phi_{N} )=&\, \bigPhi\left( \Phi_1, \ldots, \Phi_{N},  \oint_{C} \frac{d z_I}{2 \pi i} \frac{\Phi_I}{z_I^{\, n_I+1}}, \oint_{C'} \frac{d z_J}{2 \pi i} \frac{\Phi_J}{z_J^{\, n_J+1}}\right) \nonumber \\
    &- \bigPhi\left( \Phi_1, \ldots, \Phi_{N},  \oint_{C} \frac{d z_J}{2 \pi i} \frac{\Phi_J}{z_J^{\, n_J+1}}, \oint_{C'} \frac{d z_I}{2 \pi i} \frac{\Phi_I}{z_I^{\, n_I+1}}\right).
\end{align}
In the above expression, $C$ is a contour which surrounds the origin and $z_1, \ldots, z_N$, while $C'$ is a contour which surrounds $C$.

If one imagines expanding out the terms on the RHS of \eqref{comm_intermediate_11}, there will be three kinds of terms in the perturbiner expansion:  (1) terms only containing $\Phi_1, \ldots, \Phi_N$, (2) terms which contain exactly one of $\Phi_I$ and $\Phi_J$, and (3) terms which contain both $\Phi_I$ and $\Phi_J$. Terms of type (1) and (2) will separately cancel out, due to the subtraction present in the above equation and the fact that the contours can be deformed away from of $z_1, \ldots, z_N$. Only terms of the type (3) will remain. Because each term in the resulting expression will contain both $\Phi_I$ and $\Phi_J$, by linearity we can pull the integrals outside of the arguments of the perturbiner and write
\begin{equation}
\begin{aligned}
    &[ \delta^{\Delta_{I},a_I }_{\bn_I, n_I}, \delta^{\Delta_{J},a_J }_{\bn_{J}, n_{J} } ] \bigPhi(\Phi_1, \ldots, \Phi_N) \\
    & = \frac{1}{(2 \pi i)^2} \left( \oint_{C} \frac{dz_I}{z_I^{\, n_I+1}}  \oint_{C'} \frac{dz_J}{z_J^{\, n_J+1}} -  \oint_{C'} \frac{dz_I}{z_I^{\, n_I+1}}  \oint_{C} \frac{dz_J}{z_J^{\, n_J+1}}  \right) \bigPhi( \Phi_1, \ldots, \Phi_N, \Phi_I, \Phi_J)
\end{aligned}
\end{equation}
where again note that in the RHS of the above equation, all of the terms which contain either none or only one of $\Phi_I$ and $\Phi_J$ will cancel due to the relative minus sign between the above integrals.

The standard contour deformation argument familiar from 2d CFT \cite{Polchinski:1998rq} then yields
\begin{equation}
    [ \delta^{\Delta_{I},a_I }_{\bn_I, n_I}, \delta^{\Delta_{J},a_J }_{\bn_{J}, n_{J} } ] \bigPhi(\Phi_1, \ldots, \Phi_N)
    =   \oint_{C } \frac{dz_I}{2 \pi i}  \mathop{\mathrm{Res}}_{z_J \to z_I} \frac{1}{z_I^{\, n_I+1}}  \frac{1}{z_J^{\, n_J+1}} \bigPhi( \Phi_1, \ldots, \Phi_N, \Phi_I, \Phi_J). \label{eq_res_1}
\end{equation}
For the final step, we need to figure out how the perturbiner expansion behaves when $z_J \to z_I$. From \eqref{sdym_recursive_1}, we can read off that the only singular term is
\begin{align}
    \lim_{z_J \to z_I}  \bigPhi( \Phi_1, &\ldots, \Phi_N, \Phi_I, \Phi_J) = \bigPhi( \Phi_1, \ldots, \Phi_N \rvert \frac{1}{z_{IJ}} [ \Phi_I, \Phi_J ]\Big\rvert_{z_J = z_I} ) .
\end{align}
We transform the above expression using the relation \eqref{sdym_same_z}, which states that the matrix commutator of two half-descendant modes with the same $z$ satisfies the $\mathfrak{s}$ algebra commutation relation:
\begin{equation}\label{eq_lim_1}
    = \frac{1}{z_{IJ}} f^{a_I a_J a_K} \bigPhi( \Phi_1, \ldots, \Phi_N \rvert  \epsilon_I \epsilon_J \Phi^{\Delta_I + \Delta_J - 1, a_K}_{\bn_I + \bn_J, z_I}  ).  
\end{equation}
We can now plug \eqref{eq_lim_1} into \eqref{eq_res_1} and deduce
\begin{equation}\label{Ls_alg_confirmation}
\boxed{ \textcolor{white}{\Bigg\rvert}
    [ \delta^{\Delta_{I} , a_I }_{\bn_I, n_I}, \delta^{\Delta_{J}, a_J }_{\bn_{J}, n_{J} } ] = f^{a_I a_J a_K} \delta^{\Delta_I + \Delta_J-1, a_K}_{\bn_I + \bn_J, n_I + n_J+1}.  } 
\end{equation}
This means that our variation \eqref{proposed_action_1} on the space of self-dual connections really does satisfy the $\Ls$ commutation relation \eqref{Lscomm}!

It is worth pausing and reflecting on a few features of the above computation. We have seen that the variation $\delta^{\Delta,a}_{\bn, n}$ corresponds to adding the full-descendant mode $\Phi^{\Delta,a}_{\bn, n}$ to the list of seed functions. Because a perturbiner expansion trivially posses a bosonic permutation symmetry, one may wonder how this algebra of variations can possibly be non-abelian. This happens because our prescription for picking the contour $C$ involved making it large enough to surround all previous seed functions, and it is only due to this picking of successive contours, which must get larger and larger to surround all the previous contours, that the non-commuting nature of these variations is introduced.

The fundamental reason that it is necessary to choose this contour prescription is that the perturbiner expansion contains $1/z_{ij}$ singularities at coincident points. Because full-descendant modes all live, in a sense, at $z = 0$, without such a prescription there would be a fundamental ambiguity regarding the order in which to take the $z_{ij} \to 0$ coincident limits.

Because the perturbiner expansion is known from equation \eqref{psi_k}, we can write, as advertised in the introduction, the $N^{\rm th}$ variation of the $\Ls$ algebra on the $\Phi = 0$ vacuum:
\begin{equation}\label{new_formula_spin1}
\begin{aligned}
    (1 + \delta^{\Delta_N,a_N}_{\bn_N, n_N}) \ldots \, & (1 +  \delta^{\Delta_2,a_2}_{\bn_2, n_2}) (1 + \delta^{\Delta_1,a_1}_{\bn_1, n_1}) \cdot (\Phi = 0) \\
    &= \bigPhi\left( \oint_{C_1} \frac{dz_1}{2 \pi i} \frac{\Phi_1}{z_1^{\; n_1+1}},  \oint_{C_2} \frac{dz_2}{2 \pi i} \frac{\Phi_2}{z_2^{\; n_2+1}}, \ldots,  \oint_{C_N} \frac{dz_N}{2 \pi i} \frac{\Phi_N}{z_N^{\; n_N+1}} \right) 
\end{aligned}
\end{equation}
where the contour $C_i$ is enclosed by $C_j$ if $i < j$, and all contours enclose the origin.

In appendix \ref{app_example}, we provide an example calculation which demonstrates how one can actually compute these expressions explicitly if desired.

\subsection{Residual gauge symmetry of the Chalmers-Siegel scalar}\label{sec26}

The Chalmers-Siegel gauge choice \eqref{A_Phi} has a set of residual gauge transformations \cite{Campiglia:2021srh} where the gauge parameter $\lambda$ is a function of $\bbu$ and $\bbw$ alone:
\begin{equation}
    \lambda = \lambda(\bbu, \bbw).
\end{equation}
A basis of such gauge transformations is given by
\begin{equation}\label{lambda_lightcone}
    \lambda^{\Delta, a}_{\bn} = \bT^a \; \bbu^\bn \; (-\bbw)^{1 - \Delta - \bn}.
\end{equation}
The matrix commutator of these gauge parameters is given by
\begin{equation}\label{lambda_s}
    [ \lambda^{\Delta_1, a}_{\bn_1}, \lambda^{\Delta_2, b}_{\bn_2} ] = f^{a b c} \lambda^{\Delta_1 + \Delta_2 - 1,c}_{\bn_1 + \bn_2}
\end{equation}
which is exactly the $\mathfrak{s}$-algebra \eqref{s_alg}.

The variation $\delta_\lambda \Phi$ one of these residual gauge transformations induces on the Chalmers-Siegel scalar can be found by solving the equation
\begin{equation}\label{resid_phi_solve}
    A_\mu[\delta_\lambda  \Phi] = \delta_\lambda A[\Phi]_\mu \hspace{0.5 cm} \implies \hspace{0.5 cm} \begin{pmatrix} 0 \\ \partial_w \delta_\lambda \Phi \\ 0 \\ \partial_u \delta_\lambda \Phi \end{pmatrix} = \begin{pmatrix} 0 \\ -[\partial_w \Phi, \lambda] - \partial_\bu \lambda \\ 0 \\- [\partial_u \Phi, \lambda] - \partial_\bw \lambda \end{pmatrix},
\end{equation}
which gives
\begin{equation}\label{gauge_var_sdym}
    \delta_\lambda \Phi = -[ \Phi, \lambda] - (u \, \partial_\bw + w \, \partial_{\bu}) \lambda.
\end{equation}
Note that in the above equation, $w \partial_\bu + u \partial_\bw = \sL_1$ \eqref{list_lorentz_gen}, and this is a differential operator which will appear often in this work.

If desired, we could change from lightcone coordinates $(u, \bbu, w, \bbw)$ to the perhaps more familiar flat Bondi coordinates $(U,R,Z,\bZ)$ via \cite{Himwich:2020rro}
\begin{equation}\label{flat_bondi_coords}
    u = U + R \, Z\,  \bZ , \hspace{0.75 cm} \bbu = R, \hspace{0.75 cm} w = R \, Z, \hspace{0.75 cm} \bbw = R \, \bZ .
\end{equation}
In these coordinates, the flat metric is
\begin{equation}
    ds^2 = 4\, (d U \, d R - R^2 \, d Z \, d \bZ) \, ,
\end{equation}
and the residual gauge transformations \eqref{lambda_lightcone} are
\begin{equation}
    \lambda^{\Delta,a}_{\bn} = \mathbf{T}^a R^{1 - \Delta} (-\bZ)^{1 - \Delta - \bn} \, .
\end{equation}
Notice $\lambda^{\Delta,a}_{\bn}$ are all anti-holomorphic gauge transformations in $\bZ$, and the more negative $\Delta$ is the more over-leading in $R$ the gauge transformations are.

\subsection{Pure gauge and ``trivial'' descendants of spin-1 primaries}\label{sec27}

Descendants of the conformal primaries $\Phi^{\Delta,a}_{\bz,z}$ are found by taking repeated $(\partial_{\bz})^\bn$ and $(\partial_z)^n$ derivatives. If $\Delta = 1, 0, -1, -2, \ldots$, naive finite dimensional $\mathfrak{sl}(2, \mathbb{R})$ representation theory suggests that $\bbn$ and $n$ should lie in the range
\begin{equation*}
    0 \leq \bbn \leq - 2 \bar{h} , \hspace{1 cm} 0 \leq n \leq - 2h,
\end{equation*}
with $-2 \bar{h} = - \Delta + 1$ and $- 2 h = -\Delta - 1$.\footnote{We note that the anti-holomorphic range $0 \leq \bbn \leq - \Delta+1$ corresponds to the $\Ls$ generators which lie in the truncated wedge subalgebra $\Ls_{\wedge}$. Furthermore, for the special cases $\Delta = 1, 0$, there are actually \textit{no} choices for $n$ which lie in the range $0 \leq n \leq - 2 h$. We emphasize that in this work, we are allowing ourselves to consider an analytically continued notion of descendant, where both integers $\bn$ and $n$ are allowed to lie outside of these ranges and even be negative.} Note that there is a mismatch between the number of antiholomorphic descendants and the number of holomorphic descendants: there seem to be ``two fewer'' holomorphic descendants, corresponding to $n = - \Delta$ and $n = - \Delta +1$. However, the explicit formula for $\Phi^{\Delta,a}_{\bz, z}$ in \eqref{conf_prim} does not seem to treat $\bbz$ and $z$ differently in any fundamental sense, and differentiation by $\partial_{\bz}$ does not act any differently from differentiation by $\partial_z$.

So, what is the source of the mismatch? The answer is that the $n = -\Delta$ descendant is \textit{pure gauge}, and the $n = - \Delta + 1$ descendant is ``trivial.''
\begin{equation}
    \begin{aligned}
        n &= - \Delta && \text{ is pure gauge } \\
        n &= - \Delta +1 && \text{ is trivial } \\
    \end{aligned}
\end{equation}
Quoting the formula for the full-descendant wave function \eqref{phi_spin_1_full_first} again, 
\begin{equation}
    \Phi^{\Delta,a}_{\bn, n} = \mathbf{T}^a \frac{1}{(1 - \Delta - n)!} ( - u  \partial_\bw - w  \partial_\bu)^{1 - \Delta - n} \, \bbu^\bn \, ( - \bbw)^{1 - \Delta - \bn}.
\end{equation}
we note that when $n = -\Delta$, the descendant wavefunction is linear in $u$ and $w$,
\begin{equation}\label{Phi_n_delta}
\begin{aligned}
    \Phi^{\Delta, a}_{\bn, (n = -\Delta)} &= \mathbf{T}^a (- u \partial_\bw - w \partial_\bu) \, \bbu^\bn (-\bbw)^{1 - \Delta - \bn} \\ 
    &= -(u \partial_\bw + w \partial_\bu ) \, \lambda^{\Delta,a}_{\bn}
\end{aligned}
\end{equation}
and when $n = -\Delta +1$, the descendant wave function is independent of $u$ and $w$ and only depends on $\bbu$ and $\bbw$.
\begin{equation}\label{Phi_n_delta1}
\begin{aligned}
    \Phi^{\Delta, a}_{\bn, (n = -\Delta+1)} &= \mathbf{T}^a \, \bbu^\bn (- \bbw)^{1 - \Delta- \bn} \\ 
    &=  \lambda^{\Delta,a}_{\bn}.
\end{aligned}
\end{equation}
Plugging \eqref{Phi_n_delta} into \eqref{A_Phi}, we then indeed see that the $n = - \Delta$ descendant is pure gauge
\begin{align}
    A[\Phi^{\Delta, a}_{\bn, (n = -\Delta)}]_\mu = \begin{pmatrix} 0 \\ \partial_w \Phi^{\Delta, a}_{\bn, (n = -\Delta)} \\ 0 \\ \partial_u  \Phi^{\Delta, a}_{\bn, (n = -\Delta)} \end{pmatrix} =  -\begin{pmatrix} 0 \\ \partial_\bu \lambda^{\Delta,a}_\bn \\ 0 \\ \partial_\bw \lambda^{\Delta,a}_\bn \end{pmatrix}=  -\partial_\mu \lambda^{\Delta,a}_\bn
\end{align}
while from \eqref{Phi_n_delta1}, the $n = -\Delta + 1$ descendant gives the zero connection
\begin{align}
    A[\Phi^{\Delta, a}_{\bn, (n = -\Delta+1)}]_\mu = \begin{pmatrix} 0 \\ \partial_w \Phi^{\Delta, a}_{\bn, (n = -\Delta+1)} \\ 0 \\ \partial_u  \Phi^{\Delta, a}_{\bn, (n = -\Delta+1)} \end{pmatrix} = \begin{pmatrix} 0 \\ 0  \\ 0 \\ 0 \end{pmatrix}
\end{align}
and hence it is ``trivial.''

Furthermore, we note that when $n > -\Delta + 1$, the descendant modes are not only trivial, but actually zero.

\begin{equation}
    \Phi^{\Delta,a}_{\bn,(n > -\Delta + 1)} = 0
\end{equation}
This will end up being related to the fact that the $\Ls$ variations on the Chalmers-Siegel scalar $\delta^{\Delta,a}_{\bn,(n > -\Delta+1)}$ will vanish.

\subsection{The large-gauge $\mathfrak{s} \subset \Ls$}\label{sec28}

In the last section, we saw that the gluon wavefunctions $\Phi^{\Delta,a}_{\bn,(n = -\Delta)}$ were pure gauge. What happens if we add one of these pure gauge wavefunctions to a pre-existing list of seed functions in a perturbiner expansion? This would propagate the pure gauge particle on a background created by all of the other particles. In this context, we might intuitively expect that ``scattering a pure gauge particle'' is equivalent to ``doing a gauge transformation.'' Indeed, this is exactly the case, as we shall now show!

Say we have a connection and vary it by a gauge transformation $\lambda$ as
\begin{equation}
    \delta_\lambda A_\mu = \underbrace{- \partial_\mu \lambda}_{0^{\rm th} \text{ order term}} - \; \underbrace{[A_\mu, \lambda]}_{1^{\rm st} \text{ order term}}.
\end{equation}
There are two terms in the variation, one $0^{\rm th}$ order in $A_\mu$ and one $1^{\rm st}$ order in $A_\mu$. There are no higher order terms. On the Chalmers-Siegel scalar, from \eqref{gauge_var_sdym}, any such (residual) gauge transformation takes the form 
\begin{equation}
\begin{aligned}
    \delta_\lambda \Phi = \;\; -\underbrace{(u \partial_\bw + w \partial_\bu )\lambda}_{0^{\rm th} \text{ order term}} \;\;\; - \underbrace{[\Phi, \lambda]}_{1^{\rm st} \text{ order term}}.
\end{aligned}
\end{equation}

We will use our explicit formula for the action of $\Ls$ on our Chalmers-Siegel scalar, \eqref{proposed_action_1}, to show that the transformation $\delta^{\Delta,a}_{\bn,(n=-\Delta)}$ equals the action of the gauge transformation $\lambda^{\Delta,a}_{\bn}$ \eqref{lambda_lightcone}. This transformation corresponds to adding the pure gauge wavefunction $\Phi^{\Delta,a}_{\bn,(n = -\Delta)}$ to the list of seed functions.

Because there is a whole $\mathfrak{s}$ algebra of such transformations by \eqref{lambda_s}, this will imply there is a $\mathfrak{s} \subset \Ls$ subalgebra which is pure gauge.

With all this in mind, let us finally calculate
\begin{equation*}
    \delta^{\Delta,a }_{\bn, (n = -\Delta)} \bigPhi( \Phi_1, \ldots, \Phi_{N} )
\end{equation*}
using the formula for the perturbiner expansion \eqref{psi_k}, \eqref{perturbinerPhi}. There will be three types of terms which will appear. The first term will simply be the pure gauge seed function itself, which is the $0^{\rm th}$ order term.
\begin{equation}\label{spin_1_0_order}
    \frac{1}{2 \pi i} \oint_C \frac{dz}{z^{1 -\Delta}} \Phi^{\Delta,a}_{\bn, z}  \;\; \subset  \;\; \delta^{\Delta,a }_{\bn, (n = -\Delta)} \bigPhi( \Phi_1, \ldots, \Phi_{N} ).
\end{equation}
The second type of term will occur when the pure gauge seed function appears as either the first or last seed function in the color ordering. These are the $1^{\rm st}$ order terms.
\begin{equation}\label{spin_1_1_order}
\begin{aligned}
    &\frac{1}{2 \pi i} \oint_C \frac{dz}{z^{1 -\Delta}}\; \frac{\Phi^{\Delta,a}_{\bn, z} }{z - z_{i_1}} \; \frac{\Phi_{i_1} \ldots \Phi_{i_k}}{z_{i_1 i_2} \ldots z_{i_{k-1} i_k}}   \;\; \subset  \;\; \delta^{\Delta,a }_{\bn, (n = -\Delta)} \bigPhi( \Phi_1, \ldots, \Phi_{N} ), \\
    &\frac{1}{2 \pi i} \oint_C \frac{dz}{z^{1 -\Delta}} \; \frac{\Phi_{i_1} \ldots \Phi_{i_k}}{z_{i_1 i_2} \ldots z_{i_{k-1} i_k}} \; \frac{\Phi^{\Delta,a}_{\bn, z} }{z_{i_k} - z}    \;\; \subset  \;\; \delta^{\Delta,a }_{\bn, (n = -\Delta)} \bigPhi( \Phi_1, \ldots, \Phi_{N} ).
\end{aligned}
\end{equation}
The third type of term will occur when the pure gauge seed function appears somewhere in the middle of the color ordering. These terms would be higher order, although we'll actually show they end up vanishing.
\begin{equation}\label{spin_1_2_order}
\begin{aligned}
    \frac{1}{2 \pi i}  \oint_C \frac{dz}{z^{1 -\Delta}} \frac{\Phi_{i_1} \ldots \Phi_{i_{m-1}}}{z_{i_1 i_2} \ldots z_{i_{m-2} i_{m-1}} } \frac{\Phi^{\Delta,a}_{\bn,z}}{(z_{i_{m-1}} - z)(z - z_{i_{m}})}&  \frac{\Phi_{i_m} \ldots \Phi_{i_{k}}}{z_{i_m i_{m+1}} \ldots z_{i_{k-1} i_{k}} } \\ & \\
    &\hspace{-1cm} \subset  \;\; \delta^{\Delta,a }_{\bn, (n = -\Delta)} \bigPhi( \Phi_1, \ldots, \Phi_{N} ).
\end{aligned}
\end{equation}

Let us now evaluate the $0^{\rm th}$, $1^{\rm st}$, and higher order terms separately. From \eqref{Phi_n_delta}, the $0^{\rm th}$ order term is
\begin{equation}
    \frac{1}{2 \pi i} \oint_C \frac{dz}{z^{1 -\Delta}} \Phi^{\Delta,a}_{\bn, z} = \Phi^{\Delta,a}_{\bn, (n = -\Delta)} = - (u \partial_\bw + w \partial_\bu) \, \lambda^{\Delta,a}_{\bn}.
\end{equation}
The $1^{\rm st}$ order interaction terms \eqref{spin_1_1_order} contain a single factor of $1/(z - z_i)$ where $z_i$ is the location of another particle insertion. (Note that $|z| > |z_i|$ because we are integrating $z$ around a contour $C$ which encloses $z_i$.) Performing a Taylor expansion of $1/(z - z_i)$, we find only a single term (the $r=0$ term below) is non-zero because $\Phi^{\Delta,a}_{\bn,n}$ is zero for $n > 1 -\Delta$. The only non-vanishing term is $\Phi^{\Delta,a}_{\bn,1-\Delta} = \lambda^{\Delta,a}_{\bn}$ by \eqref{Phi_n_delta1}, giving
\begin{equation}\label{pure_gauge_eq_2}
\begin{aligned}
    \frac{1}{2 \pi i}  \oint_C \frac{dz}{z^{1 -\Delta}} \frac{1}{z - z_i} \Phi^{\Delta,a}_{\bn, z} &= \frac{1}{2 \pi i} \oint_C \frac{dz}{z^{1 -\Delta}} \frac{1}{z} \sum_{r = 0}^\infty \left( \frac{z_i}{z}\right)^r \Phi^{\Delta,a}_{\bn, z}\\
    &= \sum_{r = 0}^\infty z_i^r \Phi^{\Delta,a}_{\bn,1-\Delta+r} \\
    &= \lambda_\bn^{\Delta,a}.
\end{aligned}
\end{equation}
Finally, we look at the higher order interaction terms \eqref{spin_1_2_order}. These terms include a factor $1/(z-z_i) \times 1/(z - z_j)$. Doing two Taylor expansions and performing a similar computation to \eqref{pure_gauge_eq_2}, we find such terms are zero.
\begin{equation}\label{pure_gauge_eq_3}
\begin{aligned}
    \frac{1}{2 \pi i}  \oint_C \frac{dz}{z^{1 -\Delta}} \frac{1}{z - z_i} \frac{1}{z - z_j} \Phi^{\Delta,a}_{\bn, z} &= \frac{1}{2 \pi i}  \oint_C \frac{dz}{z^{1 -\Delta}} \frac{1}{z^2} \sum_{r = 0}^\infty \left( \frac{z_i}{z} \right)^r \sum_{s = 0}^\infty \left( \frac{z_j}{z} \right)^s \Phi^{\Delta,a}_{\bn, z} \\
    &= \sum_{r = 0}^\infty \sum_{s = 0}^\infty z_i^r z_j^s \Phi^{\Delta, a}_{\bn, 2 - \Delta + r + s} = 0.
\end{aligned}
\end{equation}

Putting this all together, we get
\begin{equation}
\begin{aligned}
    \delta^{\Delta,a }_{\bn, (n = -\Delta)} & \bigPhi( \Phi_1, \ldots, \Phi_{N} ) \\
    &= -(u \partial_\bw + w \partial_\bu) \, \lambda^{\Delta, a}_{\bn} + \lambda^{\Delta, a}_{\bn} \bigPhi( \Phi_1, \ldots, \Phi_{N} ) - \bigPhi( \Phi_1, \ldots, \Phi_{N} ) \lambda^{\Delta, a}_{\bn}.
\end{aligned}
\end{equation}

We have therefore derived, organically, that when we scatter a pure gauge particle off of a background field, only the $0^{\rm th}$ and $1^{\rm st}$ order terms involving the original background field survive in the perturbiner expansion. All higher order terms are zero. The above equation can be rewritten as
\begin{equation}\label{s_alg_pure_gauge}
    \delta^{\Delta,a }_{\bn, (n = -\Delta)} \bigPhi( \Phi_1, \ldots, \Phi_{N} ) = \delta_{ \lambda^{\Delta,a}_{\bn} }\bigPhi( \Phi_1, \ldots, \Phi_{N} ).
\end{equation}
As promised, the LHS equals the $\Ls$ change induced on $\Phi$ via the scattering of a pure gauge particle, defined by \eqref{proposed_action_1}, and the RHS equals the action of a gauge transformation, defined by \eqref{gauge_var_sdym}. In fact, we can ``factor out'' the perturbiner notation from both sides of the above equation if we like, and simply write
\begin{equation}
    \delta^{\Delta,a }_{\bn, (n = -\Delta)} \Phi = \delta_{ \lambda^{\Delta,a}_{\bn} } \Phi =  - (u \, \partial_\bw + w \, \partial_{\bu}) \lambda^{\Delta,a}_{\bn} - [ \Phi, \lambda^{\Delta,a}_{\bn}].
\end{equation}

Therefore, we have shown that the $\Ls$ transformations with $n = -\Delta$ are pure gauge. A similar computation shows that $\Ls$ transformations with $n = - \Delta +1$ are trivial, simply amounting to the addition of a function of $(\bbu,\bbw)$:
\begin{equation}
    \delta^{\Delta,a }_{\bn, (n = -\Delta+1)} \Phi = \Phi^{\Delta,a }_{\bn, (n = -\Delta+1)} = \bT^a \bbu^\bn (-\bbw)^{1-\Delta-\bn}.
\end{equation}
Finally, if $n > -\Delta+1$, another straightforward analogous computation shows the action of the $\Ls$ transformations on the Chalmers-Siegel scalar will actually vanish:
\begin{equation}
    \delta^{\Delta,a }_{\bn, (n > -\Delta+1)} \Phi = 0.
\end{equation}

\subsection{The recursion operator $\mR$ and $\Ls$ in SDYM}\label{sec29}

If $\Phi$ solves the equation of motion \eqref{eom_siegel}, a linearized perturbation $\delta \Phi$ around $\Phi$ will satisfy
\begin{equation}
    \Box \, \delta \Phi - [ \partial_u \Phi, \partial_w \delta \Phi] - [ \partial_u \delta \Phi, \partial_w  \Phi] = 0.
\end{equation}
There exists an operator $\mR$, called the ``recursion operator,'' which maps the space of linearized perturbations around $\Phi$ into itself, meaning $\mR \delta \Phi$ will be a function which satisfies the above equation just as $\delta \Phi$ does.

Let us give some motivation for the definition of this operator. Recall that there are two ways to write $A_\mu$ in lightcone gauge in SDYM, one with the Chalmers-Siegel scalar $\Phi$, \eqref{A_Phi}, and one with the Yang $J$-matrix, \eqref{AJ},
\begin{equation}\label{sdym_two_ways}
    \begin{pmatrix}
        A_u \\
        A_\bu \\
        A_w \\
        A_\bw
    \end{pmatrix} = 
    \begin{pmatrix}
        0 \\
        \partial_w \Phi \\
        0 \\
        \partial_u \Phi
    \end{pmatrix} =
    \begin{pmatrix}
        0 \\
        J^{-1} \partial_\bu J\\
        0 \\
        J^{-1} \partial_\bw J
    \end{pmatrix}
\end{equation}
where $J$ must satisfy the constraint $\partial_u ( J^{-1} \partial_\bu J ) - \partial_w (  J^{-1} \partial_\bw J) = 0$.

It is natural to consider the set of variations $\delta_\Lambda  J = J \Lambda$, $\delta_\Lambda J^{-1} = -\Lambda J^{-1}$ where $\Lambda$ is some $\mathfrak{g}$-valued spacetime function. Such a transformation will vary the connection as
\begin{equation}\label{quasi_sdym}
    \begin{pmatrix}
        \delta_\Lambda A_u \\
        \delta_\Lambda  A_\bu \\
        \delta_\Lambda  A_w \\
        \delta_\Lambda  A_\bw
    \end{pmatrix} = 
    \begin{pmatrix}
        0 \\
        \partial_\bu \Lambda + [ A_\bu, \Lambda]  \\
        0\\
        \partial_\bw \Lambda + [ A_\bw, \Lambda] 
    \end{pmatrix}.
\end{equation}
However, not all choices of $\Lambda$ will preserve the constraint equation. In fact, plugging $J + \delta_\Lambda J$ into the constraint, we find that the set of allowed $\Lambda$'s must satisfy
\begin{equation}
    \Box \Lambda + \partial_u ([J^{-1} \partial_\bu J, \Lambda]) - \partial_w ( [ J^{-1} \partial_\bw J, \Lambda] ) = 0.
\end{equation}
Using \eqref{sdym_two_ways}, this then becomes
\begin{equation}\label{Lambda_constraint}
    \Box  \Lambda - [ \partial_u \Phi, \partial_w \Lambda] - [ \partial_u \Lambda, \partial_w  \Phi] = 0
\end{equation}
which is exactly the e.o.m.\! for linearized perturbations of $\Phi$! In other words, we are able to enact the ``quasi gauge transformation'' \eqref{quasi_sdym} as long as the constraint \eqref{Lambda_constraint} is satisfied. Notice that, in general, $\Lambda = \Lambda(u,\bbu,w,\bbw)$. In the special case that $\Lambda = - \lambda(\bbu, \bbw)$, this ``quasi gauge transformation'' coincides with an actual residual gauge transformation \eqref{resid_phi_solve}. 

So each linear perturbation to $\Phi$ can be used to build a ``quasi-gauge transformation'' in the sense of \eqref{quasi_sdym}.

Now, what change on $\Phi$ does the quasi-gauge transformation \eqref{quasi_sdym} induce? Let us call this variation $\delta_\Lambda \Phi$. Clearly, it must satisfy
\begin{equation}
    \begin{aligned}
        \partial_w \delta_\Lambda \Phi &= \partial_\bu \Lambda + [\partial_w \Phi, \Lambda]  \, ,\\
        \partial_u \delta_\Lambda \Phi &= \partial_\bw \Lambda + [\partial_u \Phi, \Lambda]  \, .
    \end{aligned}
\end{equation}
This is exactly the defining equation of the recursion operator. To be completely explicit, $\mR \delta \Phi$ is defined as the solution to the pair of differential equations
\begin{equation}\label{recursion_1}
\begin{aligned}
    \partial_w ( \mR \delta \Phi) &= \partial_\bu \delta \Phi + [ \partial_w \Phi, \delta \Phi ] \, , \\
    \partial_u ( \mR \delta \Phi) &= \partial_\bw \delta \Phi + [ \partial_u \Phi, \delta \Phi ]\,    .
\end{aligned}
\end{equation}
meaning $\mR \delta \Phi$ is the change in $\Phi$ induced by acting with $\Lambda = \delta \Phi$ as a ``quasi-gauge transformation'' on the $J$-matrix.

The compatibility of the two equations \eqref{recursion_1} follows from the fact that $\delta \Phi$ is a linear perturbation of $\Phi$:
\begin{equation}
    \partial_u ( \partial_w ( \mR \delta \Phi) ) - \partial_w ( \partial_u ( \mR \delta \Phi) ) = \Box \delta \Phi - [ \partial_u \delta \Phi, \partial_w \Phi ] - [ \partial_u \Phi, \partial_w \delta \Phi ] = 0.
\end{equation}
Furthermore, $\mR \delta \Phi$ is guaranteed to solve the linearized e.o.m.\! by 
\begin{equation}
    \Box \, \mR \delta \Phi -[\partial_u \Phi, \partial_w \mR \delta \Phi] - [\partial_u \mR \delta \Phi, \partial_w \Phi] = [\Box \Phi - [\partial_u \Phi, \partial_w \Phi], \delta \Phi] = 0.
\end{equation}

Note that in this paper we used perturbiners to write down a natural set of $\Phi$'s which solve the equations of motion, as well as a set of $\delta \Phi$'s which solve the linearized e.o.m.\! on the $\Phi$ background. These are of course
\begin{align}
    \Phi = \bigPhi(\Phi_1, \ldots, \Phi_N )\, , \hspace{1 cm} \delta \Phi = \bigPhi(\Phi_1, \ldots, \Phi_N \rvert \Phi_I ).
\end{align}
In Appendix C.2 of \cite{mypaper} (theorem C.3) we give a proof that the action of $\mR$ on the above $\delta \Phi$ is given by
\begin{equation}\label{Recursionsdymz}
    \mR \; \bigPhi(\Phi_1, \ldots, \Phi_N \rvert \Phi_I ) = - z_I \bigPhi(\Phi_1, \ldots, \Phi_N \rvert \Phi_I ).
\end{equation}
This means that, denoting $\delta^{\Delta, a}_{\bn, n} \Phi$ as the action of the $\Ls$ algebra on $\Phi$ defined in \eqref{proposed_action_1} by a contour integral, the above equation implies
\begin{equation}
    \mR (\delta^{\Delta, a}_{\bn, n}  \Phi) = - \delta^{\Delta, a}_{\bn, n-1} \Phi \, .
\end{equation}
Therefore, the recursion operator decrements $n$ by 1. One could say that the recursion operator traverses the ``loop part'' of the loop algebra $\Ls$. Repeatedly acting with $\mR$ generates the tower of variations
\begin{equation}
    \delta^{\Delta, a}_{\bn, n} \Phi \hspace{0.25 cm} \overset{-\mR\;}{\longrightarrow} \hspace{0.25 cm} \delta^{\Delta, a}_{\bn, n-1} \Phi \hspace{0.25 cm} \overset{-\mR\;}{\longrightarrow} \hspace{0.25 cm} \delta^{\Delta, a}_{\bn, n-2} \Phi \hspace{0.25 cm} \overset{-\mR\;}{\longrightarrow} \hspace{0.25 cm}  \ldots .
\end{equation}

Recall that the variations $\delta^{\Delta, a}_{\bn, (n=-\Delta)}$ correspond to pure gauge transformations by \eqref{s_alg_pure_gauge}. Therefore, all $\Ls$ transformations with $n < -\Delta$ can be created by repeatedly acting with $\mR$ on a pure gauge variation.

The general structure of the $\Ls$ transformations within the context of the recursion operators should be understood as follows. First note that, strictly speaking, the function $\mR \delta \Phi$, defined as the solution to the pair of differential equations in \eqref{recursion_1}, is only determined up to the addition of a function of $(\bbu, \bbw)$. This is a technicality which is rarely noted in the literature, although \eqref{Recursionsdymz} happily picks out for us a privileged representative among the set of all possible solutions. In any case, if one acts the recursion operator on the function 0, one could end up with either 0 again or a trivial function which only depends on $(\bbu,\bbw)$. Acting with $\mR$ on a trivial function turns out to give a pure gauge variation, and acting with $\mR$ again gives a non pure gauge variation. The rest of the tower will also be non pure gauge.
\begin{equation}
    \begin{matrix}
        \scriptstyle \text{zero} \\
        {\scriptstyle (n = -\Delta + 2)}
    \end{matrix}
    \hspace{0.25 cm} \overset{-\mR\;}{\longrightarrow} \hspace{0.25 cm}
    \begin{matrix}
        \scriptstyle \text{trivial} \\
        {\scriptstyle (n = -\Delta + 1)}
    \end{matrix}
    \hspace{0.25 cm} \overset{-\mR\;}{\longrightarrow} \hspace{0.25 cm}
    \begin{matrix}
        \scriptstyle \text{pure gauge} \\
        {\scriptstyle (n = -\Delta )}
    \end{matrix}
    \hspace{0.25 cm} \overset{-\mR\;}{\longrightarrow} \hspace{0.25 cm}
    \begin{matrix}
        \scriptstyle \text{non pure gauge} \\
        {\scriptstyle (n = -\Delta -1)}
    \end{matrix}
    \hspace{0.25 cm} \overset{-\mR\;}{\longrightarrow} \hspace{0.25 cm}
    \ldots .
\end{equation}

\section{SDG and the $\Lw$ algebra}

In this section we give an analysis of self-dual gravity and $\Lw$ mirroring our previous analysis for self-dual Yang Mills and $\Ls$.

\subsection{Equations of motion for SDG}\label{sec31}

All self-dual metrics can be written as\footnote{The proof of this fact can be found in Appendix B of \cite{mypaper}.} 
\begin{equation}
    ds^2 = 4 \big( d u \, d \bbu - d w \, d \bbw + (\partial_{w}^2 \phi ) \, d \bbu^2 + (2 \partial_u \partial_w \phi) \, d \bbu d \bbw + (\partial_u^2 \phi) \, d \bbw^2  \big)
\end{equation}
where $\phi$ is a scalar field referred to as Plebański's second scalar. We shall refer to metrics written in this way as being in ``Plebański gauge.'' These metrics can be expressed as
\begin{equation}\label{metric_g_h}
    g[\phi]_{\mu \nu} = \eta_{\mu \nu} + h[\phi]_{\mu \nu}
\end{equation}
where the metric perturbation is given by
\begin{equation}\label{h_phi}
    h[\phi]_{\mu \nu} = \begin{pmatrix} 
    h[\phi]_{uu} & h[\phi]_{u\bu} & h[\phi]_{uw} & h[\phi]_{u \bw} \\
    h[\phi]_{\bu u} & h[\phi]_{\bu \bu} & h[\phi]_{\bu w} & h[\phi]_{\bu \bw} \\
    h[\phi]_{wu} & h[\phi]_{w \bu} & h[\phi]_{ww} & h[\phi]_{w \bw} \\
    h[\phi]_{\bw u} & h[\phi]_{\bw \bu} & h[\phi]_{\bw w} & h[\phi]_{\bw \bw} 
    \end{pmatrix}
    \equiv
    \begin{pmatrix} 
    0 & 0 & 0 & 0 \\
    0 & 4 \partial_w^2 \phi & 0 & 4 \partial_u \partial_w \phi \\
    0 & 0 & 0 & 0 \\
    0 & 4 \partial_u \partial_w \phi & 0 & 4 \partial_u^2 \phi
    \end{pmatrix}.
\end{equation}
Importantly, \eqref{metric_g_h} is an exact non-linear equation, and $h_{\mu \nu}$ need not be small. Due to the form of \eqref{h_phi}, the following exact equation for the inverse metric also holds,
\begin{equation}
    g[\phi]^{\mu \nu} = \eta^{\mu \nu} - h[\phi]^{\mu \nu}\, ,
\end{equation}
where
\begin{equation}
    h[\phi]^{\mu \nu} =\eta^{\mu \alpha} \eta^{\nu \beta} h[\phi]_{\alpha \beta} = g^{\mu \alpha} g^{\nu \beta} h[\phi]_{\alpha \beta} .
\end{equation}
Furthermore,
\begin{align}
     \eta^{\mu \nu} h[\phi]_{\mu \nu} &= 0 = g^{\mu \nu} h[\phi]_{\mu \nu} \\
     \eta^{\mu \nu} \partial_\mu h[\phi]_{\nu \rho} &= 0 =  g^{\mu \nu} \partial_\mu h[\phi]_{\nu \rho} \\
     \det( g[\phi] ) &= 16 =  \det( \eta )
\end{align}
and these equations imply that Plebański gauge is a non-linear version of transverse traceless gauge.

Note that if a function of the form $f_1(\bbu, \bbw) + u f_2 (\bbu, \bbw) + w f_3(\bbu,\bbw)$ is added to $\phi$, the metric $h[\phi]_{\mu \nu}$ does not change. We call these functions ``trivial.''

$\phi$ is required to satisfy Plebański's second heavenly equation, which is
\begin{equation}\label{pleb_2}
    \Box \phi - \{ \partial_u \phi, \partial_w \phi \} = 0,  
    \hspace{1 cm} \{ f, g \} = \pdv{f}{u} \, \pdv{g}{w} - \pdv{f}{w} \, \pdv{g}{u}.
\end{equation}
This is the equation of motion of the action
\begin{equation}\label{Ssdg}
    S_{\rm SDG}(\phi, \bphi) = \int d^4 x \, \bphi \, ( \Box \phi - \{ \partial_u \phi, \partial_w \phi\} )
\end{equation}
where the Lagrange multiplier $\bphi$ has the e.o.m.
\begin{equation}
    \Box \bphi - \{ \partial_u \bphi, \partial_w \phi \} - \{ \partial_u \phi, \partial_w \bphi \} = 0
\end{equation}
which we also note is the linearized e.o.m.\! for $\phi + \epsilon \, \bphi$. While $\bphi$ has been introduced here as a Lagrange multiplier, it should really be understood as characterizing linearized ASD perturbations on the SD background $\phi$. For example, if a single ASD graviton is added to the spacetime, then the ASD component of the curvature tensor $\Psi_{ABCD}$ is proportional to $\bphi$. (See equation 3.17 of \cite{mypaper}.)

There is a second formulation of SDG using anti-self-dual 2-forms, which the reader should feel free to skip on a first pass. If $A, B = 1, 2$ and $\dot{A}, \dot{B} = \dot{1}, \dot{2}$ are spinor indices\footnote{We use conventions $\varepsilon^{1 2} = - \varepsilon_{12} = \varepsilon^{\dot 1 \dot 2} = - \varepsilon_{\dot 1 \dot 2} = 1
$} and $\theta^{A \dot A} = \theta^{A \dot A}_\mu dx^\mu$ are a basis of tetrad 1-forms satisfying
\begin{equation}\label{ds_tetrad}
     ds^2 = \frac{1}{2} \varepsilon_{A B} \varepsilon_{\dot A \dot B} \theta^{A \dot A} \theta^{B \dot B} = \theta^{1 \dot 1} \theta^{2 \dot 2} - \theta^{1 \dot 2} \theta^{2 \dot 1} \, ,
\end{equation}
then there is a convenient parameterization of these 1-forms as
\begin{equation}\label{tetrad}
\begin{aligned}
    \theta^{1 \dot 1} &= 2 d \bbu \, ,  \\
    \theta^{1 \dot 2} &= 2 d \bbw \, , \\
    \theta^{2 \dot 1} &= 2( d w - (\partial_u \partial_w \phi ) d \bbu - (\partial_u^2 \phi) d \bbw ) \, ,\\
    \theta^{2 \dot 2} &= 2 (d u + ( \partial_w^2 \phi) d \bbu + (\partial_u \partial_w \phi) d \bbw ) \, .
\end{aligned}
\end{equation}
Plugging \eqref{tetrad} into \eqref{ds_tetrad}, one recovers the metric in Plebański gauge.

With this basis of 1-forms, we can define a basis of three ASD 2-forms $\Sigma^{AB} = \frac{1}{2} \Sigma^{AB}_{\mu \nu} d x^\mu \wedge dx^\nu$, with $\Sigma^{AB} = \Sigma^{(AB)}$, by
\begin{equation}
    \Sigma^{AB} \equiv \varepsilon_{\dot A \dot B} \theta^{A \dot A} \wedge \theta^{B \dot B}.
\end{equation}
For our given tetrad \eqref{tetrad},
\begin{equation}\label{sigmas}
\begin{aligned}
    \Sigma^{11}[\phi] &=  2 \; d \bbw \wedge d \bbu \, ,\\
    \Sigma^{12}[\phi] &= du \wedge d \bbu - d w \wedge d \bbw \, ,\\
    \Sigma^{22}[\phi] &= 2 ( du + (\partial_w^2 \phi) d\bbu + ( \partial_\bbu \partial_w \phi) d \bbw ) \wedge ( dw - (\partial_u \partial_w \phi) d\bbu - (\partial_u^2 \phi) d \bbw ) \, .
\end{aligned}
\end{equation}

Notice that only $\Sigma^{22}[\phi]$  depends on $\phi$, while $\Sigma^{11}[\phi]$ and $\Sigma^{22}[\phi]$ are independent of $\phi$. All three 2-forms are closed if the e.o.m.\! \eqref{pleb_2} is satisfied:
\begin{equation}
    \dd \Sigma^{AB} = 0.
\end{equation}
It turns out that it is possible to reformulate Einstein gravity in terms of these ASD 2-forms \cite{capovilla1991self,Capovilla:1993cvm}, and it is possible to find a tetrad $\theta^{A \dot A}$ such that $\dd \Sigma^{AB} = 0$ if and only if the metric is self-dual. In this work, we think of the 2-forms as being analogous to the Yang $J$-matrix in SDYM.

\subsection{Abstract definitions of the $\myw_{1 + \infty}$ and $\Lw$ algebras}\label{sec32}

Abstractly, elements of the $\myw_{1+\infty}$ algebra are polynomials in two variables $(u,w)$ where the Lie bracket is the Poisson bracket, given in \eqref{pleb_2}. The generators of this algebra are denoted
\begin{equation}\label{w_gens}
    \ww^{\Delta}_{\bn} \equiv u^{2 - \Delta - \bn} w^{\bn} \, .
\end{equation}
A straightforward computation yields the commutation relation
\begin{equation}\label{w_comm}
    \{ \ww^{\Delta_1}_{\bn_1}, \ww^{\Delta_2}_{\bn_2} \} = \big( \bbn_2 (2 - \Delta_1) - \bbn_1 (2 - \Delta_2) \big) \ww^{\Delta_1 + \Delta_2}_{\bn_1 + \bn_2 - 1} \, .
\end{equation}
When $\Delta$ and $\bbn$ range over all integers, these generators are contained within $\myw_{1+\infty}$.
\begin{equation}
    \ww^{\Delta}_{\bn} \in \myw_{1+\infty} \hspace{0.5 cm} \text{if} \hspace{0.25 cm} \Delta, \bbn \in \mathbb{Z}.
\end{equation}
When the range of $\Delta$ and $\bbn$ are restricted such that the polynomials have non-negative degree, we say they are elements of the wedge subalgebra $\myw_{\wedge} \subset \myw_{1+\infty}$.
\begin{equation}
    \ww^{\Delta}_{\bn} \in \myw_{\wedge} \hspace{0.5 cm} \text{if} \hspace{0.25 cm} \Delta \leq 2, \hspace{0.5 cm} 0 \leq \bbn \leq 2 - \Delta .
\end{equation}

Generators of the loop algebra $\Lw$ are denoted by appending an extra integer label $n$ to the $\myw_{1+\infty}$ generators: 
\begin{alignat}{3}
    \ww^{\Delta}_{\bn, n} & \in \Lw & \hspace{1 cm} & \Delta, \bbn, n \in \mathbb{Z}.
\end{alignat}
The commutation relations of $\Lw$ are then defined to be\footnote{If we make the substitutions $\Delta = - 2 p + 4$, $\bbn = -\frac{\Delta-2}{2} - m$, and $n_{\rm here} = - \frac{\Delta+2}{2} - n_{\rm there}$ we recover the conventions of \cite{Strominger:2021mtt}.}
\begin{align}\label{Lw_comm}
    [ \ww^{\Delta_1}_{\bn_1, n_1}, \ww^{\Delta_2}_{\bn_2, n_2} ] &\equiv \big( \bbn_2 (2 - \Delta_1) - \bbn_1 (2 - \Delta_2) \big) \ww^{\Delta_1 + \Delta_2}_{\bn_1 + \bn_2 - 1, n_1 + n_2+1}.
\end{align}

\subsection{Spin-2 conformal primary modes, half-descendants, and full-descendants in Plebański gauge}\label{sec33}

Just like in the spin-1 case from section \ref{sec23}, we define the spin-2 conformal primary modes of the Plebański scalar, paramterized by $\Delta$, $\bbz$, $z$, as
\begin{equation}\label{conf_prim_grav}
    \phi^\Delta_{\bz, z}(X) \equiv (q(\bbz,z) \cdot X)^{2 - \Delta}.
\end{equation}
Using the 6 Lorentz generators which act on spin-2 fields via the Lie dervatives
\begin{equation}\label{lie_der_2}
\begin{aligned}
    \mL_n h_{\mu \nu} &= (\ell_n)_\mu^{\;\;\; \alpha} h_{\alpha \nu} + (\ell_n)_\nu^{\;\;\; \alpha} h_{\mu \alpha} + \sL_n h_{\mu \nu} \, , \\
    \bmL_n h_{\mu \nu} &= (\Bell_n)_\mu^{\;\;\; \alpha} h_{\alpha \nu} + (\Bell_n)_\nu^{\;\;\; \alpha} h_{\mu \alpha} + \bL_n h_{\mu \nu} \, ,
\end{aligned}
\end{equation}
the modes \eqref{conf_prim_grav} transform like 2d conformal primaries with weights $h = \frac{\Delta+2}{2}$, $\bar{h} = \frac{\Delta-2}{2}$, up to a pure-gauge term for $\mL_1$:
\begin{align}
    \mL_n \, h[\phi^{\Delta}_{\bz, z}]_{\mu \nu} &= \left( \frac{\Delta +2}{2} (n+1) z^n + z^n \partial_z \right) h[\phi^{\Delta}_{\bz, z}]_{\mu \nu}  - 8 \, \delta_{n,1}\partial_{(\mu}   A[\phi^{\Delta}_{\bz, z}]_{\nu)} \, , \\
    \bmL_n \, h[\phi^{\Delta}_{\bz, z}]_\mu &= \left( \frac{\Delta -2}{2} (n+1) \bbz^n + \bbz^n \partial_\bz \right) h[\phi^{\Delta}_{\bz, z}]_{\mu \nu} \, .
\end{align}
The Plebański gauge breaks manifest Lorentz invariance, and in appendix \ref{app_lorentz} we explain how all 6 Lorentz generators act on the Plebański scalar $\phi$. Conveniently, both $\mL_{-1}$ and $\bmL_{-1}$ act through the simple formula
\begin{equation}
    \mL_{-1} h[\phi]_{\mu \nu} = h[ \sL_{-1} \phi ]_{\mu \nu}, \hspace{1 cm} \bmL_{-1} h[\phi]_{\mu \nu} = h[ \bL_{-1} \phi ]_{\mu \nu}.
\end{equation}

Because the modes $\phi^{\Delta}_{\bz, z}$ transform like conformal primaries under the Lorentz algebra, the generators $\sL_{-1}$ and $\bL_{-1}$ simply act via differentiation by $\partial_z$ and $\partial_\bz$.
\begin{align}
    \sL_{-1} \phi^{\Delta}_{\bz, z} = \partial_z \phi^{\Delta}_{\bz, z}\, , \hspace{1 cm}
    \bL_{-1} \phi^{\Delta, a}_{\bz, z} = \partial_\bz \phi^{\Delta}_{\bz, z}\, .
\end{align}
We can now trade the continuous parameters $(\bbz, z) \in \mathbb{C}^2$ for two integer parameters $(\bbn, n) \in \mathbb{Z}^2$ by Taylor expanding around $\bbz = 0$, $z = 0$. If we first Taylor expand in $\bbz$ while holding $z$ fixed, we get the half-descendants
\begin{equation}
    \phi^{\Delta}_{\bn, z} \equiv \left(\frac{\Gamma(\Delta-2+\bbn)}{\Gamma(\Delta-2)}\right) \, (\bL_{-1})^{\bn} \phi^\Delta_{\bz = 0, z}
\end{equation}
where the prefactor above has been chosen such that the half-descendants have the simple functional form
\begin{align}\label{phi_half_descendant}
    \phi^{\Delta}_{\bn, z} &= (u - \bbw z )^{2 - \Delta - \bn} (w - \bbu z)^\bn.
\end{align}
We could also consider the above formula to be the definition of the half-desecendant modes, as this allows us to take $\bbn$ to be any integer, including a negative integer.

Because the half-descendants are equal to the $\ww^\Delta_{\bn}$ monomials \eqref{w_gens} ``shifted'' by $z$, the Poisson bracket of two half-descendants with the same $z$ reproduces the $\myw_{1+\infty}$ algebra exactly:
\begin{equation}\label{sdg_same_z}
    \{ \phi^{\Delta_1}_{\bn_1, z}, \phi^{\Delta_2}_{\bn_2, z} \} = \big( \bbn_2 (2 - \Delta_1) - \bbn_1 (2 - \Delta_2) \big) \phi^{\Delta_1 + \Delta_2}_{\bn_1 + \bn_2 - 1, z}.
\end{equation}
This formula will be useful in section \ref{sec35}.

We finally define the full-descendants $\phi^{\Delta}_{\bn, n}$, Taylor expanding in both $\bbz$ and $z$, through the use of the Cauchy integral formula
\begin{equation}\label{full_descendant_contour}
    \phi^{\Delta}_{\bn, n} \equiv \oint \frac{dz}{2 \pi i} \frac{\phi^{\Delta}_{\bn,z}}{z^{1 + n}} \, .
\end{equation}
In appendix \ref{app_full_descendant}, we calculate the functional form of $\phi^{\Delta}_{\bn,n}$ and find it to be
\begin{equation}\label{Phi_spin_2_expr}
    \phi^{\Delta}_{\bn, n} = \frac{1}{(2 - \Delta - n)!} ( - u  \partial_\bw - w  \partial_\bu)^{2 - \Delta - n} \, \bbu^\bn \, ( - \bbw)^{2 - \Delta - \bn}.
\end{equation}
Notice that when $n > 2 - \Delta$, the above full-descendant mode actually vanishes:
\begin{equation}
    \phi^{\Delta}_{\bn,n} = 0 \hspace{0.5 cm} \text{ if } \hspace{0.5 cm} n > 2-\Delta.
\end{equation}
This is because $\phi^{\Delta}_{\bn,z}$ is a polynomial in $z$ with degree $2 -\Delta$ (see \eqref{phi_half_descendant}) and so if we act on it with $(\partial_z)^n$, for $n > 2 - \Delta$, the polynomial will be annihilated.

\subsection{Review of perturbiner expansion in SDG}\label{sec34}

Let us now review the perturbiner expansion of the Plebański scalar, which was the main subject of \cite{mypaper}. Just as in the spin-1 case from section \ref{sec24}, we shall define the seed functions $\phi_i$ to be a tiny parameter $\epsilon_i$ times a half-descendant mode
\begin{equation}
    \phi_i \equiv \epsilon_i \phi^{\Delta_i}_{\bn_i, z_i}
\end{equation}
and denote the perturbiner expansion of a list of these seed functions as
\begin{equation}\label{bigphi1}
    \bigphi ( \phi_1, \ldots, \phi_N ) \equiv \;\;\;\; \begin{matrix}\text{full perturbiner expansion of Plebański} \\ \text{scalar $\phi$ with seed functions }\phi_1, \ldots, \phi_N \end{matrix} 
\end{equation}
where the use of the large font should be noted. The perturbiner expansion solves the full non-linear equation of motion \eqref{pleb_2} assuming that each infinitesimal parameter squares to zero, $(\epsilon_i)^2 = 0$, but the product of distinct infinitesimal parameters do not square to zero, so $\epsilon_i \epsilon_j \neq 0$ for $i \neq j$.

We further define
\begin{equation}
    \bigphi^{(k)}( \phi_{i_1}, \ldots, \phi_{i_k}) \equiv \; \begin{matrix}\text{sum of terms in } \bigphi(\phi_1, \ldots, \phi_N ) \\ \text{containing } \phi_{i_1}, \ldots, \phi_{i_k} \end{matrix}
\end{equation}
so that we can organize the terms in the perturbiner expansion according to which seed functions are contained within it, writing it as the sum
\begin{equation}
    \bigphi\,(\phi_1, \ldots, \phi_N) = \sum_{k = 1}^N \sum_{ \substack{\{i_1, \ldots, i_k\}   \subset \{1, \ldots, N\} }  } \bigphi^{(k)}( \phi_{i_1}, \ldots, \phi_{i_k}).
\end{equation}

We now define a few objects that will allow us to write down a simple expression for the perturbiner. Define a differential operator $\partial_\mu^{(i)}$ which only acts on the seed function $\phi_i$, but not on $\phi_j$ for $i \neq j$, via
\begin{equation}
    \partial_\mu^{(i)} \phi_i \equiv \partial_\mu \phi_i, \hspace{1 cm} \partial_\mu^{(i)} \phi_j \equiv 0 \hspace{0.5 cm} \text{ for } i \neq j.
\end{equation}
For example
\begin{equation}
    \partial_\mu^{(2)} ( \phi_1 \phi_2 \phi_3 ) = \phi_1 ( \partial_\mu \phi_2 ) \phi_3.
\end{equation}
We then define another differential operator $D^{ij}$ to be
\begin{equation}
    D^{ij} \equiv \partial_u^{(i)} \partial_w^{(j)} - \partial_u^{(j)} \partial_w^{(i)}
\end{equation}
which is essentially the Poisson bracket but it only acts on $\phi_i$ and $\phi_j$.

We will now explain how the perturbiner expansion can be expressed using a certain collection of graphs. Denote  $\mT_N$ to be the collection of connected ``marked tree graphs'' with $N$ or fewer nodes, where each node has a distinct integer label in the set $\{1, 2, \ldots, N\}$ and can connect to arbitrarily many other nodes. Some exampled of graphs contained in $\mathcal{T}_9$ are drawn below.
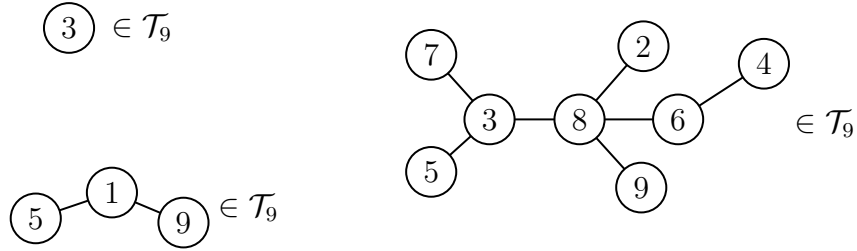
\begin{center}
    \input{figures/gravity_graph_2}
\end{center}
In order to translate a graph into a term in the perturbiner expansion, each node $i$ corresponds to a seed function $\phi_i$ and each edge between $i$ and $j$ corresponds to the operator $D^{ij}/z_{ij}$. So, in particular, two nodes $i$ and $j$ connected by an edge corresponds to the term $\frac{D^{ij}}{z_{ij}} \phi_i \phi_j$.

\begin{center}
    \input{figures/gravity_graph_18}
\end{center}

More generally, each graph $\bt \in \mathcal{T}_n$ is associated with a term $\phi_{\bt}$, given by
\begin{equation}\label{phibt}
    \phi_\bt \equiv \left( \prod_{e_{ij} \, \in \, \text{edges of } \bt} \frac{D^{ij}}{z_{ij}}\right) \left( \prod_{k \, \in \, \text{nodes of } \bt } \phi_k \right) \hspace{0.5 cm} \text{ for } \bt \in \mT_n.
\end{equation}
An example of one such term is
\begin{center}
    \input{figures/gravity_graph_3}
\end{center}
The complete perturbiner expansion then, remarkably, turns out to be equal to the sum over all such tree diagrams
\begin{equation}\label{gravity_perturbiner}
    \bigphi( \phi_1, \ldots, \phi_n ) = \sum_{\bt \in \mT_n} \phi_{\bt}.
\end{equation}
This sum over tree diagrams is analogous to the sum over color orderings in SDYM.

Amazingly, the tree formula for the perturbiner expansion also satisfies a recursion relation. In particular, $\phi^{(k)}$ can be computed from $\phi^{(k-1)}$ via 
\begin{equation}\label{sdg_recursive}
    \bigphi^{(k)} ( \phi_1, \ldots, \phi_k ) = \sum_{i = 1}^{k-1} \bigphi^{(k-1)}  ( \phi_1, \ldots, \phi_{k-1} ) \bigg\rvert_{\displaystyle{  \phi_i \mapsto \frac{1}{z_{ik}} \{ \phi_i, \phi_k\} }  }.
\end{equation}
The proofs of equations \eqref{gravity_perturbiner} and \eqref{sdg_recursive} can be found in \cite{mypaper} as Theorem 2.1 and Theorem A.1, respectively. The above formula is highly reminiscent of a Ward identity in a 2d CFT, and in the next subsection we will get some mileage from this observation.

For a simple example of how one can use the recursive formula, when $N = 2$ the perturbiner expansion is
\begin{equation}
    \begin{aligned}
        \bigphi(\phi_1, \phi_2) &= \bigphi^{(1)}(\phi_1) +  \bigphi^{(1)}(\phi_2) +  \bigphi^{(2)}(\phi_1, \phi_2) \\
        &= \phi_1 + \phi_2 + \frac{1}{z_{12}}\{ \phi_1, \phi_2 \}.
    \end{aligned}
\end{equation}

\subsection{$\Lw$ action on spacetime metric}\label{sec35}

We have now defined all of the ingredients necessary to explain how the $\Lw$ algebra acts on the Plebański scalar $\phi$.

Let us notate the linear change in the perturbiner associated to adding in a single seed function by
\begin{align}
    \bigphi( \phi_1, \ldots, \phi_N \rvert \phi_{I} ) &\equiv \bigphi(\phi_1, \ldots, \phi_N, \phi_{I} ) - \bigphi(\phi_1, \ldots, \phi_N).
\end{align}

We denote $ \delta^{\Delta_{I}}_{\bn_I, n_I}$ to be the infinitesimal change in the perturbiner expansion which results from integrating the $z_I$ variable of the added half-descendant seed function $\phi^{\Delta_I}_{\bn_I, z_I}$ in a contour surrounding all previous insertions, weighted by $1/z_I^{n_I + 1}$:
\begin{equation}\label{lw_action_phi}
    \delta^{\Delta_{I} }_{\bn_I, n_I} \bigphi( \phi_1, \ldots, \phi_{N} ) \equiv \frac{1}{2 \pi i} \oint_{C} \frac{d z_I}{z_I^{\, n_I+1}} \bigphi\left(\phi_1, \ldots, \phi_{N} \rvert \phi_I \right).
\end{equation}
We run the same argument as in section \ref{sec25} and find that the commutator of this action is
\begin{equation}
\begin{aligned}
    &[ \delta^{\Delta_{I} }_{\bn_I, n_I}, \delta^{\Delta_{J} }_{\bn_{J}, n_{J} } ] \bigphi(\phi_1, \ldots, \phi_N) \\
    & = \frac{1}{(2 \pi i)^2} \left( \oint_{C} \frac{dz_I}{z_I^{\, n_I+1}}  \oint_{C'} \frac{dz_J}{z_J^{\, n_J+1}} -  \oint_{C'} \frac{dz_I}{z_I^{\, n_I+1}}  \oint_{C} \frac{dz_J}{z_J^{\, n_J+1}}  \right) \bigphi( \phi_1, \ldots, \phi_N, \phi_I, \phi_J) \\
    &=   \oint_{C } \frac{dz_I}{2 \pi i}  \mathop{\mathrm{Res}}_{z_J \to z_I} \frac{1}{z_I^{\, n_I+1}}  \frac{1}{z_J^{\, n_J+1}} \bigphi( \phi_1, \ldots, \phi_N, \phi_I, \phi_J). \label{eq_res}
\end{aligned}
\end{equation}
To find the residue above, we must determine how the perturbiner expansion behaves when $z_J \to z_I$. From \eqref{sdg_recursive}, we can immediately read off that the only singular term is
\begin{align}
    \lim_{z_J \to z_I}  \bigphi( \phi_1, &\ldots, \phi_k, \phi_I, \phi_J) = \bigphi( \phi_1, \ldots, \phi_k \rvert \frac{1}{z_{IJ}} \{ \phi_I, \phi_J \}\Big\rvert_{z_J = z_I} ) .
\end{align}
Using the relation \eqref{sdg_same_z}, which states that the Poisson bracket of two half-descendant modes with the same $z$ satisfies the $\myw_{1 + \infty}$ commutation relation, we know this is
\begin{equation}\label{eq_lim}
    = \frac{1}{z_{IJ}} \Big( \bbn_J (2 - \Delta_I) - \bbn_I (2 - \Delta_J) \Big) \bigphi( \phi_1, \ldots, \phi_k \rvert  \epsilon_I \epsilon_J \phi^{\Delta_I + \Delta_J}_{\bn_I + \bn_J - 1, z_I}  ), 
\end{equation}
and we can then plug \eqref{eq_lim} into \eqref{eq_res} and deduce
\begin{equation}\label{check_lw_comm}
\boxed{ \textcolor{white}{\Bigg\rvert}
    [ \delta^{\Delta_{I} }_{\bn_I, n_I}, \delta^{\Delta_{J} }_{\bn_{J}, n_{J} } ] = \Big( \bbn_J (2 - \Delta_I) - \bbn_I (2 - \Delta_J) \Big) \delta^{\Delta_I + \Delta_J }_{\bn_I + \bn_J-1, n_I + n_J+1}.  } 
\end{equation}
This means that our variation \eqref{lw_action_phi} on the space of self-dual metrics really does satisfy the $\Lw$ commutation relation \eqref{Lw_comm}.

We can also express the action of an arbitrary number of $\Lw$ transformations on the flat space $\phi = 0$ vacuum as
\begin{equation}\label{new_formula_spin2}
\begin{aligned}
    (1 + \delta^{\Delta_N}_{\bn_N, n_N}) \ldots \, & (1 +  \delta^{\Delta_2}_{\bn_2, n_2}) (1 + \delta^{\Delta_1}_{\bn_1, n_1}) \cdot (\phi = 0) \\
    &= \bigphi\left( \oint_{C_1} \frac{dz_1}{2 \pi i} \frac{\phi_1}{z_1^{\; n_1+1}},  \oint_{C_2} \frac{dz_2}{2 \pi i} \frac{\phi_2}{z_2^{\; n_2+1}}, \ldots,  \oint_{C_N} \frac{dz_N}{2 \pi i} \frac{\phi_N}{z_N^{\; n_N+1}} \right) 
\end{aligned}
\end{equation}
where contour $C_j$ encloses $C_i$ if $j>i$. Because the perturbiner expansion is known exactly, the above formula can be expanded out and be used to calculate the metric explicitly if desired. See appendix \ref{app_example} for example.

\subsection{Residual diffeomorphisms of Plebański gauge}\label{sec36}

The set of residual diffeomorphisms of Plebański gauge were written down by Campiglia and Nagy in \cite{Campiglia:2021srh}. We begin by quoting their result, and will then compute the Lie brackets of their vector fields, finding them to generate an algebra we denote $\myw_{\infty} \ltimes \mathfrak{f}$. See also \cite{Nagy:2022xxs}.

There are two families of residual diffeomorphisms, each parameterized by their own scalar functions of $\bbu, \bbw$, which we denote as $\alpha$ and $\beta$, respectively.
\begin{equation}
    \alpha = \alpha(\bbu, \bbw), \hspace{1 cm} \beta = \beta(\bbu, \bbw).
\end{equation}
We also remind the reader of the definition of the Lorentz generator $\sL_1$ from \eqref{list_lorentz_gen},
\begin{equation}
    \sL_1 = u \partial_\bw + w \partial_\bu \, .
\end{equation}

The first family of diffeomorphisms is parameterized by $\alpha$ via
\begin{align}
    \sF[\alpha]^\mu \partial_\mu \equiv (\partial_\bu \alpha) \partial_u - (\partial_\bw \alpha ) \partial_w.
\end{align}
These diffeomorphisms induce a variation $\delta_\alpha \phi$ on the Plebański scalar
\begin{equation}
    \mathcal{L}_{\sF[\alpha]} g[\phi]_{\mu \nu} = h[\delta_\alpha \phi]_{\mu \nu}
\end{equation}
which is given by
\begin{equation}\label{delta_alpha_phi}
    \delta_\alpha \phi = \frac{1}{2} ( \sL_1 )^2 \alpha + \sF[\alpha]^\mu \partial_\mu \phi.
\end{equation}

The second family of residual diffeomorphisms, parameterized by the function $\beta$, are
\begin{align}
    \sW[\beta]^\mu \partial_\mu &\equiv \big( \sL_1 \partial_\bu \, \beta \big) \, \partial_u - (\partial_\bw \beta) \partial_\bu - \big( \sL_1 \partial_\bw \, \beta \big) \, \partial_w + (\partial_\bu \beta) \partial_\bw
\end{align}
and induce a change on the Plebański scalar $\delta_\beta \phi$
\begin{equation}
    \mathcal{L}_{\sW[\beta]} g[\phi]_{\mu \nu} = h[\delta_\beta \phi]_{\mu \nu}
\end{equation}
equal to
\begin{equation}\label{delta_beta_phi}
    \delta_\beta \phi = \frac{1}{6} (\sL_1)^3 \beta + \sW[\beta]^\mu \partial_\mu \phi.
\end{equation}

Let us now compute the Lie bracket of these vector fields. In order to express the answers cleanly, we define the starred Poisson bracket, using variables $\bbu, \bbw$, via
\begin{equation}
    \{ f, g \}^* \equiv \pdv{f}{\bbu} \pdv{g}{\bbw} - \pdv{f}{\bbw} \pdv{g}{\bbu}.
\end{equation}
The Lie brackets somewhat miraculously turn out to be
\begin{equation}
\begin{aligned}
    [\, \sW[\beta_1], \sW[\beta_2] \,] &= \sW[ \{ \beta_1, \beta_2 \}^* ] \, ,\\
    [\, \sW[\beta] \, ,\,  \sF[\alpha]\, \,] &= \sF[ \{ \beta, \alpha \}^* ] \, ,\\
    [\, \sF[\alpha_1] , \sF[\alpha_2] \,] &= 0 \, .
\end{aligned}
\end{equation}

If we define a basis of polynomial functions for $\alpha$ and $\beta$
\begin{equation}
    \alpha^\Delta_\bn \equiv \bbu^\bn (-\bbw)^{2 - \Delta - \bn} , \hspace{1 cm} \beta^\Delta_\bn \equiv \bbu^\bn  (-\bbw)^{2 - \Delta - \bn} ,
\end{equation}
and their corresponding vector fields
\begin{equation}\label{F_W_def}
    \sF^\Delta_\bn \equiv \sF[\alpha^\Delta_\bn] \, , \hspace{1 cm} \sW^\Delta_\bn \equiv \sW[\beta^\Delta_\bn] \, ,
\end{equation}
then the Lie brackets can be expressed as
\begin{equation}
\begin{aligned}
    [\, \sW^{\Delta_1}_{\bn_1}, \sW^{\Delta_2}_{\bn_2} \,] &= \big( \bbn_2 (2 - \Delta_1) - \bbn_1 (2 - \Delta_2) \big) \sW^{\Delta_1 + \Delta_1}_{\bn_1 + \bn_2 - 1}, \\
    [\, \sW^{\Delta_1}_{\bn_1} \, ,\,  \sF^{\Delta_2}_{\bn_2} \,] &= \big( \bbn_2 (2 - \Delta_1) - \bbn_1 (2 - \Delta_2) \big) \sF^{\Delta_1 + \Delta_1}_{\bn_1 + \bn_2 - 1} , \\
    [\, \sF^{\Delta_1}_{\bn_1} , \sF^{\Delta_1}_{\bn_1} \,] &= 0 \, .
\end{aligned}
\end{equation}

Thus we see that the $W^{\Delta}_{\bn}$ vector fields generate a $\myw_{\infty}$ algebra of diffeomorphisms while the $F^{\Delta}_{\bn}$ vectors fields generate an abelian algebra of diffeomorphisms that we will name the ``$\mathfrak{f}$ algebra.'' The full algebra of diffeomorphisms is $\myw_{\infty} \ltimes \mathfrak{f}$. ($\myw_{\infty}$ is simply $\myw_{1+\infty}$ where we remove the constant commuting element with $\Delta = 2$, $\bbn = 0$.)

Here are some special cases of vector fields. The four spacetime translations are
\begin{equation}
    F^1_0 = -\partial_w, \hspace{0.5 cm} F^1_1 = \partial_u, \hspace{0.5 cm} W^1_1 = -\partial_\bu , \hspace{0.5 cm} W^1_1 =\partial_\bw,
\end{equation}
and the three anti-holomorphic Lorentz generators are
\begin{equation}
    W^0_m = -2 \, \bL_{1-m} \hspace{1 cm} \text{for }m = 0,1,2.
\end{equation}
The three holomorphic Lorentz generators are not part of the algebra.

We also have an infinite number of supertranslation and superrotation vector fields, denoted $\xi_f$ and $\xi_Y$. If we switch to flat Bondi coordinates $(U,R,Z,\bZ)$, defined in \eqref{flat_bondi_coords}, then to leading order in $1/R$ these vector fields are parameterized by $f = f(Z, \bZ)$, $Y^Z= Y^Z(Z)$, and $Y^\bZ = Y^\bZ(\bZ)$ as
\begin{align}
    \xi_f &= f \; \partial_U - \frac{1}{R} (\partial_\bZ f  )\, \partial_Z - \frac{1}{R} (\partial_Z f )\, \partial_\bZ + (\partial_Z \partial_\bZ f  )\, \partial_R + \ldots \, ,  \\
    \xi_Y &= Y^Z \, \partial_Z + Y^\bZ \partial_\bZ + \frac{U}{2} (\partial_Z Y^Z + \partial_\bZ Y^\bZ) \, \partial_U - \frac{R}{2} (\partial_Z Y^Z + \partial_\bZ Y^{\bZ})\, \partial_R + \ldots \, .
\end{align}
An explicit computation yields, for all $\bbn \in \mathbb{Z}$,
\begin{align}
    \sF^1_\bn &= \xi_f && \hspace{-3cm} \text{for } f = (-\bZ)^{1- \bn} ,\\
    \sW^1_\bn &= \xi_f  &&\hspace{-3cm}\text{for } f = -Z (-\bZ)^{1- \bn}, \\
    \sW^0_\bn &= \xi_Y && \hspace{-3cm}\text{for } Y^Z = 0, \;\; Y^\bZ = 2 (-\bZ)^{2 - \bn},
\end{align}
where the above vector fields are equal to leading order in $1/R$ \cite{Campiglia:2021srh}.

Thus, we see that the $\mathfrak{f}$ algebra of $F$ vector fields generalizes the set of antiholomorphic supertranslations and the $\myw_{\infty}$ algebra of $W$ vector fields generalizes the set of antiholomorphic superrotations. Beyond the supertranslations and superrotations at $\Delta = 1$ and $\Delta = 0$, however, $\myw_{\infty} \ltimes \mathfrak{f}$ also contains two infinite towers of overleading diffeomorphisms as $\Delta$ becomes more negative \cite{Nagy:2022xxs}.

\subsection{Pure gauge and ``trivial'' descendants of spin-2 primaries}\label{sec37}
The graviton conformal primaries with integer dimensions $\Delta = 2, 1, 0, -1, \ldots$ have anti-holomorphic and holomorphic weights given by $-2 \bar{h} = - \Delta + 2$ and $-2 h = - \Delta - 2$. Repeating the spin-1 discussion from section \ref{sec27} for the spin-2 case, we notice that it seems as though there should be ``four fewer'' holomorphic descendants than antiholomorphic descendants in the finite dimensional $\mathfrak{sl}(2, \mathbb{R})$ representations, which seems strange given that differentiation by $\partial_z$ and $\partial_\bz$ act in the same way on $\phi^{\Delta}_{\bz,z}$. We shall show that these ``four fewer'' descendants are either pure diffeo or trivial, decomposing as
\begin{equation}
\begin{aligned}
        n &= - \Delta - 1 && \text{ is pure diffeo, } \\
        n &= - \Delta  && \text{ is pure diffeo, } \\
        n &= - \Delta +1 && \text{ is trivial, } \\
        n &= - \Delta +2 && \text{ is trivial.}
\end{aligned}
\end{equation}
We already know that $\phi^{\Delta}_{\bn,n}$ can be expressed as
\begin{equation}
    \phi^{\Delta}_{\bn, n} = \frac{1}{(2 - \Delta - n)!} ( - u  \partial_\bw - w  \partial_\bu)^{2 - \Delta - n} \, \bbu^\bn \, ( - \bbw)^{2 - \Delta - \bn}
\end{equation}
from \eqref{Phi_spin_2_expr}. For the four cases $n = - \Delta -1$, $n= -\Delta$, $n= -\Delta+1$, $n= -\Delta +2$, this formula becomes
\begin{equation}\label{spin_2_list}
\begin{aligned}
    \phi^{\Delta, a}_{\bn, (n = -\Delta-1)} &= - \tfrac{1}{6}( u \partial_\bw + w \partial_\bu)^3 \, \bbu^{\bn} (-\bbw)^{2 - \Delta - \bn} \, , \\
    \phi^{\Delta, a}_{\bn, (n = -\Delta)} &= \tfrac{1}{2}(u \partial_\bw + w \partial_\bu)^2 \, \bbu^\bn (-\bbw)^{2 - \Delta - \bn} \, ,\\
    \phi^{\Delta, a}_{\bn, (n = -\Delta+1)} &= -(u \partial_\bw + w \partial_\bu) \, \bbu^{\bn} (-\bbw)^{2 - \Delta - \bn} \, , \\
    \phi^{\Delta, a}_{\bn, (n = -\Delta+2)} &= \bbu^\bn (-\bbw)^{2 - \Delta - \bn} \, .
\end{aligned}
\end{equation}
Using \eqref{delta_alpha_phi}, \eqref{delta_beta_phi}, we then indeed see that the $n = - \Delta-1$ and $n = - \Delta$ wavefuctions are linearized pure gauge
\begin{align}
    h[\phi^{\Delta}_{\bn, (n = -\Delta-1)}]_{\mu \nu} &= -  \mathcal{L}_{W^\Delta_\bn} \eta_{\mu \nu} \\
    \hspace{1 cm} h[\phi^{\Delta}_{\bn, (n = -\Delta)}]_{\mu \nu} &= + \mathcal{L}_{F^\Delta_\bn} \eta_{\mu \nu}
\end{align}
and the $n = - \Delta+1,  - \Delta+2$ wavefunctions are trivial, because in these cases the Plebański scalar does not contain any powers of $u$ or $w$ to the second power or higher.
\begin{align}
    h[\phi^{\Delta}_{\bn, (n = -\Delta+1)}]_{\mu \nu} &= 0 \\
    h[\phi^{\Delta}_{\bn, (n = -\Delta+2)}]_{\mu \nu} &= 0
\end{align}

\subsection{The $\myw_{\infty} \ltimes \mathfrak{f}$ large-diffeo  algebra within $\Lw$}\label{sec38}

In the last section we saw that the graviton wave functions $\phi^{\Delta}_{\bn, (n  = - \Delta-1)}$, $\phi^{\Delta}_{\bn, (n  = - \Delta)}$ were pure linearized diffs. If we add one of these pure diffeomorphism wave-functions to a pre-existing list of seed functions in a perturbiner expansion, we should expect that this will have the effect of performing said diffeomorphism on the background spacetime created by all of the pre-existing seed functions. We will now show that this is indeed the case, in exact analogy with what we saw for SDYM in section \ref{sec28}.

We already know that if we vary the metric by a residual diffeomorphism of Plebański gauge, $F[\alpha]^\mu$ or $W[\beta]^\mu$, the Plebański scalar varies as
\begin{equation}
\begin{aligned}
    \delta_\alpha \phi = \underbrace{\frac{1}{2} ( \sL_1 )^2 \alpha}_{0^{\rm th} \text{ order}} + \underbrace{\sF[\alpha]^\mu \partial_\mu \phi}_{1^{\rm st} \text{ order}}, \hspace{1.25 cm} \delta_\beta \phi = \underbrace{\frac{1}{6} (\sL_1)^3 \beta}_{0^{\rm th} \text{ order}} + \underbrace{\sW[\beta]^\mu \partial_\mu \phi}_{1^{\rm st} \text{ order}}.
\end{aligned}
\end{equation}
We will now use our explicit formula for the action of $\Lw$ on the Plebański scalar, \eqref{lw_action_phi}, to show that the transformations $\delta^{\Delta}_{\bn,(n=-\Delta-1)}$ and $\delta^{\Delta}_{\bn,(n=-\Delta)}$ equal the action of the diffeomorphisms $W^{\Delta}_{\bn}$ and $F^{\Delta}_{\bn}$. These $\Lw$ transformations corresponds to adding the pure diff wavefunctions $\phi^{\Delta}_{\bn, (n  = - \Delta-1)}$ or $\phi^{\Delta}_{\bn, (n  = - \Delta)}$ to the list of seed functions.

\begin{figure}
    \centering
\input{figures/three_graphs}
    \caption{    \label{fig:three_trees} Three types of trees which arise when we add in a pure-diff seed function. The first type of tree is just the pure-diff node itself. The second type of tree will have the pure-diff node connected to exactly one other node. The third type of tree will have the pure-diff node connected to two or more other nodes.}
\end{figure}
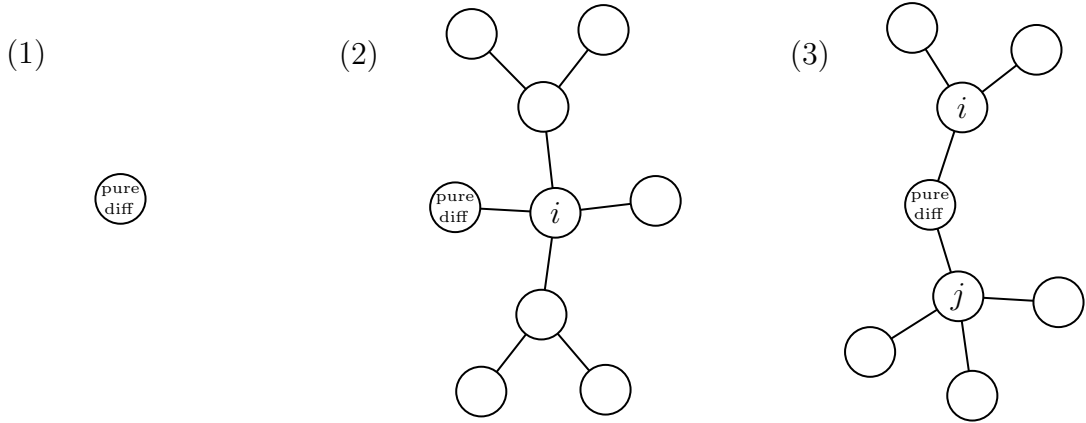

We now calculate
\begin{equation*}
    \delta^{\Delta}_{\bn, (n = - \Delta-1)} \bigphi( \phi_1, \ldots, \phi_N)
\end{equation*}
using the tree formula \eqref{gravity_perturbiner}. (The analogous calculation of $\delta^{\Delta}_{\bn, (n = - \Delta)}$ follows along similar lines.) There will be three types of trees which will contribute to the above quantity, all of which are shown in figure \ref{fig:three_trees}.

The first type of tree (1) is just the single node representing the pure-diff wavefunction. In the second type of tree (2), the pure-diff node only connects to a single other node. In the third type of tree (3), the pure-diff node connects to two or more other nodes.

Of course, the ``background field'' created by the pre-existing seed functions $\phi_1, \ldots, \phi_N$ is made out of the sum of trees not containing the pure-diff node. Tree (1) can be thought of as a $0^{\rm th}$ order variation which does not depend on the original background. Likewise, the sum of the (2) trees will produce a variation which is $1^{\rm st}$ order in background. The sum of the (3) trees would in principle produce terms that would be higher order in the background, but we will show these trees evaluate to zero.

The term corresponding to tree (1) is
\begin{equation}
    \frac{1}{2 \pi i} \oint_C \frac{dz}{z^{-\Delta}} \phi^{\Delta}_{\bn, z} = \phi^{\Delta}_{\bn, (n = - \Delta-1)} = - \frac{1}{6} (\sL_1)^3 \beta^\Delta_\bn.
\end{equation}
For a tree of type (2), let's assume the pure-diff node attaches to a node labelled $i$. We now look at the part of the tree-term corresponding to the connection between the pure-diff node and $i$, which is
\begin{equation}\label{eq_379}
\begin{aligned}
    \frac{1}{2 \pi i} \oint_C \frac{dz}{z^{-\Delta}} \frac{1}{z - z_i} \{ \phi^{\Delta}_{\bn, z} , \phi_i \} &= \frac{1}{2 \pi i} \oint_C \frac{dz}{z^{-\Delta}} \frac{1}{z} \sum_{r = 0}^\infty \frac{z_i^r}{z^r} \{ \phi^{\Delta}_{\bn, z} , \phi_i \} \\
    &= \sum_{r = 0}^\infty z_i^r \{ \phi^\Delta_{\bn,(n = - \Delta +r)}, \phi_i \} \\
    &= \{ \phi^{\Delta}_{\bn, (n = - \Delta)}, \phi_i \} + z_i \{ \phi^{\Delta}_{\bn, (n = - \Delta+1 )}, \phi_i \}.
\end{aligned}
\end{equation}
In the above line, only the $r = 0$ and $r = 1$ terms in the sum contributed. All $r > 2$ terms are zero because $\phi^{\Delta}_{\bn,n}$ is zero for $n > -\Delta + 2$. The potential $r = 2$ term evaluates to 0 because the wave function $\phi^{\Delta}_{\bn, (n = - \Delta +2)}$ is independent of $u$ and $w$ from \eqref{spin_2_list}, and thus is annihilated by the $\partial_u$ and $\partial_w$ derivatives of the Poisson bracket.

Now, in the full term there are other seed functions present, coming from the other nodes of the tree. Say a particular tree $\bt \in \mT_N$ under consideration has nodes $i_1, \ldots, i_k$, which implies, from \eqref{phibt},
\begin{equation*}
    \phi_{\bt} \propto \phi_{i_1} \ldots \phi_{i_k}.
\end{equation*}
Momentarily abusing notation and treating $\phi^{\Delta}_{\bn, z}$ as the ``$i = 0$'' seed function, the full tree term arising from \eqref{eq_379} is
\begin{equation}\label{tree_term_arising}
\begin{aligned}
    \sum_{j = 1}^k D^{0 \; i_j} \phi^{\Delta}_{\bn,(n = - \Delta)} \phi_{\bt} + \sum_{j = 1}^k D^{0 \; i_j} z_{i_j} \phi^{\Delta}_{\bn,(n = - \Delta+1)} \phi_{\bt}  \;\; \subset \;\; 
    \delta^{\Delta}_{\bn, (n = - \Delta-1)} \bigphi( \phi_1, \ldots, \phi_N).
\end{aligned}
\end{equation}
Now, using the the elementary property 
\begin{equation}
    - z_{i_j} \partial_w \phi_{i_j} = \partial_\bu \phi_{i_j}, \hspace{1.5 cm} - z_{i_j} \partial_u \phi_{i_j} = \partial_\bw \phi_{i_j},
\end{equation}
(which just follows from \eqref{phi_half_descendant}) and
\begin{equation}
    \phi^{\Delta}_{\bn, (n = - \Delta)} = \frac{1}{2} (\sL_1)^2 \beta^\Delta_\bn \, , \hspace{1 cm} \phi^{\Delta}_{\bn, (n = - \Delta + 1)} = - ( \sL_1) \beta^\Delta_\bn \, ,
\end{equation}
we can rewrite \eqref{tree_term_arising} as
\begin{equation}
\begin{aligned}
    \partial_u\left(  \frac{1}{2} (\sL_1)^2 \beta^\Delta_\bn\right) & \partial_w \phi_\bt - \partial_w\left(  \frac{1}{2} (\sL_1)^2 \beta^\Delta_\bn\right) \partial_u \phi_\bt + \partial_u \left( \sL_1 \beta^{\Delta}_\bn \right) \partial_\bu \phi_\bt - \partial_w \left( \sL_1 \beta^{\Delta}_\bn \right) \partial_\bw \phi_\bt \\ & \\ &= -(W^\Delta_\bn)^\mu \partial_\mu \phi_\bt  \;\;\; \subset \;\;\;
    \delta^{\Delta}_{\bn, (n = - \Delta-1)} \bigphi( \phi_1, \ldots, \phi_N).
\end{aligned}
\end{equation}
Therefore, the effect of summing up all type (2) graphs is to act with $-(W^\Delta_\bn)^\mu \partial_\mu$ on the Plebański scalar pre-adding-in the pure diff node.

Finally, we come to the graphs of type (3). Say we have a pure-diff node connected to two other nodes $i$ and $j$. We look at the subset of the tree-term corresponding to these two connections, which is
\begin{equation}
\begin{aligned}
    \frac{1}{2 \pi i} \oint_C \frac{dz}{z^{-\Delta}} \frac{D^{0i}}{z - z_i} \frac{D^{0j}}{z-z_j}   \phi^{\Delta}_{\bn, z} \phi_i \phi_j &= \frac{1}{2 \pi i} \oint_C \frac{dz}{z^{-\Delta}} \frac{1}{z^2} \sum_{r = 0}^\infty \frac{z_i^r}{z^r} \sum_{s = 0}^\infty \frac{z_j^s}{z^s} D^{0i} D^{0j} \phi^{\Delta}_{\bn, z} \phi_i \phi_j\\
    &= \sum_{r = 0}^\infty \sum_{s = 0}^\infty z_i^r z_j^s D^{0 i} D^{0 j} \phi^{\Delta}_{\bn,(n =  - \Delta +1 + r + s)} \phi_i \phi_i \\
    &= 0
\end{aligned}
\end{equation}
and turns out to be zero. This computation is similar to \eqref{eq_379}. Note that the only possibly non-zero term is killed because $\phi^\Delta_{\bn, n = -\Delta+1}$ is linear in $u$ and $w$ and is annihilated by the two $\partial_u$, $\partial_w$ derivatives from $D^{0i} D^{0j}$. 

Futhermore, based on the previous computation it is clear that if the pure-diff node had connected to three or more other nodes, which would introduce more factors like $1/(z - z_i)$, the associated tree terms would also be zero.

Combining the terms from the trees (1), (2), and (3), we have
\begin{equation}
    \delta^{\Delta}_{\bn, (n = - \Delta-1)} \bigphi( \phi_1, \ldots, \phi_N) = - \frac{1}{6} (\sL_1)^3 \beta^\Delta_\bn - (W^\Delta_\bn)^\mu \partial_\mu \bigphi( \phi_1, \ldots, \phi_N).
\end{equation}
A similar computation yields, for the $F$ vector fields,
\begin{equation}
    \delta^{\Delta}_{\bn, (n = - \Delta)} \bigphi( \phi_1, \ldots, \phi_N) = \frac{1}{2} (\sL_1)^2 \alpha^\Delta_\bn + (F^\Delta_\bn)^\mu \partial_\mu \bigphi( \phi_1, \ldots, \phi_N).
\end{equation}

We can rewrite these equations as
\begin{equation}
\begin{aligned}
    \delta^{\Delta}_{\bn, (n = - \Delta-1)} \bigphi( \phi_1, \ldots, \phi_N) = - \delta_{\beta^\Delta_\bn}  \bigphi( \phi_1, \ldots, \phi_N), \\
    \delta^{\Delta}_{\bn, (n = - \Delta)} \bigphi( \phi_1, \ldots, \phi_N) = + \delta_{\alpha^\Delta_\bn}  \bigphi( \phi_1, \ldots, \phi_N).
\end{aligned}
\end{equation}
We then see that the RHS of the above equations are exactly those given by the diffeomorphisms $W^\Delta_\bn$ and $F^\Delta_\bn$ from \eqref{delta_beta_phi} and \eqref{delta_alpha_phi}, which is exactly what we intended to show. We can at this stage discard the notation of the perturbiner expansion, and simply write
\begin{equation}
\begin{aligned}
    \delta^{\Delta}_{\bn, (n = - \Delta-1)} \phi &= -\delta_{\beta^\Delta_{\bn}} \phi \, ,\\
    \delta^{\Delta}_{\bn, (n = - \Delta)} \phi &= + \delta_{\alpha^\Delta_{\bn}} \phi \, .
\end{aligned}
\end{equation}
We could also perform a similar computation on the $n = - \Delta + 1$ and $n = - \Delta+2$ variations, finding
\begin{equation}
    \begin{aligned}
        \delta^{\Delta}_{\bn, (n = - \Delta+1)} \phi &= \phi^{\Delta}_{\bn, (n = - \Delta+1)} = - (u \partial_\bw + w \partial_\bu) \bbu^\bn (-\bbw)^{2 - \Delta - \bn},\\
    \delta^{\Delta}_{\bn, (n = - \Delta+2)} \phi &= \phi^{\Delta}_{\bn, (n = - \Delta+2)} = \bbu^\bn (- \bbw)^{2 - \Delta - \bn} ,
    \end{aligned}
\end{equation}
which are indeed trivial transformations. Performing the computation for $n > - \Delta + 2$, we further find these $\Lw$ transformations are zero on the Plebański scalar.
\begin{equation}
    \delta^{\Delta}_{\bn, (n > - \Delta+2)} \phi = 0
\end{equation}

\subsection{The recursion operator $\mR$ and $\Lw$ in SDG}\label{sec39}

In SDG, the recursion operator $\mR$ maps the space of linearized perturbations of the Plebański scalar $\phi$ into itself. In particular, if $\delta \phi$ is a linearized perturbation of $\phi$, $\mR \delta \phi$ is defined as the solution to the pair of differential equations
\begin{equation}
\begin{aligned}
    \partial_u ( \mR \delta \phi) &= \partial_\bw \delta \phi + \{ \partial_u \phi, \delta \phi \}, \\
    \partial_w ( \mR \delta \phi) &= \partial_\bu \delta \phi + \{ \partial_w \phi, \delta \phi \}.
\end{aligned}
\end{equation}

In SDYM, we motivated the definition of the recursion operator by looking at all of the allowed ``quasi-gauge transformations'' of the Yang $J$-matrix. Here we will give an analogous interpretation of the recursion operator in SDG. Just as there were two formulations of SDYM (with the $\Phi$ scalar and the Yang $J$-matrix) there are also two formulations of SDG (with the $\phi$ scalar and the ASD 2-forms $\Sigma^{AB}$).

In the convenient choice of 2-forms from equation \eqref{sigmas}, we had that $\Sigma^{11}$ and $\Sigma^{12}$ were independent of $\phi$ and only $\Sigma^{22}$ depended on $\phi$.

A true diffeomorphism by a vector field $\xi$ would act on all three 2-forms by $\delta \Sigma^{AB} = \mathcal{L}_\xi \Sigma^{AB}$. However, because $\Sigma^{11}$ and $\Sigma^{12}$ are independent of $\phi$, with analogy with the quasi-gauge transformations of the Yang $J$-matrix, one might consider transformations which only affect $\Sigma^{22}$ but leave $\Sigma^{11}$, $\Sigma^{12}$ untouched. It turns out that, via explicit computation, one can verify that
\begin{equation}
    \Sigma^{22}[\phi + \mR \delta \phi] - \Sigma^{22}[\phi] = \mL_{\mX_{\delta \phi}} \Sigma^{22}[\phi],
\end{equation}
where $\mX_f$ denotes the spacetime Hamiltonian vector field generated by the spacetime function $f = f(u, \bbu, w, \bbw)$,
\begin{equation}
    \mX_f \equiv \{\cdot, f \} = -(\partial_u f ) \partial_w + (\partial_w f) \partial_u.
\end{equation}
This variation corresponds to the ``quasi-diffeo''
\begin{equation}\label{2_form_var_1}
\begin{aligned}
    \delta \Sigma^{11} &= 0 \\
    \delta \Sigma^{12} &= 0 \\
    \delta \Sigma^{22} &= \mL_{\mX_{\delta \phi}} \Sigma^{22}.
\end{aligned}
\end{equation}
where the linearized perturbation $\delta \phi$ plays the role of the Hamiltonian generator $f$. Note that the quasi-diffeo preseves the self-duality condition $\dd \Sigma^{AB} = 0$.

In other words, the change in $\phi$ incurred by adding $\mR \delta \phi$ to it is exactly the same as the change incurred by acting with a quasi-diffeomorphism corresponding to the vector field $\mX_{\delta \phi}$. 

For the special case of vector fields which can be expressed as $\sF[\alpha]$, these quasi-diffeomorphisms are also actual diffeomorphisms, with $f = \sL_1 \alpha$.

Using perturbiners, we have in this paper provided a natural set $\phi$'s which solve the equations of motion, as well as a set of $\delta \phi$'s which solve the linearized e.o.m.\! on the $\phi$ background. These are
\begin{align}
    \phi = \bigphi(\phi_1, \ldots, \phi_N )\, , \hspace{1 cm} \delta \phi = \bigphi(\phi_1, \ldots, \phi_N \rvert \phi_I ).
\end{align}
In theorem 3.1 of \cite{mypaper}, we prove that the action of $\mR$ on the above $\delta \phi$ is given by
\begin{equation}\label{R_z_sdg}
    \mR \bigphi(\phi_1, \ldots, \phi_N \rvert \phi_I ) = - z_I \bigphi(\phi_1, \ldots, \phi_N \rvert \phi_I ).
\end{equation}
This means that, denoting $\delta^{\Delta}_{\bn, n}$ as the action of the $\Lw$ algebra on the Plebański scalar, we have
\begin{equation}
    \mR (\delta^{\Delta}_{\bn, n}  \phi) = - \delta^{\Delta}_{\bn, n-1} \phi.
\end{equation}
Just as in SDYM, the recursion operator traverses the ``loop part'' of the loop algebra $\Lw$. Repeatedly acting with $\mR$, we generate a tower of variations
\begin{equation}
    \delta^{\Delta}_{\bn, n} \phi \hspace{0.25 cm} \overset{-\mR\;}{\longrightarrow} \hspace{0.25 cm} \delta^{\Delta}_{\bn, n-1} \phi \hspace{0.25 cm} \overset{-\mR\;}{\longrightarrow} \hspace{0.25 cm} \delta^{\Delta}_{\bn, n-2} \phi \hspace{0.25 cm} \overset{-\mR\;}{\longrightarrow} \hspace{0.25 cm}  \ldots .
\end{equation}

In section \ref{sec38}, we saw that the generators $\delta^{\Delta}_{\bn, (n=-\Delta-1)}$ and  $\delta^{\Delta}_{\bn, (n=-\Delta)}$ correspond to pure diffeo transformations, while $\delta^{\Delta}_{\bn, (n=-\Delta+2)}$ and $\delta^{\Delta}_{\bn, (n=-\Delta+1)}$ are trivial, and $\delta^{\Delta}_{\bn, (n>\Delta+2)}$ are zero. The tower of variations therefore looks like
\begin{equation}
\begin{aligned}
    &\begin{matrix}
        \scriptstyle \text{zero} \\
        {\scriptstyle (n = -\Delta + 3)}
    \end{matrix}
    \hspace{0.25 cm} \overset{-\mR\;}{\longrightarrow} \hspace{0.25 cm}
    \begin{matrix}
        \scriptstyle \text{trivial} \\
        {\scriptstyle (n = -\Delta + 2)}
    \end{matrix}
    \hspace{0.25 cm} \overset{-\mR\;}{\longrightarrow} \hspace{0.25 cm}
    \begin{matrix}
        \scriptstyle \text{trivial} \\
        {\scriptstyle (n = -\Delta +1)}
    \end{matrix}
    \hspace{0.25 cm} \overset{-\mR\;}{\longrightarrow} \hspace{0.25 cm}
     \begin{matrix}
        \scriptstyle \text{pure diffeo } \mathfrak{f} \\
        {\scriptstyle (n = -\Delta )}
    \end{matrix}
    \hspace{0.25 cm} \\& \hspace{0.75 cm} \overset{-\mR\;}{\longrightarrow}  \hspace{0.25 cm} \begin{matrix}
        \scriptstyle \text{pure diffeo } \myw_{\infty} \\
        {\scriptstyle (n = -\Delta -1 )}
    \end{matrix}
    \hspace{0.25 cm} \overset{-\mR\;}{\longrightarrow} \hspace{0.25 cm}
    \begin{matrix}
        \scriptstyle \text{not pure diffeo} \\
        {\scriptstyle (n = -\Delta -2 )}
    \end{matrix}
    \hspace{0.25 cm} \overset{-\mR\;}{\longrightarrow} \hspace{0.25 cm}
    \ldots  .
\end{aligned}
\end{equation}

\section*{Acknowledgements}

The author has benefited from enlightening conversations with Tim Adamo, Adam Ball, Roland Bittleston, Wei Bu, Eduardo Casali, Erin Crawley, Dan Freed, Alfredo Guevara, Elizabeth Himwich, Dan Kapec, Lionel Mason, Jacob McNamara, Walker Melton, Hirosi Ooguri, Yorgo Pano, Sabrina Pasterski, Andrea Puhm, Atul Sharma, David Skinner, Simone Speziale, Andrew Strominger, Ana-Maria Raclariu, Romain Ruzziconi Adam Tropper, and Tianli Wang. The author gratefully acknowledges support from the Sivian Fund at the Institute for Advanced Study and DOE grant DE-SC0009988, along with the Princeton Gravity Initiative at Princeton University.

\appendix

\section{A reader's manual for \cite{mypaper}}\label{app_manual}

In \cite{mypaper} we developed the theory for the perturbiner expansion of the Plebański scalar $\phi$ in self-dual gravity. In \cite{mypaper}, however, we defined the seed functions $\phi_i$ of the pertubiner expansion to be \textit{plane waves} while in this paper we defined them to be \textit{half-descendant modes}. 
\begin{equation}
\begin{aligned}
    \phi_i^{\text{in paper \cite{mypaper}}} &= \epsilon_i e^{i \omega_i q(\bz_i, z_i) \cdot X} &= \epsilon_i \exp( i \omega_i (u + z_i \bbz_i \bbu - \bbz_i w - z_i \bbw) )  \\
    \phi_i^{\text{in this paper}} &= \epsilon_i \phi^{\Delta_i}_{\bn_i, z_i} &= \epsilon_i (u - \bbw z_i )^{2 - \Delta_i - \bn_i} (w - \bbu z_i)^{\bn_i}\hspace{0.763cm}
\end{aligned}
\end{equation}
This may seem like a big change, but happily, all of the results and proofs we need to use in this paper work equally well for both plane waves and half-descendant modes. The reason turns out to be simple. All the relevant proofs we need to use in \cite{mypaper} only require that the seed functions satisfy the following two relations:
\begin{equation}
    \partial_\bu \phi_i = - z_i \partial_w \phi_i, \hspace{1 cm} \partial_{\bw} \phi_i = - z_i \partial_u \phi_i.
\end{equation}
Given that the above formula is satisfied for both the plane wave seed functions and the half-descendant mode seed functions, it turns out that all of the proofs from \cite{mypaper} directly port over to this paper.

We will now list all the results we need from \cite{mypaper}, where they are used in this paper, and where they are proven in \cite{mypaper}.

\begin{itemize}
\item The first fact we need is that the perturbiner expansion of the Plebański scalar can be written as a sum over marked tree graphs. This statement is equation \eqref{gravity_perturbiner} in this paper and Theorem 2.1 in \cite{mypaper}.
\item The second result we need is the recursive formula for the perturbiner expansion of the Plebański scalar. This statement is equation \eqref{sdg_recursive} in this paper and Theorem A.1 in \cite{mypaper}.
\item The third fact we need is that the recursion operator $\mR$, when acting on the linear perturbation of a perturbiner expansion corresponding to the insertion of a single extra seed function $\phi_I$ in the list of seed functions, simply multiplies the perturbation by $-z_I$. This statement is equation \eqref{R_z_sdg} in this paper and Theorem 3.1 in \cite{mypaper}.
\end{itemize}

We will now list all of the analogous results we need for self-dual Yang Mills, which all reside in Appendix C of \cite{mypaper}. The Chalmers-Siegel scalar seed functions in each paper are given by
\begin{equation}
\begin{aligned}
    \Phi_i^{\text{in paper \cite{mypaper}}} &= \epsilon_i \bT^{a_i} e^{i \omega_i q(\bz_i, z_i) \cdot X} &= \epsilon_i \bT^{a_i} \exp( i \omega_i (u + z_i \bbz_i \bbu - \bbz_i w - z_i \bbw) ) , \\
    \Phi_i^{\text{in this paper}} &= \epsilon_i \Phi^{\Delta_i}_{\bn_i, z_i} &= \epsilon_i \bT^{a_i} (u - \bbw z_i )^{1 - \Delta_i - \bn_i} (w - \bbu z_i)^{\bn_i}.\hspace{0.76cm}
\end{aligned}
\end{equation}
Both sets of seed functions satisfy
\begin{equation}
    \partial_\bu \Phi_i = - z_i \partial_w \Phi_i, \hspace{1 cm} \partial_{\bw} \Phi_i = - z_i \partial_u \Phi_i,
\end{equation}
and this is all that is needed to port the proofs of \cite{mypaper} into this paper.

\begin{itemize}
    \item The first fact we need is that the perturbiner expansion of the Chalmers-Siegel scalar can be written as a sum over color orderings. This statement is equation \eqref{psi_k} in this paper and Theorem C.1 in \cite{mypaper}.
    \item The second result we need is the recursive formula for the perturbiner expansion of the Chalmers-Siegel scalar. This statement is equation \eqref{sdym_recursive_1} in this paper and Theorem C.2 in \cite{mypaper}.
    \item The third fact we need is that the recursion operator $\mR$, when acting on the linear perturbation of a perturbiner expansion corresponding to the insertion of a single extra seed function $\Phi_I$, simply multiplies the perturbiner by $-z_I$. This statement is equation \eqref{Recursionsdymz} in this paper and Theorem C.3 in \cite{mypaper}.
\end{itemize}

\section{Computing the full-descendant modes}\label{app_full_descendant}

In this appendix we will derive the following explicit expression for the full-descendant integer mode in SDYM
\begin{equation}
\begin{aligned}
     \Phi^{\Delta,a}_{\bn, n}&= \oint \frac{dz}{2 \pi i} \frac{1}{z^{1 + n}} \bT^a (u - \bbw z )^{1 - \Delta - \bn} (w - \bbu z)^\bn \\
     &= \frac{1}{(1 - \Delta - n)!} \bT^a (- u \partial_{\bw} - w \partial_{\bu})^{1 - \Delta - n} \bbu^\bn \, ( - \bbw)^{1 - \Delta - \bn}
\end{aligned}
\end{equation}
and SDG
\begin{equation}
\begin{aligned}
    \phi^{\Delta}_{\bn, n}&= \oint \frac{dz}{2 \pi i} \frac{1}{z^{1 + n}} (u - \bbw z )^{2 - \Delta - \bn} (w - \bbu z)^\bn\\
    &= \frac{1}{(2 - \Delta - n)!} ( - u \partial_{\bw} - w \partial_{\bu} )^{2 - \Delta - n} \, \bbu^\bn \, ( - \bbw)^{2 - \Delta - \bn}.
\end{aligned}
\end{equation}
We will do this by showing the following integral equality holds:
\begin{equation}
    \begin{aligned}
        \frac{1}{2 \pi i} &\oint \frac{dz}{z^{1 + n}} (-w + \bbu z)^{\bn} (u - \bbw z)^{p-\bn} \\
        &= \frac{1}{(p-n)!} ( -u \partial_{\bw} - w \partial_{\bu} )^{p - n} (\bbu)^{\bn} (-\bbw)^{p - \bn}.
    \end{aligned}
\end{equation}
We now begin the computation. We start by noting the following differential identity,
\begin{equation}
    (\partial_{\bz})^\bn\Big\rvert_{\bz = 0} (a \bbz + b)^p = \frac{p!}{(p - \bbn)!} a^\bn b^{p - \bbn}
\end{equation}
which follows by the binomial theorem. We now define a modified differential operator which essentially just strips the factorials off of the above equation, by
\begin{equation}
    \widetilde{(\partial_{\bz})^\bn }\Big\rvert_{\bz = 0} (a \bbz + b)^p \equiv a^\bn \;  b^{p - \bn}.
\end{equation}
This differential operator can also be expressed with a countour integral
\begin{equation}
    \widetilde{(\partial_{\bz})^\bn }\Big\rvert_{\bz = 0} (a \bbz + b)^p = \lim_{\epsilon \to 0} \oint \frac{d \bbz}{\bbz^{1 + \bn}} \frac{\bbn! (p + \epsilon - \bbn)!}{(p + \epsilon)!} (a \bbz + b)^{p+\epsilon}
\end{equation}
if desired. This differential operator is convenient to define because it can be used to compute the half-descendant modes from the conformal primary modes via the identity
\begin{equation}
\begin{aligned}
    \widetilde{(\partial_{\bz})^\bn }\Big\rvert_{\bz = 0} \big( q(\bbz, z) \cdot X\big)^p &= \widetilde{(\partial_{\bz})^\bn }\Big\rvert_{\bz = 0}  \big(u + z \bbz \bbu - w \bbz - \bbw z\big)^p \\
    &= (-w + \bbu z)^{\bn} (u - \bbw z)^{p - \bn}.
\end{aligned}
\end{equation}
In any case, using this operator, we now perform the computation.
\begin{equation}
\begin{aligned}
    \frac{1}{2 \pi i} &\oint \frac{dz}{z^{1 + n}} (-w + \bbu z)^{\bn} (u - \bbw z)^{p-\bn} \\&= \widetilde{(\partial_{\bz})^\bn }\Big\rvert_{\bz = 0} \frac{1}{2 \pi i} \oint \frac{dz}{z^{1 + n}} \big(u + z \bbz \bbu - w \bbz - \bbw z\big)^p \\
    &= \widetilde{(\partial_{\bz})^\bn }\Big\rvert_{\bz = 0}  \; \frac{1}{n!} \; (\partial_z)^n \Big\rvert_{z = 0} \; \big(  z (\bbz \bbu- \bbw) + (u - w \bbz  ) \big)^p \\
    &= \widetilde{(\partial_{\bz})^\bn }\Big\rvert_{\bz = 0}  \; \frac{p!}{n! (p - n)!} (u - \bbz w)^{p - n} ( - \bbw + \bbu \bbz )^n \\
    &= \widetilde{(\partial_{\bz})^\bn }\Big\rvert_{\bz = 0}  \; \frac{1}{(p-n)!} ( -u \partial_{\bw} - w \partial_{\bu} )^{p - n} ( - \bbw + \bbu \bbz )^p  \\
    &= \frac{1}{(p-n)!} ( -u \partial_{\bw} - w \partial_{\bu} )^{p - n} (\bbu)^{\bn} (-\bbw)^{p - \bn} .
\end{aligned}
\end{equation}

\section{An example computation of an explicit $\Ls$ action}\label{app_example}

In order to showcase how one can explicitly compute $\Ls$ variations in SDYM, we shall provide an example computation here, computing two sequential $\Ls$ variations acting on the vacuum field $\Phi = 0$. We will also verify in this case that the commutation relation \eqref{Ls_alg_confirmation} holds, without appealing to the contour-deformation-based argument. We begin by rewriting equation \eqref{phi_spin_1_full_first} for the full-descendant modes here as
\begin{equation}
    \Phi^{\Delta,a}_{\bn,n} = \frac{1}{(1 - \Delta - n)!} (- \sL_1)^{1 - \Delta - n} \lambda^{\Delta,a}_\bn.
\end{equation}
From \eqref{new_formula_spin1}, a single $\Ls$ variation on the vacuum is given by
\begin{equation}
\begin{aligned}
    (1 + \delta^{\Delta,a}_{\bn, n} )\cdot (\Phi = 0) &=  \bigPhi( \epsilon_1 \oint \frac{dz}{2 \pi i} \frac{\Phi^{\Delta,a}_{\bn,z}}{z^{1 + n}} ) \\
    &= \epsilon_1 \oint \frac{dz}{2 \pi i} \frac{\Phi^{\Delta,a}_{\bn,z}}{z^{1 + n}}  \\
    &= \epsilon_1 \Phi^{\Delta,a}_{\bn, n}.
\end{aligned}
\end{equation}
The action of two $\Ls$ variations on the vacuum $\Phi = 0$ is given by
\begin{equation}\label{two_Ls_vacuum}
\begin{aligned}
    &(1 + \delta^{\Delta_2,b}_{\bn_2, n_2} ) (1 + \delta^{\Delta_1,a}_{\bn_1, n_1} ) \cdot (\Phi = 0) \\
    &= \bigPhi( \epsilon_1  \oint_{C_1} \frac{dz_1}{2 \pi i} \frac{\Phi^{\Delta_1, a}_{\bn_1, z_1}}{z_1^{1 + n_1}} , \epsilon_2 \oint_{C_2} \frac{dz_2}{2 \pi i} \frac{\Phi^{\Delta_2, b}_{\bn_2, z_2}}{z_2^{1 + n_2}} )\\
    &= \epsilon_1  \oint_{C_1} \frac{dz_1}{2 \pi i} \frac{\Phi^{\Delta_1, a}_{\bn_1, z_1}}{z_1^{1 + n_1}} +  \epsilon_2 \oint_{C_2} \frac{dz_2}{2 \pi i} \frac{\Phi^{\Delta_2, b}_{\bn_2, z_2}}{z_2^{1 + n_2}} + \frac{\epsilon_1 \epsilon_2 }{(2 \pi i)^2}\oint_{C_2} \frac{dz_2}{z_2^{1 + n_2}} \oint_{C_1} \frac{dz_1}{z_1^{1 + n_2}}  \frac{[\Phi^{\Delta_1, a}_{\bn_1, z_1} ,  \Phi^{\Delta_2, b}_{\bn_2, z_2}]}{z_1 - z_2}  \\
    &= \epsilon_1 \Phi^{\Delta_1,a}_{\bn_1,n_1}+ \epsilon_2 \Phi^{\Delta_2,b}_{\bn_2,n_2} - \frac{\epsilon_1 \epsilon_2}{(2 \pi i)^2} \oint_{C_2} \frac{dz_2}{z_2^{1 + n_2}}   \oint_{C_1} \frac{dz_1}{z_1^{1 + n_1}} \sum_{k = 0}^\infty \frac{1}{z_2} \frac{z_1^k}{z_2^k} [\Phi^{\Delta_1,a}_{\bn_1,z_1},\Phi^{\Delta_2,b}_{\bn_2,z_2}] \\
    &= \epsilon_1 \Phi^{\Delta_1,a}_{\bn_1,n_1}+ \epsilon_2 \Phi^{\Delta_2,b}_{\bn_2,n_2} - \epsilon_2 \epsilon_2 \sum_{k = 0}^{(1 - \Delta_2 - n_2)-1} [ \left( \oint_{C_1} \frac{dz_1}{2 \pi i} \frac{\Phi^{\Delta_1, a}_{\bn_1, z_1}}{z_1^{1 + n_1 - k}} \right), \left(\oint_{C_2} \frac{dz_2}{2 \pi i} \frac{\Phi^{\Delta_2, b}_{\bn_2, z_2}}{z_2^{2 + n_2 + k}}  \right) ] \\
    &= \epsilon_1 \Phi^{\Delta_1,a}_{\bn_1,n_1}+ \epsilon_2 \Phi^{\Delta_2,b}_{\bn_2,n_2} - \epsilon_1 \epsilon_2 \sum_{k = 0}^{(1 - \Delta_2 - n_2)-1} [\Phi^{\Delta_1,a}_{\bn_1,n_1-k},\Phi^{\Delta_2,b}_{\bn_2,n_2+k+1}] 
\end{aligned}
\end{equation}
where $C_1$ enclosed the origin and $C_2$ enclosed $C_1$. 

We will now use this formula to verify that our equation for the $\Ls$ commutator holds, using the elementary identity
\begin{equation}
\begin{aligned}
    [\delta^{\Delta_1,a}_{\bn_1, n_1}, \delta^{\Delta_2,b}_{\bn_2, n_2}]  &= (1 + \delta^{\Delta_1,a}_{\bn_1, n_1} ) (1 + \delta^{\Delta_2,b}_{\bn_2, n_2} ) - (1 + \delta^{\Delta_2,b}_{\bn_2, n_2} ) (1 + \delta^{\Delta_1,a}_{\bn_1, n_1} ) \, .
\end{aligned}
\end{equation}
For convenience in the following steps, we define
\begin{equation}
    m_1 \equiv 1 - \Delta_1 - n_1, \hspace{1 cm} m_2 \equiv 1 - \Delta_2 - n_2 \, .
\end{equation}
From \eqref{two_Ls_vacuum}, the commutator becomes
\begin{equation}
\begin{aligned}
    [\delta^{\Delta_1,a}_{\bn_1, n_1}, & \, \delta^{\Delta_2,b}_{\bn_2, n_2}] \cdot (\Phi = 0) = \epsilon_1 \epsilon_2 \left( \sum_{k = 0}^{m_2 - 1} [\Phi^{\Delta_1,a}_{\bn_1,n_1-k},\Phi^{\Delta_2,b}_{\bn_2,n_2+k+1}] - \sum_{k = 0}^{m_1 - 1} [\Phi^{\Delta_2,b}_{\bn_2,n_2-k},\Phi^{\Delta_1,a}_{\bn_1,n_1+k+1}] \right) \\
    &= \epsilon_1 \epsilon_2\sum_{k = -m_1}^{m_2 - 1} [\Phi^{\Delta_1,a}_{\bn_1,n_1-k},\Phi^{\Delta_2,b}_{\bn_2,n_2+k+1}] \\
    &=  \epsilon_1 \epsilon_2\sum_{k = 0}^{m_1 + m_2 - 1} \frac{1}{(m_1 + m_2 -1 -k )!} \frac{1}{k!} [(-\sL_1)^{m_1+m_2 -1 - k} \lambda^{\Delta_1,a}_{\bn_1}, (-\sL_1)^{k} \lambda^{\Delta_2,b}_{\bn_2} ] \\
    &= \epsilon_1 \epsilon_2\frac{1}{(m_1 + m_2 -1)!} (-\sL_1)^{m_1 + m_2 - 1} [\lambda^{\Delta_1,a}_{\bn_1},\lambda^{\Delta_2,b}_{\bn_2} ] \\
    &= \epsilon_1 \epsilon_2\frac{1}{(1 - (\Delta_1 + \Delta_2) - (n_1 + n_2))!} (-\sL_1)^{1 - (\Delta_1 + \Delta_2) - (n_1 + n_2)} f^{abc} \lambda^{\Delta_1 + \Delta_2 - 1,c}_{\bn_1 + \bn_2} \\
    &= \epsilon_1 \epsilon_2 \, f^{abc} \Phi^{\Delta_1 + \Delta_2 -1,c}_{\bn_1 + \bn_2, n_1 + n_2 -1} \textcolor{white}{\frac{1}{1}} \\ 
    &= f^{abc} \delta^{\Delta_1 + \Delta_2 - 1, c}_{\bn_1 + \bn_2, n_1 + n_2 + 1} \cdot (\Phi = 0).  \textcolor{white}{\frac{1}{1}}
\end{aligned}
\end{equation}
The above expression correctly reproduces the commutator \eqref{Ls_alg_confirmation}, as expected.

\section{Lorentz symmetry in lightcone gauge for SDYM and SDG}\label{app_lorentz}

While SDYM is manifestly Lorentz invariant, our gauge choice of $A_u = A_w = 0$ is not. Therefore, Lorentz symmetry is not manifest in the action $S_{\rm SDYM}(\bPhi, \Phi)$ because certain Lorentz transformation break the gauge choice.

First, it is worth finding which Lorentz symmetries are preserved in the action \eqref{Ssdym}. This can be done with the following observation. If we take the Lorentz generator $\ell_{-1}$ from \eqref{ell_matrices} and raise the indices, we find that its components are
\begin{equation}
    (\ell_{-1})^{wu} = -(\ell_{-1})^{uw} = \frac{1}{2}, \hspace{0.5 cm} \text{rest are }0.
\end{equation}
This means we can rewrite $S_{\rm SDYM}(\bPhi, \Phi)$ as
\begin{equation}
    S_{\rm SDYM}(\bPhi, \Phi) = \int d^4 x \Tr \bPhi (  \Box\, \Phi + (\ell_{-1})^{\mu \nu} [\partial_\mu \Phi, \partial_\nu \Phi ] ).
\end{equation}
In this form, it is clear that the generator $\ell_{-1}$ is ``singled out'' by our gauge choice. If one varies the fields by
\begin{align}\label{deltaM1}
    \delta_M \Phi = M_\mu^{\;\;\nu} X^\mu \partial_\nu \Phi, \hspace{1 cm} \delta_M \bPhi = M_\mu^{\;\;\nu} X^\mu \partial_\nu \bPhi
\end{align}
where $M$ is some anitsymmetric matrix, then one can show that
\begin{equation}
    \delta_M S_{\rm SDYM}(\Phi, \bPhi) = 0 \hspace{0.25 cm} \text{ if and only if } \hspace{0.25 cm} [\ell_{-1}, M] = 0.
\end{equation}
Therefore, \eqref{deltaM1} is a symmetry of $S_{\rm SDYM}$ if $M$ commutes with $\ell_{-1}$, which means that $\Bell_{-1}$, $\Bell_0$, $\Bell_1$, and $\ell_{-1}$ are all manifest Lorentz symmetries. In fact, a direct computation of the Lie derivatives \eqref{lie_der_A} yields
\begin{align}
    \mL_{-1} A[\Phi]_\mu = A[\sL_{-1} \Phi]_\mu, \hspace{1 cm} \bmL_n A[\Phi]_\mu = A[\bL_n \Phi]_\mu \text{ for } n = -1, 0, 1
\end{align}
so 4 of the 6 Lorentz symmetries are unbroken by this gauge.

For $\mL_0$ and $\mL_{1}$ Lorentz symmetry is encoded is a more complex way. For $\mL_0$,
\begin{equation}
    \mL_{0} A[\Phi]_\mu = A[\sL_{0} \Phi + \Phi]_\mu.
\end{equation}
In fact, if one accompanies $\delta_0 \Phi = \sL_0 \Phi + \Phi$ with $\delta_0 \bPhi = \sL_0 \bPhi - \bPhi$, this is another off-shell symmetry of the action.

The story is more involved for $\mL_1$, which breaks our gauge condition:
\begin{equation}
    \mL_{1} A[\Phi]_\mu = A[\sL_{1} \Phi]_\mu - 2 (\ell_0)_\mu^{\;\;\; \nu} \partial_\nu \Phi.
\end{equation}
In order to act by this Lorentz generator on $\Phi$, we need to subtract off a compensating gauge transformation which re-inforces our gauge $A_u = A_w = 0$. It turns out that this gauge parameter is $\Phi$ itself, which acts as
\begin{equation}
    \delta_{\Phi} A_\mu =  - \partial_\mu \Phi - [A_\mu , \Phi].
\end{equation}
In full,
\begin{align}
    (\mL_1 + \delta_\Phi) A[\Phi]_\mu = A[\delta_1 \Phi]_\mu \implies \begin{pmatrix}
        0 \\
        \sL_1 \partial_w \Phi - \partial_{\bu} \Phi - [\partial_w \Phi, \Phi] \\
        0 \\
        \sL_1 \partial_u \Phi - \partial_\bw \Phi - [\partial_u \Phi, \Phi]
    \end{pmatrix} = \begin{pmatrix}
        0 \\
        \partial_w \delta_1 \Phi \\
        0 \\
        \partial_u \delta_1 \Phi 
    \end{pmatrix}.
\end{align}
The above equation gives a pair of differential equations which can be used to solve for $\delta_1 \Phi$. However, these differential equations are compatible only if $\Phi$ is on-shell:
\begin{align}
    \partial_u (\partial_w \delta_1 \Phi ) - \partial_w (\partial_u \delta_1 \Phi) &= - 2 (\Box \Phi - [ \partial_u \Phi, \partial_w \Phi] ) = 0.
\end{align}
Therefore, $\delta_1 \Phi$ does not exist as a local off-shell Lorentz symmetry of the action.

In conclusion, only 5 of the 6 Lorentz generators are off-shell symmetries of the action while 1 Lorentz generator is an on-shell symmetry. The local symmetries are
\begin{alignat}{4}
    &\bar{\delta}_{-1} \Phi = \bL_{-1} \Phi && \quad \quad \bar{\delta}_{0} \Phi = \bL_{0} \Phi && \quad \quad \bar{\delta}_{1} \Phi = \bL_{1} \Phi  \\
    &\bar{\delta}_{-1} \bPhi = \bL_{-1} \bPhi && \quad \quad \bar{\delta}_{0} \bPhi = \bL_{0} \bPhi && \quad \quad \bar{\delta}_{1} \bPhi = \bL_{1} \bPhi \\
    &\delta_{-1} \Phi = \sL_{-1} \Phi && \quad \quad \delta_{0} \Phi = \sL_{0} \Phi + \Phi && \\
    &\delta_{-1} \bPhi = \sL_{-1} \bPhi && \quad \quad \delta_{0} \bPhi = \sL_{0} \bPhi - \bPhi. && \quad \quad 
\end{alignat}

The story is very similar for the metric field in SDG. The $\bmL_{-1}$, $\bmL_0$, $\bmL_1$, and $\mL_{-1}$ generators \eqref{lie_der_2} are unbroken by the gauge choice \eqref{h_phi}:
\begin{align}
    \mL_{-1} h[\phi]_{\mu \nu} = h[\sL_{-1} \phi]_{\mu \nu}, \hspace{1 cm} \bmL_n h[\phi]_{\mu \nu} = h[\bL_n \phi]_{\mu \nu} \text{ for } n = -1, 0, 1.
\end{align}
The Lorentz generator $\mL_0$ acts via
\begin{equation}
    \mL_{0} h[\phi]_{\mu \nu} = h[\sL_{0} \phi + 2\phi]_{\mu \nu}.
\end{equation}
All of the 5 off-shell symmetries of the action \eqref{Ssdg} are
\begin{alignat}{4}
    &\bar{\delta}_{-1} \phi = \bL_{-1} \phi && \quad \quad \bar{\delta}_{0} \phi = \bL_{0} \phi && \quad \quad \bar{\delta}_{1} \phi = \bL_{1} \phi  \\
    &\bar{\delta}_{-1} \bphi = \bL_{-1} \bphi && \quad \quad \bar{\delta}_{0} \bphi = \bL_{0} \bphi && \quad \quad \bar{\delta}_{1} \bphi = \bL_{1} \bphi \\
    &\delta_{-1} \phi = \sL_{-1} \phi && \quad \quad \delta_{0} \phi = \sL_{0} \phi +2\phi && \\
    &\delta_{-1} \bphi = \sL_{-1} \bphi && \quad \quad \delta_{0} \bphi = \sL_{0} \bphi -2\bphi \, . && \quad \quad 
\end{alignat}
The 6th Lorentz generator, $\mL_1$, breaks our choice of gauge:
\begin{equation}
    \mL_1 h[\phi]_{\mu \nu} = h[\sL_1 \phi]_{\mu \nu} + 8 \partial_{(\mu} {\ell_0}_{\nu)}^{\;\;\; \alpha} A[\phi]_\alpha -  16 \partial_\alpha {\ell_0}_{(\mu}^{\;\;\; \alpha} {A[\phi]}_{\nu)} - 4 \partial_{(\mu} A[\phi]_{\nu)}.
\end{equation}
It turns out we can restore the gauge choice by acting the diffeomorphism generated by $A[\phi]_\mu$, which we raise into $A[\phi]^\mu$ with the inverse metric $g[\phi]^{\mu \nu}$. Recalling the definition of the Lie derviative 
\begin{equation}
    \mL_\xi g_{\mu \nu} = \xi^\alpha \partial_\alpha g_{\mu \nu} + g_{\mu \alpha} \partial_\nu \xi^\alpha + g_{\alpha \nu} \partial_\mu \xi^\alpha
\end{equation}
we record the only four non-zero components of the following tensor
\begin{equation}
\begin{aligned}
    \tfrac{1}{4} (\mL_1 - \mL_{4 A[\phi]}) g[\phi]_{\bu \bu} &= \sL_1 \partial_w^2 \phi - 2 \partial_{\bu} \partial_w \phi + 2 \partial_w^3 \phi \partial_u \phi - 2 \partial_w \phi \partial_u \partial_w^2 \phi \\
    \tfrac{1}{4} (\mL_1 - \mL_{4 A[\phi]}) g[\phi]_{\bu \bw} &= \sL_1 \partial_u \partial_w \phi - \partial_u \partial_{\bu} \phi - \partial_w \partial_\bw \phi + 2 \partial_u \phi \partial_u \partial_w^2 \phi - 2 \partial_w \phi \partial_u^2 \partial_w \phi \\
    \tfrac{1}{4}  (\mL_1 - \mL_{4 A[\phi]}) g[\phi]_{\bw \bw} &= \sL_1 \partial_u^2 \phi - 2 \partial_u \partial_\bw \phi + 2 \partial_u \phi \partial_u^2 \partial_w \phi -2 \partial_w \phi \partial_u^3 \phi \, .
\end{aligned}
\end{equation}
$\delta_1 \phi$ is then defined by
\begin{equation}
    (\mL_1 - \mL_{4 A[\phi]}) g[\phi]_{\mu \nu} = g[ \delta_1 \phi]_{\mu \nu} 
\end{equation}
or
\begin{equation}
\begin{aligned}
    4 \partial_w^2 \delta_1 \phi &= (\mL_1 - \mL_{4 A[\phi]}) g[\phi]_{\bu \bu} \\
    4 \partial_u \partial_w \delta_1 \phi &= (\mL_1 - \mL_{4 A[\phi]}) g[\phi]_{\bu \bw} \\
    4 \partial_u^2 \delta_1 \phi &= (\mL_1 - \mL_{4 A[\phi]}) g[\phi]_{\bw \bw} \, .
\end{aligned}
\end{equation}
This is a system of three differential equations must be solved to obtain $\delta_1 \phi$. These differential equations are only compatible on-shell, just as in the spin-1 case:
\begin{align}
    \partial_u (\partial_w^2 \delta_1 \phi ) - \partial_w( \partial_u \partial_w \delta_1 \phi) &= -2 \partial_w ( \Box \phi - \{ \partial_u \phi, \partial_w \phi \} ) = 0 \\
    \partial_w (\partial_u^2 \delta_1 \phi ) - \partial_u( \partial_u \partial_w \delta_1 \phi) &= 2 \partial_u ( \Box \phi - \{ \partial_u \phi, \partial_w \phi \}  ) = 0 \, .
\end{align}
Therefore, $\mL_1$ is only an on-shell symmetry of the Plebański scalar, not an off-shell symmetry.

\bibliography{sd_bib.bib}
\bibliographystyle{jhep}

\end{document}

%% file: figures/pic_that_goes_hard.tex
\tikzset{every picture/.style={line width=0.75pt}} 

\begin{tikzpicture}[x=0.75pt,y=0.75pt,yscale=-1,xscale=1]

\draw   (119,74) -- (379,74) -- (379,222) -- (119,222) -- cycle ;
\draw    (249,148) -- (253.03,132.9) ;
\draw [shift={(253.8,130)}, rotate = 104.93] [fill={rgb, 255:red, 0; green, 0; blue, 0 }  ][line width=0.08]  [draw opacity=0] (7.14,-3.43) -- (0,0) -- (7.14,3.43) -- (4.74,0) -- cycle    ;
\draw    (253.8,130) -- (263.24,120.88) ;
\draw [shift={(265.4,118.8)}, rotate = 136.01] [fill={rgb, 255:red, 0; green, 0; blue, 0 }  ][line width=0.08]  [draw opacity=0] (7.14,-3.43) -- (0,0) -- (7.14,3.43) -- (4.74,0) -- cycle    ;
\draw    (265.4,118.8) -- (280.73,115.61) ;
\draw [shift={(283.67,115)}, rotate = 168.25] [fill={rgb, 255:red, 0; green, 0; blue, 0 }  ][line width=0.08]  [draw opacity=0] (7.14,-3.43) -- (0,0) -- (7.14,3.43) -- (4.74,0) -- cycle    ;
\draw    (283.67,115) -- (300.85,121.3) ;
\draw [shift={(303.67,122.33)}, rotate = 200.14] [fill={rgb, 255:red, 0; green, 0; blue, 0 }  ][line width=0.08]  [draw opacity=0] (7.14,-3.43) -- (0,0) -- (7.14,3.43) -- (4.74,0) -- cycle    ;
\draw    (303.67,122.33) -- (312.72,135.72) ;
\draw [shift={(314.4,138.2)}, rotate = 235.92] [fill={rgb, 255:red, 0; green, 0; blue, 0 }  ][line width=0.08]  [draw opacity=0] (7.14,-3.43) -- (0,0) -- (7.14,3.43) -- (4.74,0) -- cycle    ;
\draw  [fill={rgb, 255:red, 0; green, 0; blue, 0 }  ,fill opacity=1 ] (246.82,148) .. controls (246.82,146.79) and (247.79,145.82) .. (249,145.82) .. controls (250.21,145.82) and (251.18,146.79) .. (251.18,148) .. controls (251.18,149.21) and (250.21,150.18) .. (249,150.18) .. controls (247.79,150.18) and (246.82,149.21) .. (246.82,148) -- cycle ;
\draw  [fill={rgb, 255:red, 0; green, 0; blue, 0 }  ,fill opacity=1 ] (312.22,138.2) .. controls (312.22,136.99) and (313.19,136.02) .. (314.4,136.02) .. controls (315.61,136.02) and (316.58,136.99) .. (316.58,138.2) .. controls (316.58,139.41) and (315.61,140.38) .. (314.4,140.38) .. controls (313.19,140.38) and (312.22,139.41) .. (312.22,138.2) -- cycle ;

\draw (250.39,65.33) node [anchor=south] [inner sep=0.75pt]   [align=left] {Space of Self-Dual Spacetimes};
\draw (249,154.9) node [anchor=north] [inner sep=0.75pt]    {$\phi =0$};

\end{tikzpicture}

%% file: figures/pic_that_goes_hard2.tex
\tikzset{every picture/.style={line width=0.75pt}} 

\begin{tikzpicture}[x=0.75pt,y=0.75pt,yscale=-1,xscale=1]

\draw   (182,53) -- (442,53) -- (442,201) -- (182,201) -- cycle ;
\draw    (312,137) -- (316.03,121.9) ;
\draw [shift={(316.8,119)}, rotate = 104.93] [fill={rgb, 255:red, 0; green, 0; blue, 0 }  ][line width=0.08]  [draw opacity=0] (7.14,-3.43) -- (0,0) -- (7.14,3.43) -- (4.74,0) -- cycle    ;
\draw    (316.8,119) -- (326.24,109.88) ;
\draw [shift={(328.4,107.8)}, rotate = 136.01] [fill={rgb, 255:red, 0; green, 0; blue, 0 }  ][line width=0.08]  [draw opacity=0] (7.14,-3.43) -- (0,0) -- (7.14,3.43) -- (4.74,0) -- cycle    ;
\draw    (328.4,107.8) -- (343.73,104.61) ;
\draw [shift={(346.67,104)}, rotate = 168.25] [fill={rgb, 255:red, 0; green, 0; blue, 0 }  ][line width=0.08]  [draw opacity=0] (7.14,-3.43) -- (0,0) -- (7.14,3.43) -- (4.74,0) -- cycle    ;
\draw  [fill={rgb, 255:red, 0; green, 0; blue, 0 }  ,fill opacity=1 ] (309.82,137) .. controls (309.82,135.79) and (310.79,134.82) .. (312,134.82) .. controls (313.21,134.82) and (314.18,135.79) .. (314.18,137) .. controls (314.18,138.21) and (313.21,139.18) .. (312,139.18) .. controls (310.79,139.18) and (309.82,138.21) .. (309.82,137) -- cycle ;
\draw  [fill={rgb, 255:red, 0; green, 0; blue, 0 }  ,fill opacity=1 ] (344.48,104) .. controls (344.48,102.79) and (345.46,101.82) .. (346.67,101.82) .. controls (347.87,101.82) and (348.85,102.79) .. (348.85,104) .. controls (348.85,105.21) and (347.87,106.18) .. (346.67,106.18) .. controls (345.46,106.18) and (344.48,105.21) .. (344.48,104) -- cycle ;
\draw    (346.67,104) -- (360.72,92.57) ;
\draw [shift={(363.05,90.67)}, rotate = 140.87] [fill={rgb, 255:red, 0; green, 0; blue, 0 }  ][line width=0.08]  [draw opacity=0] (7.14,-3.43) -- (0,0) -- (7.14,3.43) -- (4.74,0) -- cycle    ;
\draw    (363.05,90.67) -- (379.7,92.67) ;
\draw [shift={(382.67,93.03)}, rotate = 186.84] [fill={rgb, 255:red, 0; green, 0; blue, 0 }  ][line width=0.08]  [draw opacity=0] (7.14,-3.43) -- (0,0) -- (7.14,3.43) -- (4.74,0) -- cycle    ;
\draw    (346.67,104) -- (365.25,111.4) ;
\draw [shift={(368.04,112.51)}, rotate = 201.71] [fill={rgb, 255:red, 0; green, 0; blue, 0 }  ][line width=0.08]  [draw opacity=0] (7.14,-3.43) -- (0,0) -- (7.14,3.43) -- (4.74,0) -- cycle    ;
\draw    (382.67,93.03) -- (385,100.85) ;
\draw [shift={(385.85,103.73)}, rotate = 253.47] [fill={rgb, 255:red, 0; green, 0; blue, 0 }  ][line width=0.08]  [draw opacity=0] (7.14,-3.43) -- (0,0) -- (7.14,3.43) -- (4.74,0) -- cycle    ;
\draw    (368.04,112.51) -- (383.16,105.05) ;
\draw [shift={(385.85,103.73)}, rotate = 153.76] [fill={rgb, 255:red, 0; green, 0; blue, 0 }  ][line width=0.08]  [draw opacity=0] (7.14,-3.43) -- (0,0) -- (7.14,3.43) -- (4.74,0) -- cycle    ;

\draw (313.39,44.33) node [anchor=south] [inner sep=0.75pt]   [align=left] {Space of Self-Dual Spacetimes};
\draw (312,143.9) node [anchor=north] [inner sep=0.75pt]    {$\phi =0$};
\draw (347.96,89.69) node  [font=\scriptsize]  {$\delta _{1}$};
\draw (383.46,117.94) node  [font=\scriptsize]  {$\delta _{1}$};
\draw (354.46,118.69) node  [font=\scriptsize]  {$\delta _{2}$};
\draw (372.46,81.69) node  [font=\scriptsize]  {$\delta _{2}$};
\draw (387.85,99.33) node [anchor=south west] [inner sep=0.75pt]  [font=\scriptsize]  {$[ \delta _{1} ,\delta _{2}]$};
\draw (114,234.94) node [anchor=west] [inner sep=0.75pt]  [font=\normalsize]  {$\left[ \delta _{\overline{n}_{1} n_{1}}^{\Delta _{1}} ,\delta _{\overline{n}_{2} n_{2}}^{\Delta _{2}}\right] =( \bar{n}_{2}( 2-\Delta _{1}) -\bar{n}_{1}( 2-\Delta _{2})) \ \delta _{\overline{n}_{1} +\overline{n}_{2} -1,\ n_{1} +n_{2} +1}^{\Delta _{1} +\Delta _{2}}$};

\end{tikzpicture}

%% file: figures/gravity_graph_2.tex
\tikzset{every picture/.style={line width=0.75pt}} 

\begin{tikzpicture}[x=0.75pt,y=0.75pt,yscale=-1,xscale=1]

\draw    (295.4,83.7) -- (344.6,83.7) ;
\draw    (344.6,83.7) -- (387.4,55.7) ;
\draw    (24.1,133.5) -- (62.1,120.5) ;
\draw    (62.1,120.5) -- (97.8,135.6) ;
\draw  [fill={rgb, 255:red, 255; green, 255; blue, 255 }  ,fill opacity=1 ] (28.2,37.7) .. controls (28.2,30.8) and (33.8,25.2) .. (40.7,25.2) .. controls (47.6,25.2) and (53.2,30.8) .. (53.2,37.7) .. controls (53.2,44.6) and (47.6,50.2) .. (40.7,50.2) .. controls (33.8,50.2) and (28.2,44.6) .. (28.2,37.7) -- cycle ;
\draw  [fill={rgb, 255:red, 255; green, 255; blue, 255 }  ,fill opacity=1 ] (11.6,133.5) .. controls (11.6,126.6) and (17.2,121) .. (24.1,121) .. controls (31,121) and (36.6,126.6) .. (36.6,133.5) .. controls (36.6,140.4) and (31,146) .. (24.1,146) .. controls (17.2,146) and (11.6,140.4) .. (11.6,133.5) -- cycle ;
\draw  [fill={rgb, 255:red, 255; green, 255; blue, 255 }  ,fill opacity=1 ] (49.6,120.5) .. controls (49.6,113.6) and (55.2,108) .. (62.1,108) .. controls (69,108) and (74.6,113.6) .. (74.6,120.5) .. controls (74.6,127.4) and (69,133) .. (62.1,133) .. controls (55.2,133) and (49.6,127.4) .. (49.6,120.5) -- cycle ;
\draw  [fill={rgb, 255:red, 255; green, 255; blue, 255 }  ,fill opacity=1 ] (85.3,135.6) .. controls (85.3,128.7) and (90.9,123.1) .. (97.8,123.1) .. controls (104.7,123.1) and (110.3,128.7) .. (110.3,135.6) .. controls (110.3,142.5) and (104.7,148.1) .. (97.8,148.1) .. controls (90.9,148.1) and (85.3,142.5) .. (85.3,135.6) -- cycle ;
\draw    (295.4,83.7) -- (250.5,83.7) ;
\draw    (250.5,83.7) -- (221.4,110.1) ;
\draw    (221.4,51.2) -- (250.5,83.7) ;
\draw  [fill={rgb, 255:red, 255; green, 255; blue, 255 }  ,fill opacity=1 ] (208.9,51.2) .. controls (208.9,44.3) and (214.5,38.7) .. (221.4,38.7) .. controls (228.3,38.7) and (233.9,44.3) .. (233.9,51.2) .. controls (233.9,58.1) and (228.3,63.7) .. (221.4,63.7) .. controls (214.5,63.7) and (208.9,58.1) .. (208.9,51.2) -- cycle ;
\draw  [fill={rgb, 255:red, 255; green, 255; blue, 255 }  ,fill opacity=1 ] (238,83.7) .. controls (238,76.8) and (243.6,71.2) .. (250.5,71.2) .. controls (257.4,71.2) and (263,76.8) .. (263,83.7) .. controls (263,90.6) and (257.4,96.2) .. (250.5,96.2) .. controls (243.6,96.2) and (238,90.6) .. (238,83.7) -- cycle ;
\draw  [fill={rgb, 255:red, 255; green, 255; blue, 255 }  ,fill opacity=1 ] (208.9,110.1) .. controls (208.9,103.2) and (214.5,97.6) .. (221.4,97.6) .. controls (228.3,97.6) and (233.9,103.2) .. (233.9,110.1) .. controls (233.9,117) and (228.3,122.6) .. (221.4,122.6) .. controls (214.5,122.6) and (208.9,117) .. (208.9,110.1) -- cycle ;
\draw    (295.4,83.7) -- (326.2,118) ;
\draw    (327.3,46.9) -- (295.4,83.7) ;
\draw  [fill={rgb, 255:red, 255; green, 255; blue, 255 }  ,fill opacity=1 ] (282.9,83.7) .. controls (282.9,76.8) and (288.5,71.2) .. (295.4,71.2) .. controls (302.3,71.2) and (307.9,76.8) .. (307.9,83.7) .. controls (307.9,90.6) and (302.3,96.2) .. (295.4,96.2) .. controls (288.5,96.2) and (282.9,90.6) .. (282.9,83.7) -- cycle ;
\draw  [fill={rgb, 255:red, 255; green, 255; blue, 255 }  ,fill opacity=1 ] (314.8,46.9) .. controls (314.8,40) and (320.4,34.4) .. (327.3,34.4) .. controls (334.2,34.4) and (339.8,40) .. (339.8,46.9) .. controls (339.8,53.8) and (334.2,59.4) .. (327.3,59.4) .. controls (320.4,59.4) and (314.8,53.8) .. (314.8,46.9) -- cycle ;
\draw  [fill={rgb, 255:red, 255; green, 255; blue, 255 }  ,fill opacity=1 ] (332.1,83.7) .. controls (332.1,76.8) and (337.7,71.2) .. (344.6,71.2) .. controls (351.5,71.2) and (357.1,76.8) .. (357.1,83.7) .. controls (357.1,90.6) and (351.5,96.2) .. (344.6,96.2) .. controls (337.7,96.2) and (332.1,90.6) .. (332.1,83.7) -- cycle ;
\draw  [fill={rgb, 255:red, 255; green, 255; blue, 255 }  ,fill opacity=1 ] (313.7,118) .. controls (313.7,111.1) and (319.3,105.5) .. (326.2,105.5) .. controls (333.1,105.5) and (338.7,111.1) .. (338.7,118) .. controls (338.7,124.9) and (333.1,130.5) .. (326.2,130.5) .. controls (319.3,130.5) and (313.7,124.9) .. (313.7,118) -- cycle ;
\draw  [fill={rgb, 255:red, 255; green, 255; blue, 255 }  ,fill opacity=1 ] (374.9,55.7) .. controls (374.9,48.8) and (380.5,43.2) .. (387.4,43.2) .. controls (394.3,43.2) and (399.9,48.8) .. (399.9,55.7) .. controls (399.9,62.6) and (394.3,68.2) .. (387.4,68.2) .. controls (380.5,68.2) and (374.9,62.6) .. (374.9,55.7) -- cycle ;

\draw (55.2,37.7) node [anchor=west] [inner sep=0.75pt]    {$\ \in \mathcal{T}_{9}$};
\draw (40.7,37.7) node    {$3$};
\draw (62.1,120.5) node    {$1$};
\draw (97.8,135.6) node    {$9$};
\draw (24.1,133.5) node    {$5$};
\draw (108.4,128.1) node [anchor=west] [inner sep=0.75pt]    {$\ \in \mathcal{T}_{9}$};
\draw (326.2,118) node    {$9$};
\draw (327.3,46.9) node    {$2$};
\draw (221.4,51.2) node    {$7$};
\draw (250.5,83.7) node    {$3$};
\draw (221.4,110.1) node    {$5$};
\draw (295.4,83.7) node    {$8$};
\draw (344.6,83.7) node    {$6$};
\draw (387.4,55.7) node    {$4$};
\draw (396,85.7) node [anchor=west] [inner sep=0.75pt]    {$\ \in \mathcal{T}_{9}$};

\end{tikzpicture}

%% file: figures/gravity_graph_18.tex
\tikzset{every picture/.style={line width=0.75pt}} 

\begin{tikzpicture}[x=0.75pt,y=0.75pt,yscale=-1,xscale=1]

\draw    (47,43.2) -- (84,43.2) ;
\draw   (22,43.2) .. controls (22,36.3) and (27.6,30.7) .. (34.5,30.7) .. controls (41.4,30.7) and (47,36.3) .. (47,43.2) .. controls (47,50.1) and (41.4,55.7) .. (34.5,55.7) .. controls (27.6,55.7) and (22,50.1) .. (22,43.2) -- cycle ;
\draw   (84,43.2) .. controls (84,36.3) and (89.6,30.7) .. (96.5,30.7) .. controls (103.4,30.7) and (109,36.3) .. (109,43.2) .. controls (109,50.1) and (103.4,55.7) .. (96.5,55.7) .. controls (89.6,55.7) and (84,50.1) .. (84,43.2) -- cycle ;

\draw (140,43.2) node [anchor=west] [inner sep=0.75pt]    {$\ \ \dfrac{D^{ij}}{z_{ij}} \phi _{i} \phi _{j}$};
\draw (34.5,43.2) node    {$i$};
\draw (96.5,43.2) node    {$j$};

\end{tikzpicture}

%% file: figures/gravity_graph_3.tex
\tikzset{every picture/.style={line width=0.75pt}} 

\begin{tikzpicture}[x=0.75pt,y=0.75pt,yscale=-1,xscale=1]

\draw    (221.2,111.81) -- (256.7,89.71) ;
\draw    (185.7,89.71) -- (221.2,111.81) ;
\draw    (185.7,89.71) -- (185.7,48.61) ;
\draw    (150.2,111.81) -- (185.7,89.71) ;
\draw  [fill={rgb, 255:red, 255; green, 255; blue, 255 }  ,fill opacity=1 ] (208.7,111.81) .. controls (208.7,104.91) and (214.3,99.31) .. (221.2,99.31) .. controls (228.1,99.31) and (233.7,104.91) .. (233.7,111.81) .. controls (233.7,118.71) and (228.1,124.31) .. (221.2,124.31) .. controls (214.3,124.31) and (208.7,118.71) .. (208.7,111.81) -- cycle ;
\draw  [fill={rgb, 255:red, 255; green, 255; blue, 255 }  ,fill opacity=1 ] (173.2,89.71) .. controls (173.2,82.81) and (178.8,77.21) .. (185.7,77.21) .. controls (192.6,77.21) and (198.2,82.81) .. (198.2,89.71) .. controls (198.2,96.61) and (192.6,102.21) .. (185.7,102.21) .. controls (178.8,102.21) and (173.2,96.61) .. (173.2,89.71) -- cycle ;
\draw  [fill={rgb, 255:red, 255; green, 255; blue, 255 }  ,fill opacity=1 ] (137.7,111.81) .. controls (137.7,104.91) and (143.3,99.31) .. (150.2,99.31) .. controls (157.1,99.31) and (162.7,104.91) .. (162.7,111.81) .. controls (162.7,118.71) and (157.1,124.31) .. (150.2,124.31) .. controls (143.3,124.31) and (137.7,118.71) .. (137.7,111.81) -- cycle ;
\draw  [fill={rgb, 255:red, 255; green, 255; blue, 255 }  ,fill opacity=1 ] (173.2,48.61) .. controls (173.2,41.71) and (178.8,36.11) .. (185.7,36.11) .. controls (192.6,36.11) and (198.2,41.71) .. (198.2,48.61) .. controls (198.2,55.51) and (192.6,61.11) .. (185.7,61.11) .. controls (178.8,61.11) and (173.2,55.51) .. (173.2,48.61) -- cycle ;
\draw  [fill={rgb, 255:red, 255; green, 255; blue, 255 }  ,fill opacity=1 ] (244.2,89.71) .. controls (244.2,82.81) and (249.8,77.21) .. (256.7,77.21) .. controls (263.6,77.21) and (269.2,82.81) .. (269.2,89.71) .. controls (269.2,96.61) and (263.6,102.21) .. (256.7,102.21) .. controls (249.8,102.21) and (244.2,96.61) .. (244.2,89.71) -- cycle ;

\draw (185.7,89.71) node    {$1$};
\draw (107,84.5) node [anchor=west] [inner sep=0.75pt]    {$\mathbf{t} =$};
\draw (150.2,111.81) node    {$2$};
\draw (185.7,48.61) node    {$3$};
\draw (221.2,111.81) node    {$4$};
\draw (358.5,84.5) node [anchor=west] [inner sep=0.75pt]    {$\phi _{\mathbf{t}} =\dfrac{D^{12}}{z_{12}}\dfrac{D^{13}}{z_{13}}\dfrac{D^{14}}{z_{14}}\dfrac{D^{45}}{z_{45}} \phi _{1} \phi _{2} \phi _{3} \phi _{4} \phi _{5}.$};
\draw (256.7,89.71) node    {$5$};

\end{tikzpicture}

%% file: figures/three_graphs.tex
\tikzset{every picture/.style={line width=0.75pt}} 

\begin{tikzpicture}[x=0.75pt,y=0.75pt,yscale=-1,xscale=1]

\draw    (292,61.41) -- (256.21,24.7) ;
\draw    (291,165.91) -- (323,203.7) ;
\draw    (261,204.63) -- (291,165.91) ;
\draw    (455,184.7) -- (499,156.7) ;
\draw    (506,206.7) -- (499,156.7) ;
\draw    (549,159.7) -- (499,156.7) ;
\draw    (499,156.7) -- (485,110.7) ;
\draw    (501,61.7) -- (476,21.7) ;
\draw    (485,110.7) -- (501,61.7) ;
\draw    (501,61.7) -- (538,32.7) ;
\draw    (298,114.7) -- (348,108.7) ;
\draw    (291,165.91) -- (298,114.7) ;
\draw    (292,61.41) -- (322,22.7) ;
\draw    (298,114.7) -- (292,61.41) ;
\draw    (248,111.91) -- (298,114.7) ;
\draw  [fill={rgb, 255:red, 255; green, 255; blue, 255 }  ,fill opacity=1 ] (235.5,111.91) .. controls (235.5,105.01) and (241.1,99.41) .. (248,99.41) .. controls (254.9,99.41) and (260.5,105.01) .. (260.5,111.91) .. controls (260.5,118.82) and (254.9,124.41) .. (248,124.41) .. controls (241.1,124.41) and (235.5,118.82) .. (235.5,111.91) -- cycle ;
\draw  [fill={rgb, 255:red, 255; green, 255; blue, 255 }  ,fill opacity=1 ] (285.5,114.7) .. controls (285.5,107.8) and (291.1,102.2) .. (298,102.2) .. controls (304.9,102.2) and (310.5,107.8) .. (310.5,114.7) .. controls (310.5,121.6) and (304.9,127.2) .. (298,127.2) .. controls (291.1,127.2) and (285.5,121.6) .. (285.5,114.7) -- cycle ;
\draw  [fill={rgb, 255:red, 255; green, 255; blue, 255 }  ,fill opacity=1 ] (335.5,108.7) .. controls (335.5,101.8) and (341.1,96.2) .. (348,96.2) .. controls (354.9,96.2) and (360.5,101.8) .. (360.5,108.7) .. controls (360.5,115.6) and (354.9,121.2) .. (348,121.2) .. controls (341.1,121.2) and (335.5,115.6) .. (335.5,108.7) -- cycle ;
\draw  [fill={rgb, 255:red, 255; green, 255; blue, 255 }  ,fill opacity=1 ] (279.5,61.41) .. controls (279.5,54.51) and (285.1,48.91) .. (292,48.91) .. controls (298.9,48.91) and (304.5,54.51) .. (304.5,61.41) .. controls (304.5,68.32) and (298.9,73.91) .. (292,73.91) .. controls (285.1,73.91) and (279.5,68.32) .. (279.5,61.41) -- cycle ;
\draw  [fill={rgb, 255:red, 255; green, 255; blue, 255 }  ,fill opacity=1 ] (278.5,165.91) .. controls (278.5,159.01) and (284.1,153.41) .. (291,153.41) .. controls (297.9,153.41) and (303.5,159.01) .. (303.5,165.91) .. controls (303.5,172.82) and (297.9,178.41) .. (291,178.41) .. controls (284.1,178.41) and (278.5,172.82) .. (278.5,165.91) -- cycle ;
\draw  [fill={rgb, 255:red, 255; green, 255; blue, 255 }  ,fill opacity=1 ] (243.71,24.7) .. controls (243.71,17.8) and (249.31,12.2) .. (256.21,12.2) .. controls (263.12,12.2) and (268.71,17.8) .. (268.71,24.7) .. controls (268.71,31.6) and (263.12,37.2) .. (256.21,37.2) .. controls (249.31,37.2) and (243.71,31.6) .. (243.71,24.7) -- cycle ;
\draw  [fill={rgb, 255:red, 255; green, 255; blue, 255 }  ,fill opacity=1 ] (68.5,107.7) .. controls (68.5,100.8) and (74.1,95.2) .. (81,95.2) .. controls (87.9,95.2) and (93.5,100.8) .. (93.5,107.7) .. controls (93.5,114.6) and (87.9,120.2) .. (81,120.2) .. controls (74.1,120.2) and (68.5,114.6) .. (68.5,107.7) -- cycle ;
\draw  [fill={rgb, 255:red, 255; green, 255; blue, 255 }  ,fill opacity=1 ] (472.5,110.7) .. controls (472.5,103.8) and (478.1,98.2) .. (485,98.2) .. controls (491.9,98.2) and (497.5,103.8) .. (497.5,110.7) .. controls (497.5,117.6) and (491.9,123.2) .. (485,123.2) .. controls (478.1,123.2) and (472.5,117.6) .. (472.5,110.7) -- cycle ;
\draw  [fill={rgb, 255:red, 255; green, 255; blue, 255 }  ,fill opacity=1 ] (488.5,61.7) .. controls (488.5,54.8) and (494.1,49.2) .. (501,49.2) .. controls (507.9,49.2) and (513.5,54.8) .. (513.5,61.7) .. controls (513.5,68.6) and (507.9,74.2) .. (501,74.2) .. controls (494.1,74.2) and (488.5,68.6) .. (488.5,61.7) -- cycle ;
\draw  [fill={rgb, 255:red, 255; green, 255; blue, 255 }  ,fill opacity=1 ] (525.5,32.7) .. controls (525.5,25.8) and (531.1,20.2) .. (538,20.2) .. controls (544.9,20.2) and (550.5,25.8) .. (550.5,32.7) .. controls (550.5,39.6) and (544.9,45.2) .. (538,45.2) .. controls (531.1,45.2) and (525.5,39.6) .. (525.5,32.7) -- cycle ;
\draw  [fill={rgb, 255:red, 255; green, 255; blue, 255 }  ,fill opacity=1 ] (463.5,21.7) .. controls (463.5,14.8) and (469.1,9.2) .. (476,9.2) .. controls (482.9,9.2) and (488.5,14.8) .. (488.5,21.7) .. controls (488.5,28.6) and (482.9,34.2) .. (476,34.2) .. controls (469.1,34.2) and (463.5,28.6) .. (463.5,21.7) -- cycle ;
\draw  [fill={rgb, 255:red, 255; green, 255; blue, 255 }  ,fill opacity=1 ] (486.5,156.7) .. controls (486.5,149.8) and (492.1,144.2) .. (499,144.2) .. controls (505.9,144.2) and (511.5,149.8) .. (511.5,156.7) .. controls (511.5,163.6) and (505.9,169.2) .. (499,169.2) .. controls (492.1,169.2) and (486.5,163.6) .. (486.5,156.7) -- cycle ;
\draw  [fill={rgb, 255:red, 255; green, 255; blue, 255 }  ,fill opacity=1 ] (442.5,184.7) .. controls (442.5,177.8) and (448.1,172.2) .. (455,172.2) .. controls (461.9,172.2) and (467.5,177.8) .. (467.5,184.7) .. controls (467.5,191.6) and (461.9,197.2) .. (455,197.2) .. controls (448.1,197.2) and (442.5,191.6) .. (442.5,184.7) -- cycle ;
\draw  [fill={rgb, 255:red, 255; green, 255; blue, 255 }  ,fill opacity=1 ] (493.5,206.7) .. controls (493.5,199.8) and (499.1,194.2) .. (506,194.2) .. controls (512.9,194.2) and (518.5,199.8) .. (518.5,206.7) .. controls (518.5,213.6) and (512.9,219.2) .. (506,219.2) .. controls (499.1,219.2) and (493.5,213.6) .. (493.5,206.7) -- cycle ;
\draw  [fill={rgb, 255:red, 255; green, 255; blue, 255 }  ,fill opacity=1 ] (536.5,159.7) .. controls (536.5,152.8) and (542.1,147.2) .. (549,147.2) .. controls (555.9,147.2) and (561.5,152.8) .. (561.5,159.7) .. controls (561.5,166.6) and (555.9,172.2) .. (549,172.2) .. controls (542.1,172.2) and (536.5,166.6) .. (536.5,159.7) -- cycle ;
\draw  [fill={rgb, 255:red, 255; green, 255; blue, 255 }  ,fill opacity=1 ] (309.5,22.7) .. controls (309.5,15.8) and (315.1,10.2) .. (322,10.2) .. controls (328.9,10.2) and (334.5,15.8) .. (334.5,22.7) .. controls (334.5,29.6) and (328.9,35.2) .. (322,35.2) .. controls (315.1,35.2) and (309.5,29.6) .. (309.5,22.7) -- cycle ;
\draw  [fill={rgb, 255:red, 255; green, 255; blue, 255 }  ,fill opacity=1 ] (248.5,204.63) .. controls (248.5,197.72) and (254.1,192.13) .. (261,192.13) .. controls (267.9,192.13) and (273.5,197.72) .. (273.5,204.63) .. controls (273.5,211.53) and (267.9,217.13) .. (261,217.13) .. controls (254.1,217.13) and (248.5,211.53) .. (248.5,204.63) -- cycle ;
\draw  [fill={rgb, 255:red, 255; green, 255; blue, 255 }  ,fill opacity=1 ] (310.5,203.7) .. controls (310.5,196.8) and (316.1,191.2) .. (323,191.2) .. controls (329.9,191.2) and (335.5,196.8) .. (335.5,203.7) .. controls (335.5,210.6) and (329.9,216.2) .. (323,216.2) .. controls (316.1,216.2) and (310.5,210.6) .. (310.5,203.7) -- cycle ;

\draw (247,108.91) node  [font=\tiny] [align=left] {\begin{minipage}[lt]{12.93pt}\setlength\topsep{0pt}
\begin{center}
pure\\ diff
\end{center}

\end{minipage}};
\draw (33.89,35.2) node    {$( 1)$};
\draw (200,36) node    {$( 2)$};
\draw (425,36) node    {$( 3)$};
\draw (484,107.7) node  [font=\tiny] [align=left] {\begin{minipage}[lt]{12.93pt}\setlength\topsep{0pt}
\begin{center}
pure\\ diff
\end{center}

\end{minipage}};
\draw (80,104.7) node  [font=\tiny] [align=left] {\begin{minipage}[lt]{12.93pt}\setlength\topsep{0pt}
\begin{center}
pure\\ diff
\end{center}

\end{minipage}};
\draw (298,114.7) node    {$i$};
\draw (501,61.7) node    {$i$};
\draw (499,156.7) node    {$j$};

\end{tikzpicture}